\documentclass[aps,amsmath,preprint,superscriptaddress,nofootinbib,prd]{revtex4-1}
\usepackage{graphicx}
\usepackage{bm}
\usepackage{color}
\usepackage{amsmath,amssymb}	
\usepackage{setspace}
\usepackage[english]{babel}
\usepackage[utf8x]{inputenc}
\usepackage{slashed}
\usepackage[normalem]{ulem}
\usepackage{mathtools}
\usepackage{enumerate}
\usepackage{multirow}
\usepackage{hyperref}
\usepackage{physics}

\newcommand{\ie}{{\it i.e.}}
\newcommand{\eg}{{\it e.g.}}

\newcommand{\Umt}{U(1)_{L_\mu-L_\tau}}

\begin{document}


\title{Muon \texorpdfstring{$g - 2$}{} and non-thermal leptogenesis in \texorpdfstring{$U(1)_{L_{\mu}-L_{\tau}}$}{} model}

\author{Shintaro~Eijima}
\email{eijima@icrr.u-tokyo.ac.jp}
\affiliation{ICRR, University of Tokyo, Kashiwa, Chiba 277-8582, Japan}

\author{Masahiro Ibe}
\email{ibe@icrr.u-tokyo.ac.jp}
\affiliation{ICRR, University of Tokyo, Kashiwa, Chiba 277-8582, Japan}
\affiliation{Kavli IPMU (WPI), UTIAS, University of Tokyo, Kashiwa, Chiba 277-8583, Japan}

\author{Kai Murai}
\email{kmurai@icrr.u-tokyo.ac.jp}
\affiliation{ICRR, University of Tokyo, Kashiwa, Chiba 277-8582, Japan}

\begin{abstract}
The gauged $U(1)_{L_{\mu}-L_{\tau}}$ symmetry is the simplest possibility to explain the observed muon $g-2$, while being consistent with the neutrino oscillations through the seesaw mechanism.
In this paper, we investigate if leptogenesis can work at the same time.
At first glance, leptogenesis seems challenging
because the right-handed neutrino masses 
are related to the $U(1)_{L_{\mu}-L_{\tau}}$ breaking scale of $10\,\text{--}\,100$\,GeV as required from the muon $g-2$.
Contrary to this expectation, we find that 
non-thermal leptogenesis with the right-handed neutrino masses of $\mathcal{O}(10^{7})$\,GeV 
is possible.
The successful scenario results in strict predictions on the neutrino oscillation parameters, which will be tested in future experiments.
\end{abstract}

\maketitle


\section{Introduction}
\label{sec:introduction}

The measurements of muon anomalous dipole moment at the Brookhaven National Laboratory~\cite{Muong-2:2006rrc} and at the Fermilab~\cite{Muong-2:2021ojo,Muong-2:2021vma} have reported the $4.2~\sigma$ deviation from the Standard Model (SM) prediction~\cite{Aoyama:2020ynm}.%
\footnote{%
For the theoretical prediction of the 
hadronic contributions to muon $g-2$, see 
Refs.~\cite{Aoyama:2012wk,Aoyama:2019ryr,Czarnecki:2002nt,Gnendiger:2013pva,Davier:2017zfy,Keshavarzi:2018mgv,Colangelo:2018mtw,Hoferichter:2019mqg,Davier:2019can,Keshavarzi:2019abf,Kurz:2014wya,Melnikov:2003xd,Masjuan:2017tvw,Colangelo:2017fiz,Hoferichter:2018kwz,Gerardin:2019vio,Bijnens:2019ghy,Colangelo:2019uex,Blum:2019ugy,Colangelo:2014qya}.
Currently, there are discrepancies between the data-driven approaches and the lattice simulations~(see Ref.~\cite{ParticleDataGroup:2022pth} and references therein).
It has been also reported that 
a recent data-driven analysis shows 
deviation from the conventional results~\cite{CMD-3:2023alj}.
}
One of the solutions to this anomaly is to introduce an extra, neutral gauge boson $Z^\prime$ associated with a gauged $U(1)$ symmetry.
Its contribution through the gauge interaction can enlarge the muon $g-2$.
Taking into account the results in search experiments of $Z'$, the simplest remaining possibility is the $\Umt$ symmetry~\cite{Foot:1990mn, He:1991qd, Foot:1994vd, Gninenko:2001hx, Baek:2001kca, Murakami:2001cs, Ma:2001md}.
See Ref.~\cite{Bauer:2018onh} for the parameter space. 

The gauge interaction of the $\Umt$ symmetry with leptons is given by
\begin{align}
    \label{eq:int_zp}
    \mathcal{L}_{Z'} 
    \supset 
    - g_{Z'} Q_{\alpha} Z'_\kappa 
    \left( 
          L^\dagger_{\alpha} \bar{\sigma}^\kappa L_{\alpha}
        - \bar{l}^\dagger_{R \alpha} \bar{\sigma}^\kappa \bar{l}_{R \alpha}
    \right),
\end{align}
where $g_{Z'}$ is the gauge coupling constant, and $Q_{e, \mu, \tau} \equiv (0, +1, -1)$ are the charge of the $\Umt$ symmetry for each flavor.
Here, $L_\alpha \equiv (\nu_{L \alpha}, l_{L \alpha})$ is the lepton doublet, and $\bar{l}_{R \alpha}$ is the singlet charged lepton with $\alpha = e, \mu, \tau$.
Note that all the fermions are described by left-handed Weyl fermions.
We follow the conventions of the spinor indices in Ref.~\cite{Dreiner:2008tw}.
The deviation of the muon $g-2$ can be explained for
$g_{Z'} \approx (3\,\text{--}\,10) \times 10^{-4}$ and the mass of $Z'$, $m_{Z'} \approx (1\,\text{--}\,20) \times 10~\text{MeV}$, while avoiding other experimental constraints~\cite{Bauer:2018onh}.
This region corresponds to the $\Umt$ breaking scale of $10 \,\text{--}\, 100\,\text{GeV}$.

Due to the $\Umt$ symmetry, neutrino oscillations can not occur.
Even if the symmetry is broken,
it is still non-trivial whether the observed neutrino oscillation parameters can be reproduced.
In Refs.\,\cite{Harigaya:2013twa,Asai:2017ryy,Asai:2018ocx} (see also Refs.~\cite{Heeck:2011wj,Araki:2012ip} for earlier works), it has been shown that the type-I seesaw mechanism~\cite{Minkowski:1977sc,Yanagida:1979as,Glashow:1979nm,Mohapatra:1979ia,GellMann:1980vs,Schechter:1980gr}
with three right-handed neutrinos
can explain the observed neutrino oscillations with $\Umt$ breaking scalar fields.
Thus, the gauged $\Umt$ symmetry can explain the muon $g-2$, while being consistent with the neutrino oscillations.

In this paper, we further investigate if leptogenesis can work while explaining the above two phenomena simultaneously.
One can naively expect some obstacles to leptogenesis in this model.
Firstly, masses of the right-handed neutrinos tend to be of the $\Umt$ breaking scale, $10 \,\text{--}\, 100$\,GeV, to reproduce the neutrino oscillations by the seesaw mechanism.  
With such light right-handed neutrinos, thermal leptogenesis~\cite{Fukugita:1986hr}, for example, cannot be achieved.
Secondly, in analogy to the electroweak symmetry breaking, the $\Umt$ symmetry seems to be restored in the early universe, and thus the universe is in the $\Umt$ symmetric phase before the freeze-out of sphaleron processes.
As we will see, leptogenesis does not work in the symmetric phase.
Therefore, to find a successful scenario of leptogenesis, 
we have to seek a setup satisfying the following conditions: 
\begin{itemize}
    \item Right-handed neutrinos have masses much larger than $10 \,\text{--}\, 100$\,GeV. 
    \item The $\Umt$ symmetry is broken even in the early universe.
\end{itemize}

We find that the first condition can be satisfied 
by a specific choice of Yukawa couplings of right-handed neutrinos to the $\Umt$ breaking fields,
which in turn results in strict predictions on the neutrino oscillation parameters.  
The second condition can also be satisfied by choosing certain couplings between the $\Umt$ breaking scalar fields and the SM Higgs boson.
Based on these outcomes, 
we demonstrate that non-thermal leptogenesis~\cite{Kumekawa:1994gx,Asaka:1999jb} can generate a sufficient amount of baryon asymmetry 
while explaining the muon $g-2$ and the neutrino oscillations at the same time.

The rest of the paper is organized as follows.
In Sec.~\ref{sec:model}, we introduce a model with the gauged $\Umt$ symmetry and review the seesaw mechanism. 
In Sec.~\ref{sec:mass spectrum},
we discuss how heavy the right-handed neutrinos can be.
In Sec.~\ref{sec:breaking},
we discuss restoration and breaking of the $\Umt$ symmetry in the early universe.
In Sec.~\ref{sec:leptogenesis},
we discuss a possibility of leptogenesis in this model.
Finally, we conclude this paper in Sec.~\ref{sec:conclusion}.

\section{Model with gauged \texorpdfstring{$\Umt$}{} symmetry}
\label{sec:model}

Let us start with the setup of the gauged $\Umt$ model
which reproduces the active neutrino mass parameters~\cite{Harigaya:2013twa,Asai:2017ryy} (see also Refs.~\cite{Heeck:2011wj,Araki:2012ip} for earlier works).
The $\Umt$ charge assignment for the doublet and the singlet leptons are given below Eq.\,\eqref{eq:int_zp}.
Three right-handed neutrinos are introduced to account for the neutrino oscillations via the type-I seesaw mechanism. 
They have the $\Umt$ charge as $(\bar{N}_e, \bar{N}_{\mu}, \bar{N}_{\tau}) = (0, -1, +1)$ in a natural way.
Note again that all the fermions are described by left-handed Weyl fermions.

To break the $\Umt$ symmetry, we introduce 
two SM singlet scalar bosons $\sigma_{1,2}$ with the $\Umt$ charge $+1$
and $+2$, respectively.
Note that the observed mixing angles among neutrinos can be reproduced by only $\sigma_1$~\cite{Harigaya:2013twa}. 
As we will see, however, 
$\sigma_2$ plays an important role in successful leptogenesis for the parameter region explaining the muon $g-2$.
Hereafter, we call the sector 
consisting of $Z'$ and $\sigma_{1,2}$ the $\Umt$ sector.
We summarize the phenomenological properties of the symmetry breaking sector in the Appendix~\ref{sec:symmetrybreakingsector}.

The Lagrangian relating to the neutrino masses is given by
\begin{align}
\label{eq:Lnu}
    \mathcal{L}_{\nu} 
    =&
    - y_L L_\alpha \Phi \bar{l}_{R \beta}
    - \lambda_\nu 
    L_\alpha 
    \tilde{\Phi} \bar{N}_{\beta} 
    - \frac{M_R}{2} \bar{N}_{\alpha} \bar{N}_{\beta}
    -h_{e \mu} \sigma_1 \bar{N}_{e} \bar{N}_{\mu}
    -h_{e \tau} \sigma_1^* \bar{N}_{e} \bar{N}_{\tau} 
    \nonumber\\
    &
    - \frac{1}{2} h_{\mu \mu} \sigma_2 \bar{N}_{\mu} \bar{N}_{\mu} 
    - \frac{1}{2} h_{\tau \tau} \sigma_2^* \bar{N}_{\tau} \bar{N}_{\tau} 
    + \mathrm{h.c.} 
    \ ,
\end{align}
where
$\tilde{\Phi} = \epsilon \Phi$ is the Higgs doublet with 
the $SU(2)$ antisymmetric tensor $\epsilon$.
Due to the $\Umt$ symmetry, the Dirac Yukawa coupling constants 
and the Majorana mass matrix for right-handed neutrinos become
\begin{equation}
    \label{eq:MR}
    y_L =
    \begin{pmatrix}
        y_e & 0 & 0 \\
        0 & y_\mu & 0 \\
        0 & 0 & y_\tau
    \end{pmatrix}
    , \quad
    \lambda_\nu =
    \begin{pmatrix}
        \lambda_e & 0 & 0 \\
        0 & \lambda_\mu & 0 \\
        0 & 0 & \lambda_\tau
    \end{pmatrix}
    , \quad
    M_{R} =
    \begin{pmatrix}
        M_{ee} & 0 & 0 \\
        0 & 0 & M_{\mu \tau} \\
        0 & M_{\mu \tau} & 0
    \end{pmatrix},
\end{equation}
where we choose $y_\alpha$, $\lambda_\alpha$, and $M_{\alpha \beta}$ are real and positive
by rotating the phases of $L$'s, $\bar{l}_R$'s, and $\bar{N}$'s.
We call $h_{\alpha \beta}$ Majorana Yukawa coupling constants.

\begin{figure}
    \centering
    \includegraphics{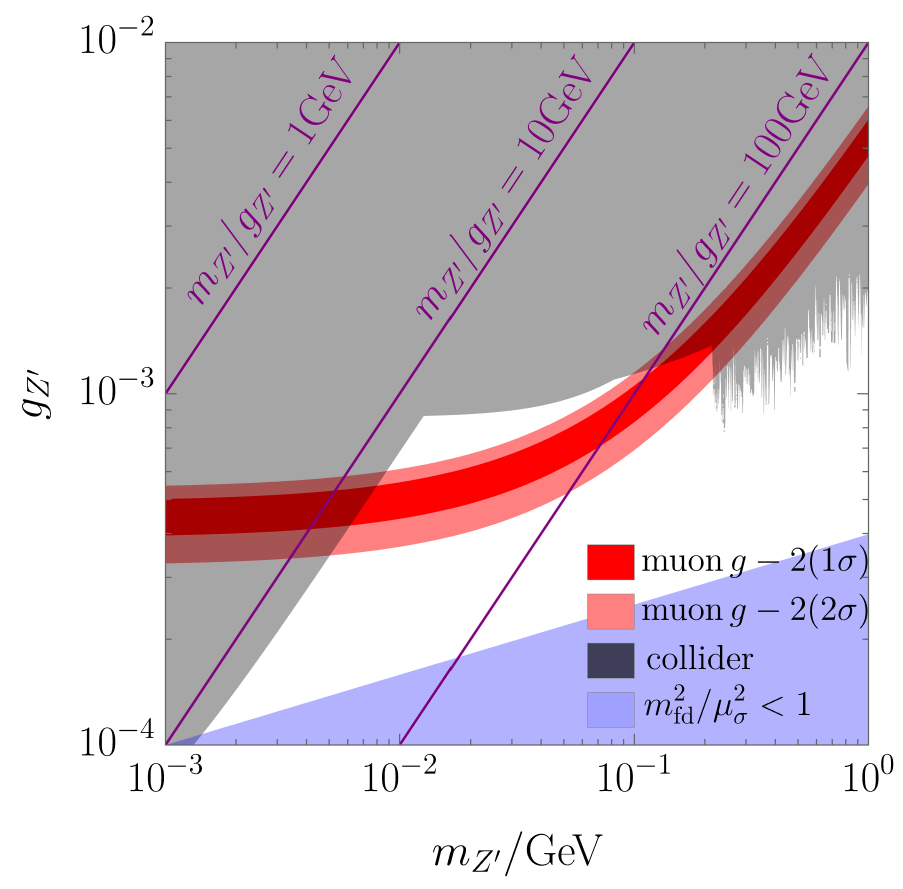}
    \caption{\sl Parameter region that explains the muon $g-2$ 
    within the $1\sigma$ (red) and the $2\sigma$ (pink) ranges~\cite{Muong-2:2021ojo}. 
    We show $m_{Z'}/g_{Z'}$ by 
    the purple lines, which indicate the size of the VEVs of the scalar fields.
    The gray shaded region is excluded by the neutrino trident production experiment~\cite{Altmannshofer:2014pba}, the neutrino-electron scattering experiments~\cite{Harnik:2012ni,Bilmis:2015lja}, and the experiment searching for $e^-e^+\to \mu^-\mu^+Z'$~\cite{BaBar:2016sci}.
    In the blue shaded region, the ratio $m_\mathrm{fd}^2/\mu_\sigma^2 < 1$ for $T=T_{\mathrm{th}}$ in Eq.\,\eqref{eq:comp} (see Sec.~\ref{subsec:restoration}).
    }    \label{fig:parameters}
\end{figure}

When the $\Umt$ charged scalar fields obtain non-vanishing expectation values, $\langle \sigma_{1,2} \rangle$, the $\Umt$ symmetry is spontaneously broken.
In the present model, both $\langle\sigma_{1,2}\rangle$ can be real positive (see the Appendix.~\ref{sec:symmetrybreakingsector}).
In this case, Eq.\,(\ref{eq:Lnu}) leads to the mass matrix of $\bar{N}_\alpha$,
\begin{align}
    \label{eq:MR2}
    M_{R, \mathrm{eff}} =
    \begin{pmatrix}
         M_{ee} & h_{e \mu} \langle \sigma_1 \rangle & h_{e \tau} \langle \sigma_1 \rangle \\
         h_{e \mu} \langle \sigma_1 \rangle & h_{\mu \mu} \langle \sigma_2 \rangle & M_{\mu \tau} \\
         h_{e \tau} \langle \sigma_1 \rangle & M_{\mu \tau} & h_{\tau \tau} \langle \sigma_2 \rangle
    \end{pmatrix}
    \ . 
\end{align}
The corresponding mass of the $Z'$ boson is given by,
\begin{align}
    m_{Z'}^2 =  2 g_{Z'}^2
    \left(\langle\sigma_1\rangle^2 + 4 \langle \sigma_2\rangle^2\right)\ .
\end{align}
To explain the deviation of the muon $g-2$, we require, at the vacuum,
\begin{align}
    \sqrt{\langle\sigma_1\rangle_0^2 + 4 \langle \sigma_2\rangle_0^2}
    \simeq
    10 \text{\,--\,} 100\,\mathrm{GeV}\ ,
\end{align}
(see Fig.\,\ref{fig:parameters}).
As we will discuss later, the temperature-dependent expectation values play important roles in successful leptogenesis.
Thus, we put the subscript $0$ on the vacuum expectation value (VEV)
to distinguish it from the temperature-dependent expectation value.
The complex scalar field $\sigma_2$ is absent in 
the minimal model in 
Refs.\,\cite{Harigaya:2013twa,Asai:2017ryy,Asai:2018ocx}.
In that case, the mass matrix 
is reduced to the 
so-called two-zero minor~\cite{Lavoura:2004tu, Lashin:2007dm}.

The active neutrino masses are given by the seesaw mechanism;
\begin{align}
    \label{eq:seesaw}
    M_\nu \simeq - M_D M_{R, \mathrm{eff}}^{-1} M_D^\mathrm{T}, 
\quad
    M_D \equiv \lambda_\nu v_\mathrm{EW}
    ,
\end{align}
where $v_\mathrm{EW}$ denotes the electroweak symmetry breaking scale, $\langle \Phi \rangle_0 = v_\mathrm{EW} \simeq 174$\,GeV.
Note that 
$\langle\sigma_{1,2}\rangle_0 \neq 0$ contribution is crucial to explain the observed neutrino oscillations since otherwise $M_{R, \mathrm{eff}}$ induces only a mixing angle $\theta_{23}$ between the active neutrinos.
As we will see in the next section, even for $\langle \sigma_{1, 2} \rangle_0 \neq 0$, 
the mass parameters $M_{ee}$ and $M_{\mu\tau}$ are severely constrained to reproduce the 
observed neutrino oscillations.

For later purposes, it is useful to consider
$\lambda_\nu^{-1} M_{R, \mathrm{eff}} \lambda_\nu^{-1}$, which is related to the low energy observables via,
\begin{align}
    \label{eq:seesaw general}
    \begin{pmatrix}
        \frac{M_{ee}}{\lambda_{e}^2} & \frac{h_{e \mu} \langle \sigma_1 \rangle_0}{\lambda_e \lambda_\mu} & \frac{h_{e \tau} \langle \sigma_1 \rangle_0}{\lambda_e \lambda_\tau} 
        \\
        \frac{h_{e \mu} \langle \sigma_1 \rangle_0}{\lambda_e \lambda_\mu} & \frac{h_{\mu \mu} \langle \sigma_2 \rangle_0}{\lambda_\mu^2} & \frac{M_{\mu \tau}}{\lambda_\mu \lambda_\tau} 
        \\
        \frac{h_{e \tau} \langle \sigma_1 \rangle_0}{\lambda_e \lambda_\tau} & \frac{M_{\mu \tau}}{\lambda_\mu \lambda_\tau} & \frac{h_{\tau \tau} \langle \sigma_2 \rangle_0}{\lambda_\tau^2}
        \\
    \end{pmatrix}
    =
    v_\mathrm{EW}^2 \times
    \mathrm{U}[ M_\nu^{d}]^{-1} \mathrm{U}^{\mathrm{T}}
    ,
\end{align}
where $M_\nu^d$ is a diagonal mass matrix defined by $M_\nu^d \equiv \mathrm{U}^\mathrm{T} M_\nu \mathrm{U}$ with the Pontecorvo-Maki-Nakagawa-Sakata (PMNS) matrix $\mathrm{U}$.
Here, $M_\nu$ is invertible, and thus all the active neutrinos become massive.
The PMNS matrix is represented by the mixing angles $\theta_{1 2}, \theta_{2 3}, \theta_{1 3}$, the Dirac CP phase $\delta$, and the Majorana CP phases $\eta_1, \eta_2$ as
\begin{align}
    \mathrm{U}
    &=
    \begin{pmatrix}
        1 &        0 &       0
        \\
        0 &  c_{2 3} & s_{2 3}
        \\
        0 & -s_{2 3} & c_{2 3}
    \end{pmatrix}
    \begin{pmatrix}
          c_{1 3}              & 0 & s_{1 3} e^{-i \delta}
        \\
                0              & 1 &       0
        \\
        - s_{1 3} e^{i \delta} & 0 & c_{1 3}
    \end{pmatrix}
    \begin{pmatrix}
          c_{1 2} & s_{1 2} & 0
        \\
        - s_{1 2} & c_{1 2} & 0
        \\
                0 &       0 & 1
    \end{pmatrix}
    \begin{pmatrix}
        1 &              0 &              0
        \\
        0 & e^{i \eta_1/2} &              0
        \\
        0 &              0 & e^{i \eta_2/2}
    \end{pmatrix}
    \nonumber \\
    &=
    \begin{pmatrix}
        c_{1 2} c_{1 3}
        &
        s_{1 2} c_{1 3}
        &
        s_{1 3} e^{-i \delta}
        \\
        - s_{1 2} c_{2 3} - c_{1 2} s_{2 3} s_{1 3} e^{i \delta}
        &
          c_{1 2} c_{2 3} -s_{1 2} s_{2 3} s_{1 3} e^{i \delta}
        &
        s_{2 3} c_{1 3}
        \\
          s_{1 2} s_{2 3} - c_{1 2} c_{2 3} s_{1 3} e^{i \delta}
        &
        - c_{1 2} s_{2 3} - s_{1 2} c_{2 3} s_{1 3} e^{i \delta}
        &
        c_{2 3} c_{1 3}
    \end{pmatrix}
    \begin{pmatrix}
        1 &              0 &              0
        \\
        0 & e^{i \eta_1/2} &              0
        \\
        0 &              0 & e^{i \eta_2/2}
    \end{pmatrix}
    \ ,
\end{align}
where $s_{i j}$ and $c_{i j}$ denote $\sin \theta_{i j}$ and $\cos \theta_{i j}$, respectively.%
\footnote{
The above PMNS matrix corresponds to the old PDG convention~\cite{ParticleDataGroup:2016lqr}.
}
The domains of the mixing angles are in $[0,\pi/2)$, while those of the CP phases are in $[0,2\pi)$. 
We should note that, in this model, there are no additional CP phases in the lepton sector other than $\delta$, $\eta_1$, and $\eta_2$.

\section{
How heavy can the right-handed neutrinos be?}
\label{sec:mass spectrum}

In this section, we consider possible ranges of the mass parameters of the right-handed neutrinos that reproduce the neutrino oscillations.
As a reference, we summarize the observed oscillation parameters in Tab.~\ref{tab:oscillation parameters}.
As the mixing angles of the active neutrinos are of $\order{1}$,
we naively expect that $M_{ee}$ and $M_{\mu\tau}$
should be $\order{\langle \sigma_{1,2} \rangle_0}$.
Thus, in order to explain the muon $g-2$
and the neutrino oscillations at the same time,  $M_{ee}$ and 
$M_{\mu\tau}$ are expected to be $10 \text{\,--\,} 100$\,GeV.

Surprisingly, however, we find that both $M_{ee}$ and $M_{\mu \tau}$ can be as large as $\order{10^{7}}$\,GeV
for special cases where 
the active neutrino masses are degenerate.
Such large right-handed neutrino mass parameters open up possibilities for leptogenesis by the decay of right-handed neutrinos.

\setlength{\tabcolsep}{2.5mm}
{\renewcommand{\arraystretch}{1.2}
\begin{table}[htbp]
\begin{minipage}{\textwidth}
    \centering
    \caption{Three-flavor oscillation parameters by NuFIT 5.1~\cite{Esteban:2020cvm}\footnote{\url{http://www.nu-fit.org/}}
    with the data on atmospheric neutrinos by the Super-Kamiokande collaboration.    }
    \label{tab:oscillation parameters}
    \begin{tabular}{ l | c c | c c }
        \hline
         & \multicolumn{2}{c|}{Normal Ordering} & \multicolumn{2}{c}{Inverted Ordering}
        \\
        \cline{2-5}
         & best fit point & 3$\sigma$ range & best fit point & 3$\sigma$ range
        \\
        \hline
        $\sin^2 \theta_{1 2}$ & 0.304 & 0.269 $\to$ 0.343 & 0.304 & 0.269 $\to$ 0.343
        \\
        $\sin^2 \theta_{2 3}$ & 0.450 & 0.408 $\to$ 0.603 & 0.570 & 0.410 $\to$ 0.613
        \\
        $\sin^2 \theta_{1 3}$ & 0.02246 & 0.02060 $\to$ 0.02435 & 0.02241 & 0.02055 $\to$ 0.02457
        \\
        $\Delta m_{2 1}^2$ [$10^{-5}$\,eV$^2$] & 7.42 & 6.82 $\to$ 8.04 & 7.42 & 6.82 $\to$ 8.04
        \\
        $\Delta m_{3 l}^2$ [$10^{-3}$\,eV$^2$] & 2.510 & 2.430 $\to$ 2.593 & $-2.490$ & $-2.574 \to -2.410$
        \\
        $\delta$ [$^\circ$] & 230 & 144 $\to$ 350 & 278 & 194 $\to$ 345
        \\
        \hline
    \end{tabular}
\end{minipage}
\end{table}
}


\subsection{Non-degenerate active neutrino masses}
\label{subsec:non-degenerate case}

First, let us consider the case where the active neutrino masses are not degenerate.
Especially, we assume that the lightest active neutrino mass $m_l$ is much smaller than $\sqrt{\Delta m^2_\mathrm{sol/atm}}$.
In this case, the right hand side of Eq.\,\eqref{eq:seesaw general} is given by
\begin{align}
\label{eq:seesaw typical}
    v_\mathrm{EW}^2 \times
    \left( \mathrm{U}[ M_\nu^{d} ]^{-1} \mathrm{U}^\mathrm{T} \right)_{\alpha \beta} 
    =
    \frac{v_\mathrm{EW}^2 }{m_l} \times 
    \left[ 
        \mathrm{U}_{\alpha l} \mathrm{U}_{\beta l}
        +
        \mathcal{O}
        \left( 
            \frac{m_l}{ \sqrt{\Delta m_\mathrm{sol/atm}^2 } }
        \right)
    \right]
    \ ,
\end{align}
where $l$ is the index corresponding to the lightest active neutrino.
For the central values of the neutrino mixing parameters in both of the normal and inverted orderings, the sizes of $\mathrm{U}_{\alpha l} \mathrm{U}_{\beta l}$ are at least of $\order{0.01}$.
In the limit of $m_l \ll \sqrt{\Delta m^2_{\mathrm{sol/atm}}}$, the first term in Eq.\,\eqref{eq:seesaw typical} is dominant.
Then, we can obtain a constraint on the mass parameters by comparing the product of $(1,1)$ and $(2,3)$ elements and that of $(1,2)$ and $(1,3)$ elements of Eq.\,\eqref{eq:seesaw general}.
We can also obtain another constraint by comparing the product of $(2,2)$ and $(3,3)$ elements and that of $(2,3)$ and $(3,2)$ elements of Eq.\,\eqref{eq:seesaw general}.
As a result, we find
\begin{align}
    \label{eq:constraint eemutau}
    \frac{M_{ee} M_{\mu\tau} }{h_{e\mu} h_{e\tau} \langle \sigma_1 \rangle_0^2}
    &\sim
    1
    \ , \\
    \label{eq:constraint mumutautau}
    \frac{M_{\mu\tau}^2 }{h_{\mu\mu}h_{\tau\tau}
    \langle \sigma_2 \rangle_0^2}
    &\sim
    1
    \ ,
\end{align}
where we roughly take all the elements of $\mathrm{U}_{\alpha l} \mathrm{U}_{\beta l} \sim 1$.
These constraints are independent of the Dirac Yukawa couplings, $\lambda_\alpha$'s.
It can be read that these constraints are valid for $m_l \lesssim 0.01 \sqrt{\Delta m_{\mathrm{sol/atm}}^2}$. 

From the second constraint, we find that $M_{\mu\tau}$ is at most of $\order{\langle \sigma_2 \rangle_0}$.
On the other hand, $M_{ee}$ can be as large as $\langle \sigma_1 \rangle_0^2 / M_{\mu \tau}$, which is much larger than $\langle \sigma_1 \rangle_0$ when $M_{\mu \tau}$ is much smaller than $\langle \sigma_1 \rangle_0$.

\subsection{Quasi-degenerate active neutrino masses in the normal ordering}
\label{subsec:degenerate in NO}

As shown above, the combinations of the elements in Eq.\,\eqref{eq:seesaw general} lead to constraints on the mass parameters.
Here, we consider the case where some elements in Eq.\,\eqref{eq:seesaw general} vanish and both $M_{e e}$ and $M_{\mu \tau}$ can be much larger than $\langle \sigma_{1,2} \rangle_0$.

First, we focus on the normal ordering.
In this case, the active neutrino masses are given by
\begin{align}
    m_1 = m_l
    \ ,
    \quad 
    m_2 = \sqrt{m_l^2 + \Delta m_\mathrm{sol}^2}
    \ ,
    \quad 
    m_3 = \sqrt{m_l^2 + \Delta m_\mathrm{atm}^2}
    \ .
\end{align}
From the CMB observations, the sum of the active neutrino masses is constrained as $\sum m_i < 0.26$\,eV at the 95\%\,C.L.~\cite{Planck:2018vyg}, which means $m_l \lesssim 0.082$\,eV.
(When the CMB lensing and the BAO are included, it becomes $\sum m_i < 0.12$\,eV, which means $m_l \lesssim 0.030$\,eV.)
As seen above, $M_{e e}$ and $M_{\mu \tau}$ are tied to $\langle \sigma_{1,2} \rangle_0$ if $m_l \ll \sqrt{\Delta m_\mathrm{sol/atm}^2}$.
To liberate
$M_{e e}$ and $M_{\mu \tau}$ from $\langle \sigma_{1,2} \rangle_0$, we consider $m_l$ comparable to $\sqrt{\Delta m_\mathrm{atm}^2}$.

As an example, we fix $m_l = 0.06$\,eV and adopt the center values for $\Delta m_\mathrm{sol/atm}^2$ and the following mixing angles:
\begin{align}
\label{eq:nu param NH}
\begin{array}{c}
    \Delta m_\mathrm{sol}^2 = 7.42 \times 10^{-5} ~\text{eV}^2
    \ , \quad
    \Delta m_\mathrm{atm}^2 = 2.510 \times 10^{-3} ~\text{eV}^2    
    \ ,
    \\
    \sin^2 \theta_{12} = 0.304
    \ , \quad 
    \sin^2 \theta_{13} = 0.02246
    \ .
\end{array}
\end{align}
We also take the rest of the oscillation parameters as
\begin{equation}
    \label{eq:mixing parameters in NO}
    \sin^2 \theta_{23} \simeq 0.565
    \ , \quad 
    \delta \simeq 268^\circ
    \ , \quad 
    \eta_1 \simeq 355^\circ
    \ , \quad 
    \eta_2 \simeq 177^\circ
    \ ,
    \end{equation}
where $\theta_{23}$ and $\delta$ are in the 2$\sigma$ ranges of the observational data~\cite{Esteban:2020cvm}.
In this case, the $(1, 2)$ and $(2, 2)$ elements of Eq.\,\eqref{eq:seesaw general} vanish:%
\footnote{%
    Here, we numerically find a parameter set in Eq.\,\eqref{eq:mixing parameters in NO} where the $(1, 2)$ and $(2, 2)$ elements of Eq.\,\eqref{eq:seesaw general} 
    are suppressed by a factor of $\order{10^{-13}}$
    compared with the other elements by varying $(\theta_{23}, \delta, \eta_1, \eta_2)$.
    Thus, strictly speaking, the $(1, 2)$ and $(2, 2)$ elements might not vanish exactly.
    In this case, the constraints on the mass parameters in Eqs.~\eqref{eq:constraint eemutau} and \eqref{eq:constraint mumutautau} are relaxed by this factor, and the mass spectrum discussed in the following can be realized.
    This is also true for the case of the inverted ordering discussed below.
}%
\begin{align}
    \label{eq:seesaw degenerate NO}
    v_\mathrm{EW}^2 \times
    \mathrm{U}[ M_\nu^{d}]^{-1} \mathrm{U}^\mathrm{T}
    \simeq
    \frac{v_\mathrm{EW}^2 }{m_l} \times
    \left(
        \begin{array}{ccc}
            0.99\, e^{-0.03 i} &                 0 & 0.05\, e^{ 1.4  i}
            \\
                             0 &                 0 & 0.88\, e^{ 3.1  i} 
            \\
             0.05\, e^{ 1.4 i} & 0.88\, e^{ 3.1 i} & 0.22\, e^{-0.05 i}
        \end{array}
    \right)
    \ .
\end{align}
This structure corresponds to $h_{e \mu} = h_{\mu \mu} = 0$.
Note also that this zero texture can be achieved with the opposite CP phases (mod $2\pi$), although
the corresponding Dirac CP phase is disfavored by observations~\cite{Esteban:2020cvm}.

With this structure of Eq.\,\eqref{eq:seesaw degenerate NO}, the mass parameters and $\langle \sigma_{1,2} \rangle_0$ are related as
\begin{align}
    \frac{M_{ee}}{\lambda_{e}^2} 
    \sim
    \frac{M_{\mu \tau}}{\lambda_\mu \lambda_\tau} 
    \sim
    \frac{h_{e \tau} \langle \sigma_1 \rangle_0}{\lambda_e \lambda_\tau} 
    \sim
    \frac{h_{\tau \tau} \langle \sigma_2 \rangle_0}{\lambda_\tau^2}
    \sim
    \frac{v_\mathrm{EW}^2}{m_l}
    \equiv
    M_s
    \simeq 
    5 \times 10^{14}\,\mathrm{GeV}
    \ .
\end{align}
From this relation, we obtain
\begin{align}
    M_{\mu \tau}
    \sim
    \lambda_\mu h_{e \tau} \langle \sigma_1 \rangle_0 
    \sqrt{\frac{M_s}{M_{e e}}}\ .
\end{align}
If $\lambda_\mu \simeq 1$, $h_{e \tau} \simeq 1$, $\langle \sigma_1 \rangle_0 \simeq 100$\,GeV, the mass parameters takes the maximum values as
\begin{align}
    \label{eq:Majorana mass upper bound}
    M_{\mu \tau}
    \sim
    10^6\,\mathrm{GeV}
    \left(
        \frac{ M_{e e} }{5 \times 10^6\,\mathrm{GeV} } 
    \right)^{-1/2}
    \ ,
\end{align}
which is a rough estimate ignoring coefficients of $\order{0.01}$ in Eq.\,\eqref{eq:seesaw degenerate NO}.
Note that, even if one of $\lambda_\alpha$'s is unity, these choices satisfy the constraints from the charged flavor violation such as $\mu \to e + \gamma$~\cite{MEG:2016leq} or $\tau \to e + \gamma$~\cite{BaBar:2009hkt} due to the smallness of the other $\lambda_\alpha$'s.

For the normal ordering,
it is also possible that both $(2,2)$ and $(3,3)$ elements in Eq.\,\eqref{eq:seesaw general} vanish with specific choices of the lightest active neutrino mass and mixing parameters.
This case is nothing but the minimal gauged $\Umt$ model without $\sigma_2$ studied in Refs.~\cite{Asai:2017ryy,Asai:2018ocx}.
For this structure, however, the mass parameters are still constrained as Eq.\,\eqref{eq:constraint eemutau} and either of $M_{e e}$ and $M_{\mu \tau}$ is smaller than the scale of $\langle \sigma_1 \rangle_0$.

At the end of this subsection, we comment on the constraints on the effective neutrino mass for the neutrinoless double beta ($0 \nu \beta \beta$) decay, 
\begin{align}
    m_{\beta \beta}
    \equiv
    \left|
        \sum_i m_i \mathrm{U}_{e i}^2    
    \right|
    \ .
\end{align}
$m_{\beta \beta}$ is bounded through the upper bound on the lifetime of the $0 \nu \beta \beta$ decay from the KamLAND-Zen~\cite{KamLAND-Zen:2022tow} and GERDA~\cite{GERDA:2020xhi} experiments as
\begin{align}
    m_{\beta \beta} 
    &<
    36 \, \text{--} \, 156\,\mathrm{meV}
    \ ,
    \\
    m_{\beta \beta} 
    &<
    79 \, \text{--} \, 180\,\mathrm{meV}
    \ ,
\end{align}
respectively.
The uncertainties of the upper bound come from the variety of nuclear matrix element calculations.
For our parameter choice, we obtain 
\begin{align}
\label{eq:mbetabetaNO}
    m_{\beta \beta}
    \simeq
    61\,\mathrm{meV}
    \ ,
\end{align}
which is consistent with the current constraints of the $0 \nu \beta \beta$ decay experiments.

\subsection{Quasi-degenerate active neutrino masses in the inverted ordering}
\label{subsec:degenerate in IO}

Next, we consider the inverted ordering.
In this case, the active neutrino masses are given by
\begin{align}
    m_1 = \sqrt{m_l^2 + \Delta m_\mathrm{atm}^2 - \Delta m_\mathrm{sol}^2}
    \ ,
    \quad 
    m_2 = \sqrt{m_l^2 + \Delta m_\mathrm{atm}^2}
    \ ,
    \quad 
    m_3 = m_l
    \ .
\end{align}
For this mass spectrum, the cosmological constraint $\sum m_i < 0.26$\,eV~\cite{Planck:2018vyg} corresponds to $m_l \lesssim 0.077$\,eV.
We again consider $m_l$ comparable to $\sqrt{\Delta m_\mathrm{atm}^2}$ and fix
\begin{align}
\begin{array}{c}
     m_l = 0.06~\text{eV}
    \ , \quad
    \Delta m_\mathrm{sol}^2 = 7.42 \times 10^{-5} ~\text{eV}^2
    \ , \quad
    \Delta m_\mathrm{atm}^2 = 2.490 \times 10^{-3} ~\text{eV}^2    
    \ ,
    \\
    \sin^2 \theta_{1 2} = 0.304
    \ , \quad 
    \sin^2 \theta_{1 3} = 0.02241
    \ .
\end{array}
\end{align}
In this case, we find that the $(1, 3)$ and $(3, 3)$ elements of Eq.\,\eqref{eq:seesaw general} vanish as
\begin{align}
    \label{eq:seesaw degenerate IO}
    v_\mathrm{EW}^2 \times
    \mathrm{U}[ M_\nu^{d}]^{-1} \mathrm{U}^\mathrm{T}
    \simeq
    \frac{v_\mathrm{EW}^2 }{m_l} \times
    \left(
        \begin{array}{ccc}
            0.78\, e^{-0.03 i} & 0.05\, e^{-1.7 i} &                 0
            \\
            0.05\, e^{-1.7  i} & 0.23\, e^{ 3.1 i} & 0.88\, e^{ 3.1 i}
            \\
                             0 & 0.88\, e^{ 3.1 i} &                 0
        \end{array}
    \right)
    \ ,
\end{align}
for
\begin{equation}
    \label{eq:mixing parameters in IO}
    \sin^2 \theta_{23} = 0.566
    \ , \quad 
    \delta \simeq 270^\circ
    \ , \quad 
    \eta_1 \simeq 355^\circ
    \ , \quad 
    \eta_2 \simeq 177^\circ
    \ ,
\end{equation}
where $\theta_{23}$ and $\delta$ are in the $1 \sigma$ ranges of the observational data~\cite{Esteban:2020cvm}.
This structure corresponds to $h_{e \tau} = h_{\tau \tau} = 0$.
As in the case of the normal ordering, this zero texture can be achieved with the opposite CP phases in spite of a disfavored Dirac CP phase.

Since the structure of Eq.\,\eqref{eq:seesaw degenerate IO} is the same as that of Eq.\,\eqref{eq:seesaw degenerate NO} except for the replacement of $\mu \leftrightarrow \tau$, we obtain the estimate of 
\begin{align}
    M_{\mu \tau}
    \sim
    \lambda_\mu h_{e \mu} \langle \sigma_1 \rangle_0
    \sqrt{\frac{M_s}{M_{e e}}}
    \ .
\end{align}
Thus, for $\lambda_\tau \simeq 1$, $h_{e \mu} \simeq 1$, $\langle \sigma_1 \rangle_0 \simeq 100$\,GeV, the mass parameters take the maximum values as in Eq.\,\eqref{eq:Majorana mass upper bound}.

For this parameter choice, the effective neutrino mass of $0 \nu \beta \beta$ decay is 
\begin{align}
\label{eq:mbetabetaIO}
    m_{\beta \beta}
    \simeq
    77\,\mathrm{meV}
    \ ,
\end{align}
which is consistent with the current constraints of the $0 \nu \beta \beta$ decay experiments.

\section{Breaking of the \texorpdfstring{$\Umt$}{} symmetry in the early universe}
\label{sec:breaking}

As we will see in the next section, successful leptogenesis requires non-vanishing expectation values of $\sigma_{1,2}$, and hence,
it is important to clarify the aspects of the $\Umt$ symmetry breaking in the early universe.

In section~\ref{sec:model}, we introduced two $\Umt$ charged scalars.
To discuss the nature of symmetry breaking, it is enough to consider a single $\Umt$ charged scalar $\sigma$.
A tree-level Lagrangian for $\sigma$ and the SM Higgs doublet $\Phi$ is given by 
\begin{align}
\begin{gathered}
    \mathcal{L}_{\sigma,\Phi} = |D_\mu \sigma|^2 + |D_\mu \Phi|^2 - V(\sigma,\Phi)\ , 
    \\
    \label{eq:pot}
    V(\sigma,\Phi)
    =
    - \mu_\sigma^2 |\sigma|^2 
    - \mu_\Phi^2 \Phi^\dagger \Phi 
    + \lambda_\sigma |\sigma|^4 
    + \lambda_\Phi \left( \Phi^\dagger \Phi \right)^2
    + \lambda_{\Phi \sigma} |\sigma|^2 (\Phi^\dagger \Phi)
    \ ,
\end{gathered}
\end{align}
where $\mu_\sigma^2$, $\mu_\Phi^2$ express mass parameters for each scalar field, $\lambda_\sigma$, $\lambda_\Phi$ are quartic self-couplings, and $\lambda_{\Phi \sigma}$ is a Higgs--$\sigma$ coupling.
In the following, we take $\mu_{\Phi}^2$, $\lambda_{\sigma}$ and $\lambda_{\Phi}$ positive.
The covariant derivative on $\sigma$
is given by, $D_\kappa = \partial_\kappa + i g_{Z^\prime} Q_{L_\mu - L_\tau} Z^\prime_\kappa$,
while $\Phi$ has no $\Umt$ charge.

At low energy, the electroweak symmetry is broken by the VEV of $\Phi$, and hence, 
the mass term of $\sigma$ around $\sigma = 0$ is given by,
\begin{align}
    - \mu_0^2 |\sigma|^2
    \equiv
    - ( \mu_\sigma^2 - \lambda_{\Phi\sigma} v_\mathrm{EW}^2 )
    |\sigma|^2
    \ .
\end{align}
Since we require that the $\Umt$ symmetry  
is broken at the vacuum, 
we take $\mu_0^2 > 0$.
The resultant VEV of $\sigma$ is given by,
\begin{align}
    \langle \sigma \rangle_0
    =
     \frac{\mu_0}{ \sqrt{2\lambda_\sigma} }
    \ ,
\end{align}
where we take $\langle \sigma \rangle_0$ real and positive without loss of generality.

\subsection{Symmetry restoration by thermal/finite density effects }
\label{subsec:restoration}

Let us consider the symmetry restoration in the early universe.
In this subsection, we neglect $\lambda_{\Phi\sigma}$ for a while, and hence, $\mu_0^2 = \mu_\sigma^2>0$.
In the early universe,
$\sigma$ obtains an effective mass term as $V_{\mathrm{eff}}= m_\mathrm{eff}^2 |\sigma|^2$, which depends on
environment such as thermal bath.
For $m_\mathrm{eff}^2>\mu_0^2$, the $\Umt$ symmetry breaking does not occur.
Therefore, we find the cosmic temperature $T = T_c$ at which the phase of the $\Umt$ symmetry changes, that is, $m_\mathrm{eff}^2|_{T_c}=\mu_0^2$.
Note that $\mu_0$ is bounded from above to explain the muon $g-2$ as
$\mu_0 = \sqrt{2 \lambda_\sigma} \langle \sigma \rangle_0 \lesssim 100\,\mathrm{GeV}$.
First, let us evaluate the effective mass squared in the environment where the $\Umt$ sector is thermalized by the $\Umt$ gauge interaction with SM particles in the thermal bath.

The rate of the gauge interaction is given by $\Gamma_g \approx (g_{Z^\prime}^4/4\pi)T$ for $T \gg m_{Z'}$.
Then, we obtain the thermalization temperature as
\begin{equation}
    \label{eq:Tth}
    T_\mathrm{th} \approx 
    6\times 10^{4} \, \mathrm{GeV} \left( \frac{g_{Z^\prime}}{10^{-3}} \right)^4, 
\end{equation}
where the interaction rate $\Gamma_g$ becomes equal to the Hubble expansion rate $H$.
For $T<T_\mathrm{th}$, the $\Umt$ sector is thermalized.
The Hubble rate is given by 
$H=\sqrt{\pi^2 g_\ast/90}T^2/M_{P}$, 
where $M_P$ is the reduced Planck scale
and $g_\ast \sim 100$ is the effective degrees of freedom of the relativistic species.

When the $\Umt$ sector is thermalized,
the thermal mass is given by 
\begin{equation}
\label{eq:meff_th}
    m_\mathrm{eff}^2
    =
    \left(
        \frac{\lambda_\sigma}{3}+\frac{g_{Z'}^2}{8}
    \right) 
    T^2
    >
    0.    
\end{equation}
At high temperatures where the thermal mass is larger than $\mu_0^2$,
the $\Umt$ symmetry is in the symmetric phase. 
Decreasing the temperature, the symmetry breaking takes place at the breaking temperature,
\begin{equation}
    \label{eq:Tbre}
    T_\mathrm{bre}=
        \left(
        \frac{\lambda_\sigma}{3}+\frac{g_{Z'}^2}{8}
    \right)^{-1/2} \mu_0
    \simeq \langle\sigma\rangle_0  = 10 \text{\,--\,} 100\,\mathrm{GeV}\ .
\end{equation}
Therefore, the $\Umt$ symmetry is in the symmetric phase for at least $T_\mathrm{bre}<T<T_\mathrm{th}$.

Next, let us move on to higher temperatures, $T>T_\mathrm{th}$, where 
the $\Umt$ sector is not thermalized and
$\sigma$ does not obtain the thermal mass.
Even in this case, 
$\sigma$ has an effective mass squared, $m_{\mathrm{eff}}^2=m_\mathrm{fd}^2$, due to the finite density effect of the non-thermalized    
$\Umt$ sector particles, which are produced from the SM thermal bath.
As derived in the Appendix~\ref{sec:density}, it is given by
\begin{align}
    m_\mathrm{fd}^2 
    = C_X
    n_{X}\langle p_X^{-1} \rangle, 
\end{align}
where $n_X$ and $p_X$ are the number density and momentum of a particle $X$ that interacts with $\sigma$, respectively, and $C_X$ is a coefficient depending on the interaction.
Here, $\langle \,\, \rangle$ expresses the averaged value over the particle distribution.

To evaluate $m_\mathrm{fd}^2$ from the self-interaction of $\sigma$, we estimate the number density of $\sigma$, $n_\sigma$.
For the production of $\sigma$, \eg, $\mu \overline{\mu} \rightarrow \sigma \sigma^\ast$, 
$n_\sigma$ follows the Boltzmann equation,
\begin{equation}
    \label{eq:Boltz}
    \frac{\mathrm{d} n_\sigma}{\mathrm{d} t}+3H n_\sigma 
    =
    \langle \sigma v \rangle n_\mu^2,
\end{equation}
where the cross section is given by $\langle \sigma v \rangle \sim g_{Z^\prime}^4/(4\pi T^2)$,
and $n_\mu = (3 \zeta(3)/2\pi^2) T^3$.
For $T>T_\mathrm{th}$, $n_\sigma$ is given by 
\begin{equation}
    \label{eq:n_sig}
    n_\sigma = \kappa \frac{g_{Z^\prime}^4}{4 \pi}M_P T^2,
\end{equation}
where $\kappa$ is a numerical $\mathcal{O}(1)$ factor. 
For the self-interaction of $\sigma$ in Eq.\,(\ref{eq:pot}), the effective mass squared can be read as
\begin{equation}
\label{eq:mfd}
    m_\mathrm{fd}^2
    \simeq 
    \lambda_{\sigma} n_\sigma \langle p_\sigma^{-1} \rangle
    \simeq
    \kappa \frac{\lambda_\sigma g_{Z^\prime}^4}{4 \pi}  M_P T,
\end{equation}
where we use $\langle p_\sigma^{-1} \rangle \simeq T^{-1}$ since $\sigma$ is produced from the SM thermal bath. 

Compared with the tree-level mass squared $ \mu_\sigma^2$,
\begin{equation}
\label{eq:comp}
    \frac{m_\mathrm{fd}^2(T)}{\mu_\sigma^2} 
    \simeq
    2 \times 10^6 \kappa 
    \left( \frac{50~\text{GeV}}{\mu_\sigma/\sqrt{2\lambda_\sigma}} \right)^2
    \left( \frac{g_{Z^\prime}}{10^{-3}} \right)^8
    \left( \frac{T}{T_\mathrm{th}} \right)
    \ ,
\end{equation}
for $T>T_\mathrm{th}$.
In Fig.\,\ref{fig:parameters}, we show the parameter region where $m_{\mathrm{fd}}^2(T_\mathrm{th})/\mu_\sigma^2 < 1$ as a blue shaded region.%
\footnote{%
We assume that $\lambda_\sigma \gg g_{Z'}^2 = \order{10^{-6}}$, and then the effective mass squared is mainly contributed by the self-interaction of $\sigma$.
In the figure, we take $\lambda_{\Phi\sigma} = 0$,
and hence $\mu_\sigma/\sqrt{2\lambda_\sigma} = m_{Z'}/\sqrt{2}g_{Z'}$.}
The figure shows that 
$m_{\mathrm{fd}}^2(T_\mathrm{th})/\mu_\sigma^2>1$ for the parameter region that explains the muon $g-2$, and hence, we find that 
the $\Umt$ symmetry is restored even at the temperature above $T_\mathrm{th}$.
As a result, we find that the symmetry is preserved at $T>T_\mathrm{bre}$ by combining the results from the thermal and finite density effects.

We should discuss here how restoration of the symmetry proceeds when the initial condition of the universe is in the $\Umt$ broken phase.
Let us suppose that the $\Umt$ charged scalar field is initially settled at $\sigma \sim \langle \sigma \rangle_0$ and the scalar potential arises as $V(\sigma) \sim m_\mathrm{fd}^2 |\sigma|^2$ just after reheating of the universe.
In this situation, we can consider that $\sigma$ rapidly moves to $\sigma = 0$ if 
\begin{equation}
    m_\mathrm{fd}^2 > H^2.
\end{equation}
By using Eq.\,\eqref{eq:mfd}, it turns out that the symmetry is restored immediately when the following condition is satisfied;
\begin{equation}
    \label{eq:T-restoration}
    T < T_\mathrm{rest} \simeq \left( \frac{\lambda_\sigma g_{Z'}^4}{4 \pi} \right)^\frac{1}{3}
    M_P
    \simeq
    10^{14}\,\mathrm{GeV} \lambda_\sigma^{1/3} \left( \frac{g_{Z'}}{10^{-3}} \right)^{4/3}. 
\end{equation}
Thus, even if the $\Umt$ symmetry is initially broken with $\sigma \sim \langle \sigma \rangle_0$, it is restored at the temperature relevant to the following discussion.

\subsection{Symmetry breaking by thermal effects}
\label{subsec:breaking}

So far, we have concentrated on the effects of the $\Umt$
gauge interaction and the self-interaction of $\sigma$,
which restores the $\Umt$ symmetry
at $T>T_{\mathrm{bre}}$.
The aspects of the $\Umt$ symmetry breaking, however, drastically change 
when the Higgs--$\sigma$ coupling $\lambda_{\Phi \sigma}$ takes a certain negative value.%
\footnote{%
    In the following, we will take $|\lambda_{\Phi \sigma}| < \order{10^{-2}}$.
    Then, the VEVs of $\Phi$ and $\sigma$ do not significantly affect each other while $\lambda_{\Phi \sigma} < 0$ plays an important role at high temperatures. 
    Moreover, $\mu_0 \simeq \mu_\sigma$ is justified for
    this range of $\lambda_{\Phi \sigma}$,
    and the blue shaded region in Fig.~\ref{fig:parameters}, $m_\mathrm{fd}^2/\mu_\sigma^2<1$, 
    is still valid.
}
In the presence of a sizable $\lambda_{\Phi \sigma}$, $\sigma$ is thermalized through the interaction with $\Phi$.
Then, the effective scalar potential of $\sigma$ obtains a contribution from the Higgs--$\sigma$ interaction,
\begin{equation}
    \label{eq:potential-negative-lambda}
    V(\sigma) 
    \sim
        - |\lambda_{\Phi \sigma}|  T^2  
    \times|\sigma|^2 
    \ ,
\end{equation}
where we omit a numerical coefficient.
For $|\lambda_{\Phi \sigma}|\gg \lambda_\sigma$, $g_{Z'}^2$, the thermal mass is dominated by this  contribution.
As a result, $\sigma$ acquires a non-zero expectation value,
\begin{align}
    \langle \sigma \rangle 
    \sim
    \sqrt{|\lambda_{\Phi \sigma}| / \lambda_\sigma} \, T
    \ ,
\end{align}
and thus
the $\Umt$ symmetry is broken even at high temperatures.

Several comments are in order.
Firstly, note that the 
relative sizes of $\lambda_{\Phi}$, $\lambda_{\sigma}$, and $|\lambda_{\Phi\sigma}|$ are constrained by the unbounded-from-below condition,
$|\lambda_{\Phi \sigma}| < 2\sqrt{\lambda_{\Phi} \lambda_{\sigma}}$.
Under this constraint, the back-reaction of $\langle \sigma \rangle$ to the Higgs potential, \ie, $\lambda_{\Phi \sigma}\langle \sigma \rangle^2 |\Phi|^2 $ is at most $\lambda_{\Phi} T^2 |\Phi|^2$.
It is subdominant compared with the top Yukawa contributions,
and thus the back-reaction does not affect the dynamics of the electroweak sector significantly.
Incidentally, we also find the thermal mass of $\Phi$ through the $\sigma$-loop is negligible  due to $|\lambda_{\Phi\sigma}|\ll \lambda_{\Phi}$ for $|\lambda_{\Phi \sigma}| \gg \lambda_{\sigma}$.

Secondly, note that $\sigma$ and $Z'$ do not affect the standard cosmology below $T < \order{m_{Z'}}$.
To see this, 
we denote the mass eigenstates of the physical components in $\Phi$ and $\sigma$ by $H$ and $S$.
The modulus of the Higgs boson $\Phi_0$ contains $H$ and $S$ as $\Phi_0 = (H \cos\theta + S \sin\theta)/\sqrt{2} $ with a mixing angle $\theta$.
We take $H$ to be the observed Higgs boson so that $m_H\simeq 125$\,GeV, while the mass of $S$, $m_S$, is smaller.%
\footnote{In the model discussed in the previous section, we have two complex scalars $\sigma_{1}$ and $\sigma_{2}$.
The discussion in the present section is given in the Appendix~\ref{sec:symmetrybreakingsector}.}
We assume $\lambda_{\sigma} \gg g_{Z'}^2$
and then $m_S \gg m_{Z'}$.
In this limit, $S$ decays into a pair of the longitudinal mode of $Z'$ 
immediately
(see the Appendix~\ref{sec:symmetrybreakingsector}).
As $Z'$ also decays immediately into a pair of neutrinos, $Z'$ and $\sigma$ do not cause 
cosmological problems for $m_S \gg m_{Z'}\gg \order{1}$\,MeV
(see Ref.~\cite{Escudero:2019gzq}).

Finally, we also comment on the experimental constraints on $|\lambda_{\Phi \sigma}|$.
In the model discussed in the previous section, we introduced two scalars, $\sigma_{1}$ and $\sigma_2$,
that couple to $\Phi$ through $\lambda_{\Phi \sigma_1}$ and $\lambda_{\Phi \sigma_2}$, respectively.
As we see in the Appendix~\ref{sec:symmetrybreakingsector},
 the upper limit on the branching fraction of Higgs invisible decay mode, $\mathrm{Br}(H\to \mathrm{invisible}) < 0.11$~\cite{Albert:2021wjs}, leads to
\begin{align}
\label{eq:invisible2scalar}
(|\lambda_{\Phi\sigma_1}|^2 + 
|\lambda_{\Phi\sigma_2}|^2)^{1/2}
< 7\times 10^{-3}\ ,
\end{align}
(see also Ref.\,\cite{Nomura:2018yej}).
 
\section{Leptogenesis in the \texorpdfstring{$\Umt$}{} model}
\label{sec:leptogenesis}

We first consider a leptogenesis scenario with a negligible $\lambda_{\Phi \sigma}$.
In this case, as seen in the previous section, the $\Umt$ symmetry is not broken at $T > T_\mathrm{bre}= 10 \,\text{--}\, 100 $\,GeV.
This temperature is lower than the temperature where sphaleron processes freeze out, $T_\mathrm{sph}\approx 130~\text{GeV}$.%
\footnote{%
    There could be a region where $T_\mathrm{bre} \gtrsim T_\mathrm{sph}$ 
    in a corner of the parameter space of $Z'$ to explain the muon $g-2$.
    In this case, the $\Umt$ symmetry breaking takes place before the sphaleron processes freeze out.
    Although this leaves a possibility of the leptogenesis scenarios at around $T_\mathrm{sph}$, it is beyond the scope of this paper.
}
Since leptogenesis requires that the sphaleron processes convert the lepton asymmetry into the baryon asymmetry, we need to consider leptogenesis in the $\Umt$ symmetric phase.
As we will see shortly, however, leptogenesis does not occur in the $\Umt$ symmetric phase.

On the other hand, when we consider a sizable $\lambda_{\Phi \sigma} < 0$, the $\Umt$ symmetry is broken in the early universe.
In this case, we find that the non-thermal leptogenesis~\cite{Kumekawa:1994gx,Asaka:1999jb} can generate sufficient lepton asymmetry to explain the observed baryon asymmetry of the universe.

\subsection{Failures of leptogenesis in the \texorpdfstring{$\Umt$}{} symmetric phase}
\label{subsec:failure}

To discuss leptogenesis in the $\Umt$ symmetric phase,
it is convenient to construct a Dirac fermion 
from the two left-handed 
Weyl fermions, 
$\bar{N}_\mu$ and 
$\bar{N}_{\tau}$,
\begin{equation}
    \label{eq:Dirac}
    \Psi_{\mu\tau} \equiv
    \begin{pmatrix}
        \bar{N}_{\mu} \\
        \bar{N}_{\tau}^\dagger \\
    \end{pmatrix},
\end{equation}
which has a $\Umt$ charge $-1$.
In the symmetric phase,
$M_{\mu\tau}$ provides 
the Dirac mass term of
$\Psi_{\mu \tau}$. 
The remaining right-handed neutrino forms a four-component Majorana
fermion,
\begin{align}
    \Psi_{e} \equiv    \begin{pmatrix}
        \bar{N}_{e} \\
        \bar{N}_{e}^\dagger \\
    \end{pmatrix},
\end{align}
whose Majorana mass is given by $M_{ee}$.
The four component lepton doublets are also given by,
\begin{align}
    \Psi_{L_{\alpha}}
    \equiv
    \begin{pmatrix}
        {L}_{\alpha} \\
        0 \\
    \end{pmatrix},
\end{align}
which satisfy $P_L\Psi_{L_{\alpha}} = \Psi_{L_{\alpha}}$ with $P_L$ being the projection operator on left-handed fermions.

In terms of them, the Lagrangian in Eq.\,\eqref{eq:Lnu} is rewritten as,
\begin{align}
    \mathcal{L}_\nu 
    =&
    \mathcal{L}_e + \mathcal{L}_{\mu \tau}
    \ ,
    \\
    \label{eq:Le}
    \mathcal{L}_{e} 
    =&
    \frac{1}{2} \overline{\Psi}_e i \slashed{\partial} \Psi_{e}
    - \frac{M_{ee}}{2} \overline{\Psi}_e \Psi_{e} 
    - (\lambda_e \overline{\Psi}_e \tilde{\Phi}P_L \Psi_{L_e}
    +\mathrm{h.c.}) ,
    \\
    \label{eq:Lpsi}
    \mathcal{L}_{\mu \tau} 
    =&
    \overline{\Psi}_{\mu\tau} i \slashed{D} \Psi_{\mu\tau} - M_{\mu \tau} \overline{\Psi}_{\mu\tau} \Psi_{\mu\tau}
    - \left( 
        \lambda_\mu \overline{\Psi}_{\mu \tau}^c \tilde{\Phi} P_L \Psi_{L_\mu} 
        + \lambda_\tau \overline{\Psi}_{\mu \tau} \tilde{\Phi} P_L \Psi_{L_\tau}
        + \mathrm{h.c.}
    \right)
    \nonumber \\
    &-  \left(
    h_{e \mu} \sigma_1 \overline{\Psi}_{e} P_L \Psi_{\mu\tau}
    + h_{e \tau} \sigma_1^\ast \overline{\Psi}_{e} P_L \Psi_{\mu\tau}^c 
    + \frac{1}{2} h_{\mu\mu} \sigma_2 \overline{\Psi}^c_{\mu\tau} P_L \Psi_{\mu\tau}
    + \frac{1}{2} h_{\tau\tau} \sigma_2^\ast \overline{\Psi}_{\mu\tau} P_L \Psi_{\mu\tau}^c
    + \mathrm{h.c.} \right)
    \ .
\end{align}
Here, the bar over the fermion denotes the Dirac conjugate, the superscript $c$ denotes the charge conjugation,%
\footnote{We define $\Psi^c \equiv - i\gamma^2 \Psi^*$ and  $\overline{\Psi}^c \equiv \overline{\Psi^c}$.}
and $P_R$ is the projection operator on right-handed fermions.
The usual lepton number%
\footnote{The usual lepton number corresponds to 
the phase rotations, $\Psi_{L_\alpha}\to e^{i\varphi} \Psi_{L_\alpha}$,  $\Psi_{e}\to e^{i\varphi \gamma_5} \Psi_{e}$, and 
$\Psi_{\mu\tau}\to e^{i\varphi \gamma_5} \Psi_{\mu \tau}$, with $\varphi \in [0,2\pi)$.
}
is violated 
 by the mass terms and Majorana Yukawa interactions in Eqs.~\eqref{eq:Le} and \eqref{eq:Lpsi}.

\begin{figure}[t]
    \begin{minipage}[b]{0.32\linewidth}
        \centering
        \includegraphics[keepaspectratio, scale=0.55]{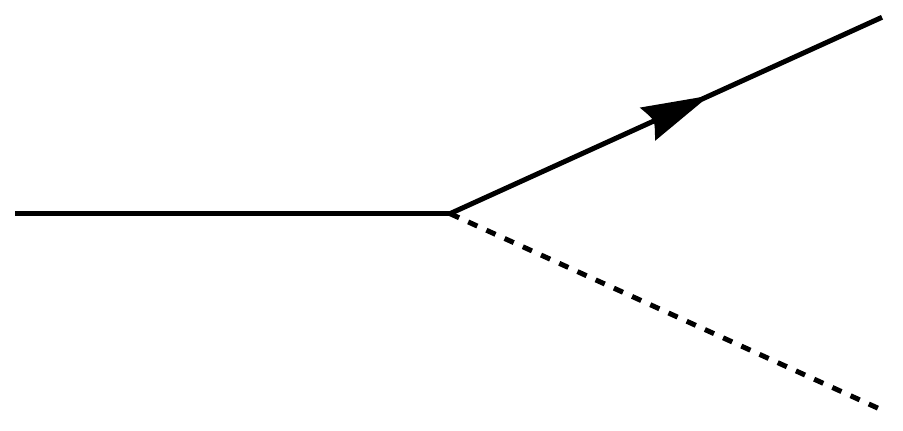}
    \end{minipage}
    \begin{minipage}[b]{0.32\linewidth}
        \centering
        \includegraphics[keepaspectratio, scale=0.55]{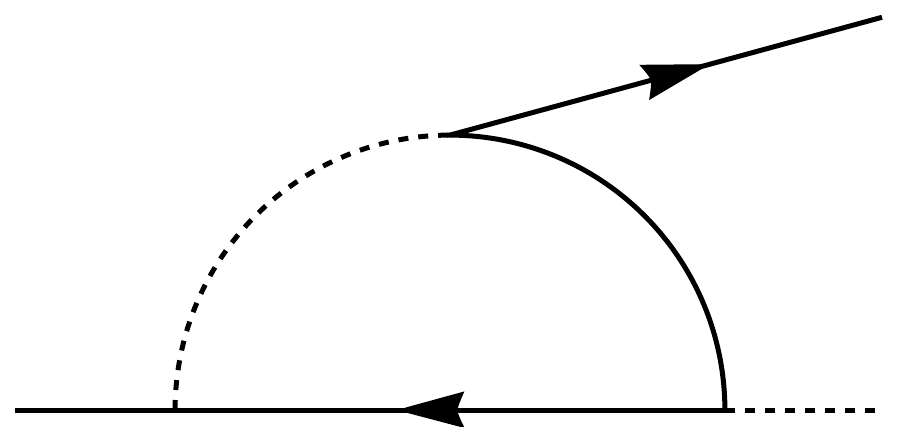}
    \end{minipage}
    \begin{minipage}[b]{0.32\linewidth}
        \centering
        \includegraphics[keepaspectratio, scale=0.55]{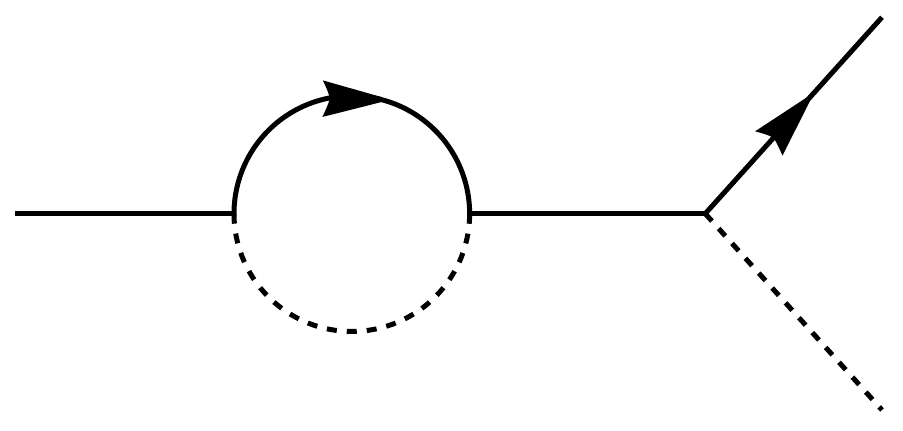}
    \end{minipage}
    \caption{\sl 
    Tree, vertex, and wave-function diagrams for the decay of right-handed neutrinos.
    The plain solid lines, solid lines with arrows, and dashed lines correspond to the right-handed neutrinos, left-handed leptons, and scalar bosons, respectively.
    }
    \label{fig:decay}
\end{figure}

As discussed in section~\ref{sec:mass spectrum},
both $M_{\mu\tau}$ and $M_{ee}$ can be as large as  $\order{10^{7}}$\,GeV.
Then, leptogenesis by the decay of the heavy right-handed neutrinos can be considered.
Let us first focus on the decay of $\Psi_{e}$.
In this case, the asymmetry is generated through the interference of the decay amplitudes given by
\begin{align}
    \mathcal{M}(\Psi_e \to \Psi_{L_e}+\Phi) &= c_{e0} + c_{e1} \mathcal{F}_e\ , \\
    \mathcal{M}(\Psi_{e} \to \Psi_{L_e}^c+\Phi^\dagger) &= c_{e0}^* + c_{e1}^* \mathcal{F}_e \ .
\end{align}
Here, $c_{e0}$ denotes the tree-level amplitude, $c_{e1}$ is the one-loop amplitude where the kinematical loop-integration function $\mathcal{F}_e$ is factored out.
The tree-level and one-loop diagrams contributing to the decay are shown in Fig.\,\ref{fig:decay}.
Note that $\Psi_{e}$ does not decay into $\Psi_{L_{\mu,\tau}}^{(c)}$ at the tree level,
and hence, we do not consider those modes.
The lepton asymmetry is proportional to
\begin{align}
    |\mathcal{M}(\Psi_{e} \to \Psi_{L_e}+\Phi)|^2 
    -|\mathcal{M}(\Psi_{e} \to \Psi_{L_e}^c+\Phi^\dagger)|^2 = - 4\mathrm{Im}[ c_{e0}c_{e1}^* ]\mathrm{Im} [\mathcal{F}_e]
    \ .
    \label{eq:imaginary part in asymmetry}
\end{align}
The imaginary part of $\mathcal{F}_e$ appears from the on-shell singularities, while the imaginary part of $c_{e0}c_{e1}^*$ depends on the interaction coefficients appearing in the diagrams.

In the $\Umt$ symmetric phase, 
the flavor-changing ($\Psi_e \leftrightarrow \Psi_{\mu\tau}$) wave-function diagrams in Fig.\,\ref{fig:decay} do not appear.
Thus, such diagrams do not contribute to the asymmetry.
The processes through the vertex diagrams with
Majorana Yukawa coupling do not result in the pairs of $\Psi_{L_\alpha}^{(c)}+\Phi^{(\dagger)}$ in the symmetric phase.
Thus, we look at the vertex diagrams with only Dirac Yukawa couplings.
However, since the Dirac Yukawa couplings are flavor diagonal, we find,
\begin{align}
    c_{e0} \propto \lambda_e^* 
    \ , \quad
    c_{e1} \propto \lambda_e^* |\lambda_e|^2 \ , 
\end{align}
and hence, $\mathrm{Im}[c_{e0}c_{e1}^*] = 0$.
Therefore, no asymmetry is generated by the decay of $\Psi_e$ in the $\Umt$ symmetric phase.

Next, we focus on the decay of $\Psi_{\mu \tau}$.
The asymmetries are obtained from the following amplitudes,
\begin{align}
    \mathcal{M}(\Psi_{\mu\tau} \to \Psi_{L_\mu}^c+\Phi^\dagger) &= c_{\mu 0} + c_{\mu 1} \mathcal{F}_\mu\ , \\
     \mathcal{M}(\Psi_{\mu\tau} \to \Psi_{L_\tau}+\Phi) &= c_{\tau 0} + c_{\tau 1} \mathcal{F}_\tau\ , \\
    \mathcal{M}(\Psi^c_{\mu\tau} \to \Psi_{L_\mu}+\Phi) &= c_{\mu 0}^* + c_{\mu 1}^* \mathcal{F}_\mu \ , \\
    \mathcal{M}(\Psi^c_{\mu\tau} \to \Psi_{L_\tau}^c+\Phi^\dagger) &= c_{\tau 0}^* + c_{\tau 1}^* \mathcal{F}_\tau \ .
\end{align}
Here, $c_{\mu0}$ and $c_{\tau0}$
are tree-level amplitudes, and 
$c_{\mu1}$ and $c_{\tau1}$ are
one-loop amplitudes where the loop-integration functions $\mathcal{F}_{\mu,\tau}$ are factored out.
In this case, the lepton asymmetry is proportional to the sum of,
\begin{align}
    |\mathcal{M}(\Psi_{\mu\tau}^c \to \Psi_{L_\mu}+\Phi)|^2 -
    |\mathcal{M}(\Psi_{\mu\tau} \to \Psi_{L_\mu}^c+\Phi^\dagger)|^2
    &=
    - 4\mathrm{Im}[ c_{\mu 0}^* c_{\mu 1} ]\mathrm{Im} [\mathcal{F}_\mu]
    \ ,
    \\
    |\mathcal{M}(\Psi_{\mu\tau} \to \Psi_{L_\tau}+\Phi)|^2
    -|\mathcal{M}(\Psi_{\mu \tau}^c \to \Psi_{L_\tau}^c+\Phi^\dagger)|^2 
    &=
    - 4\mathrm{Im}[ c_{\tau 0}c_{\tau 1}^* ]\mathrm{Im} [\mathcal{F}_\tau]
    \ .
\end{align}

As in the case of the decay of $\Psi_e$, we look at the vertex diagrams with only Dirac Yukawa couplings, and we find 
\begin{align}
\label{eq:cmu}
    c_{\mu 0} \propto \lambda_\mu 
    \ , \quad\
    c_{\mu 1} \propto \lambda_\mu |\lambda_\tau|^2  \ ,
    \\
\label{eq:ctau}
    c_{\tau 0} \propto \lambda_\tau^* 
    \ , \quad\
    c_{\tau 1} \propto \lambda_\tau^* |\lambda_\mu|^2 \ .
\end{align}
Note that, in the one-loop vertex diagrams of the decay mode into $\Psi_{L_\mu}$, only $\Psi_{L_\tau}^c$ appears as a virtual state, and vice versa.
From Eqs.\,\eqref{eq:cmu} and \eqref{eq:ctau}, we find that the decay of $\Psi_{\mu\tau}$ does not generate the lepton asymmetry.
As a result, the decay of right-handed neutrinos in the $\Umt$ symmetric phase does not generate the lepton asymmetry.
This consequence coincides with that in previous work~\cite{Adhikary:2006rf}.\footnote{In the reference, thermal leptogenesis has been argued with Dirac Yukawa couplings and Majorana masses for right-handed neutrinos under the exact $L_\mu-L_\tau$ symmetry.}

For $M_{\mu\tau}$ and $M_{ee}=\mathcal{O}(10)$\,GeV leptogenesis via right-handed neutrino oscillations can be considered~\cite{Akhmedov:1998qx,Asaka:2005pn}.
In the $\Umt$ symmetric phase, however,
the right-handed neutrinos have different $\Umt$ charges, and hence,
the oscillations among them do not occur.
Thus, leptogenesis via right-handed neutrino oscillations cannot work. 

\subsection{Leptogenesis in the \texorpdfstring{$\Umt$}{} broken phase}
\label{subsec: leptogenesis in broken}

Below let us discuss leptogenesis in the $\Umt$ broken phase.
As we have seen in Sec.\,\ref{subsec:breaking}, the $\Umt$ symmetry can be broken by the thermal effects 
for $\lambda_{\Phi \sigma}<0$
with $|\lambda_{\Phi \sigma}| \gg \lambda_{\sigma}$, $g_{Z'}^2$.
Hereafter, we assume that both $\sigma_{1}$ 
and $\sigma_{2}$ obtain non-vanishing expectation values in the early universe.
In the broken phase, 
$\Umt$ charge is no longer conserved,
and hence, leptogenesis by the right-handed neutrinos may take place.
To discuss leptogenesis in the broken phase, we take the Majorana mass eigenstates of the right-handed neutrinos $\Psi'_I$ ($I=1,2,3$) 
with the mass eigenvalues $M_I$
(see Eqs.\,\eqref{eq:Takagi factorization} and \eqref{eq: Diagonal Majorana}).
Due to the nonzero $\langle \sigma_{1,2} \rangle$, one degree of freedom in $\sigma_{1,2}$ is absorbed into the gauge boson $Z'$, and the rest remains as physical degrees of freedom, $S_1, S_2$, and $P$ (see the Appendix~\ref{sec:symmetrybreakingsector}).

As the simplest possibility, let us discuss the non-thermal leptogenesis where the inflaton mainly decays into the right-handed neutrinos~\cite{Kumekawa:1994gx,Asaka:1999jb}.
The lepton asymmetry is generated 
by the subsequent decay of the right-handed neutrinos into the SM leptons and the Higgs bosons.
In this scenario, the reheating temperature after inflation is much lower than all three right-handed neutrino masses, $T_R \ll M_I$.
This condition can be satisfied for $M_{e e}, M_{\mu \tau} \gg T_R$.
Such large mass parameters are consistent with the observed neutrino oscillations in the cases in Secs.\,\ref{subsec:degenerate in NO} and \ref{subsec:degenerate in IO}.

We also assume that the decay rates of the right-handed neutrinos are much larger than that of the inflaton, \ie, $\Gamma_{\mathrm{D},I} \gg \Gamma_\mathrm{inf}$.
In this case, since the Hubble expansion rate at the reheating is also much smaller than the decay rates of the right-handed neutrinos, 
we may approximate that $\langle \sigma_{1,2} \rangle \propto T$ is time-independent to discuss leptogenesis.

The yields of the right-handed neutrinos from the inflaton decay are given by
\begin{align}
    Y_I
    \equiv
    \frac{n_{I}}{s} 
    \simeq    
    \left. 
    \left(
    \frac{\rho_{R}}{s} \times \frac{n_\mathrm{inf}}{\rho_\mathrm{inf}}
    \right)
    \right|_{T = T_R} \times f_I
    \ .
\end{align}
Here, $s$ is the entropy density,
$\rho_{R,\mathrm{inf}}$ are the energy densities of the radiation and the inflaton, respectively,  and $n_{\mathrm{inf},I}$ are the number densities of the inflaton and each right-handed neutrino $\Psi'_I$.
The parameter $f_I$ is the number of the right-handed neutrinos expected at the decay of one inflaton.
Here, we assume that at least one of $f_I$ is of $\order{1}$.
In the second equality, we have used $\rho_{R}\simeq \rho_\mathrm{inf}$ at the reheating time. 
By noting $\rho_R/s \simeq T_R$, $n_\mathrm{inf}/\rho_\mathrm{inf} \simeq 1/m_\mathrm{inf}$ with $m_\mathrm{inf}$ being the inflaton mass, 
the yields amount to
\begin{align} 
    Y_I
    \simeq 
    \frac{T_R}{m_\mathrm{inf}} \times
    f_I \ .
\end{align}
As the right-handed neutrinos immediately decay after the production, the yield of the lepton asymmetry is given by
\begin{align}
\label{eq:lepton asymmetry}
    Y_L
    \equiv 
    \frac{\Delta n_L}{s} 
    =
    \sum_{I}
    \tilde{\epsilon}_I
     \times \frac{n_I}{s} 
     =
     \frac{T_R}{m_\mathrm{inf}} 
    \sum_I 
    \tilde{\epsilon}_I
    f_I\ ,
\end{align}
where $\Delta n_L$ 
is the difference between the number densities of $(L_\alpha, \bar{l}_{R \alpha}^\dagger)$ and $(L_\alpha^\dagger, \bar{l}_{R \alpha})$, and $\tilde{\epsilon}_I$ denotes the asymmetry parameters of each right-handed neutrino defined by
\begin{align}
    \tilde{\epsilon}_I
    =
    \sum_\alpha
    \frac{\Gamma(\Psi_I' \to \Psi_{L_\alpha} + \tilde{\Phi}) - \Gamma(\Psi_I' \to \Psi_{L_\alpha}^c + \tilde{\Phi}^\dagger)}{\Gamma_{\mathrm{D},I}}
    \ .
\end{align}
We decompose $\tilde{\epsilon}_I$ as
\begin{align}
    \tilde{\epsilon}_I
    &=
    \frac{\sum_\alpha [\Gamma(\Psi_I' \to \Psi_{L_\alpha} + \tilde{\Phi}) - \Gamma(\Psi_I' \to \Psi_{L_\alpha}^c + \tilde{\Phi}^\dagger)]}
    {\sum_\alpha [\Gamma(\Psi_I' \to \Psi_{L_\alpha} + \tilde{\Phi}) + \Gamma(\Psi_I' \to \Psi_{L_\alpha}^c + \tilde{\Phi}^\dagger)]}
    \times 
    \frac{\sum_\alpha[\Gamma(\Psi_I' \to \Psi_{L_\alpha} + \tilde{\Phi}) + \Gamma(\Psi_I' \to \Psi_{L_\alpha}^c + \tilde{\Phi}^\dagger)]}{\Gamma_{\mathrm{D},I}}
    \nonumber \\
    &\equiv
    \epsilon_I \times 
    \mathrm{Br}(\Psi'_I \to \Psi_L^{(c)} + \tilde{\Phi}^{(\dagger)})
    \ ,
\end{align}
where $\epsilon_I$ denotes the asymmetry parameters of each right-handed neutrino when $\Psi'_I$ decays into $\Psi_{L} + \tilde{\Phi}$ and $\Psi_L^c + \tilde{\Phi}^\dagger$ with the branching ratio,
$\mathrm{Br}(\Psi'_I \to \Psi_L^{(c)} + \tilde{\Phi}^{(\dagger)})$. 
Here, we sum over the flavor of leptons for both $\epsilon_I$ and the branching ratio.

To evaluate the lepton asymmetry, we present the relevant interactions of the right-handed neutrinos in the mass basis.
The couplings to $Z'$ are given by the covariant derivative as
\begin{align}
        \mathcal{L}'_\mathrm{kin}
    &=
    \frac{1}{2} \overline{\Psi'}_I 
    i\gamma^\mu
    \left( 
        \partial_\mu \delta_{I J}
        -
        i g_{Z'} \gamma^5 Z'_\mu Q'_{I J}
    \right) 
    \Psi'_J 
    \ ,
\end{align}
where $Q'_{I J}$ can be read from Eq.\,\eqref{eq:covariant D in mass basis}.
The Dirac Yukawa matrix, $\lambda'_\nu$, is given by
\begin{align}
    \mathcal{L}'_\mathrm{DY}
    &=
    - \lambda'_{\nu \alpha I}
    \overline{\Psi'}_I \tilde{\Phi} P_L \Psi_{L_\alpha}
    - \lambda^{\prime *}_{\nu \alpha I}
    \overline{\Psi}_{L_\alpha} P_R \tilde{\Phi}^\dagger \Psi'_I
    \ .
\end{align}
The breaking scalars couple with the right-handed neutrinos through the Majorana Yukawa term as
\begin{align}
    \mathcal{L}'_\mathrm{MY}
    &=
    - \sum_{X = S_1, S_2, P} \overline{\Psi'}_I F_{I J}^{X} X \Psi'_J
\end{align}
with
\begin{align}
    F_{I J}^X
    \equiv 
    \frac{1}{2\sqrt{2}}
    \left( 
        f^X_{I J} - i \gamma^5 g^X_{I J}
    \right)
    \ ,
\end{align}
for $X=S_1,S_2,P$.
See the Appendix~\ref{App:mass basis} for the explicit form of these couplings.

First, we evaluate the branching ratio, $\mathrm{Br}(\Psi'_I \to \Psi_{L}^{(c)} + \tilde{\Phi}^{(\dagger)})$.
In addition to the decay into the SM lepton and Higgs, the right-handed neutrino can also decay into another lighter right-handed neutrino and the breaking scalar: $\Psi'_I \to \Psi'_J + X$, or another lighter right-handed neutrino and the $\Umt$ gauge field: $\Psi'_I \to \Psi'_J + Z'$.
Here, we neglect $1 \to 3$ decay processes because their rates will be subdominant in the total decay rate due to the phase space suppression.
The tree decay rates of $\Psi'_I \to \Psi_L^{(c)} + \tilde{\Phi}^{(\dagger)}$, $\Psi'_I \to \Psi'_J + X$, and $\Psi'_I \to \Psi'_J + Z'$ are given by
\begin{align}
    \label{eq: tree decay PsiL Phi}
    \Gamma^{\Psi_L \Phi}_{I}
    &=
    \frac{M_I}{8 \pi} [\lambda^{\prime \dagger}_\nu \lambda'_\nu]_{I I}
    \ ,
    \\
    \label{eq: tree decay Psi' X}
    \Gamma^{\Psi'_J X}_{I}
    &=
    \frac{M_I}{128 \pi} 
    \left( 1 - r_J^2 \right)
    \left[ 
        (1 + r_J)^2 (f^{X}_{I J})^2 + (1 - r_J)^2 (g^{X}_{I J})^2
    \right]
    \Theta(M_I - M_J)
    \ ,
    \\
    \label{eq: tree decay Psi' Z'}
    \Gamma^{\Psi'_J Z'}_{I}
    &=
    \frac{g_{Z'}^2 M_I}{64 \pi} 
    \frac{\left( 1 - r_J^2 \right)^3}{r_{Z'}^2}
    |Q'_{I J}|^2 \Theta(M_I - M_J)
    \ ,
\end{align}
where $r_J \equiv M_J/M_I$, $r_{Z'} \equiv m_{Z'}/M_I$. 
Since $T_R \ll M_I$,
$m_{Z'}$ and $m_X$ evaluated at $T_R$ satisfy
$r_{Z'}\ll 1$ and $m_X/M_I \ll 1$ and we used these limits.
From these decay rates, we obtain the branching ratio as
\begin{align}
    \mathrm{Br}(\Psi'_I \to \Psi_{L}^{(c)} + \tilde{\Phi}^{(\dagger)})
    \simeq
    \frac{
        \Gamma^{\Psi_L \Phi}_{I}
    }{
        \Gamma_{\mathrm{tree}, I}
    }
    \equiv
    \frac{
        \Gamma^{\Psi_L \Phi}_{I}
    }{
        \Gamma^{\Psi_L \Phi}_{I}
        +
        \sum_{J,X} \Gamma^{\Psi'_J X}_{I}
        +
        \sum_J \Gamma^{\Psi'_J Z'}_{I}
    }
    \ ,
\end{align}
where $\Gamma_{\mathrm{tree},I}$ is the sum of Eqs.\,\eqref{eq: tree decay PsiL Phi}, \eqref{eq: tree decay Psi' X}, and \eqref{eq: tree decay Psi' Z'}.
Here, we used $\Gamma_{\mathrm{D}, I} 
\simeq \Gamma_{\mathrm{tree}, I}$.
Note that $\mathrm{Br}(\Psi'_I \to \Psi_{L}^{(c)} + \tilde{\Phi}^{(\dagger)}) = 1$ for the lightest right-handed neutrino, $I = 1$.

\begin{figure}[t]
    \begin{minipage}[b]{0.3\linewidth}
        \centering
        \includegraphics[width=0.95\linewidth]{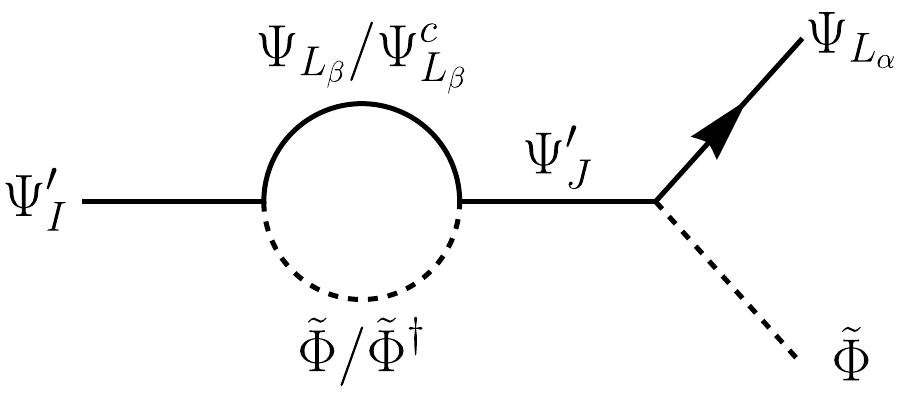}
        \\ {\footnotesize (a)}
    \end{minipage}
    \begin{minipage}[b]{0.3\linewidth}
        \centering
        \includegraphics[width=0.95\linewidth]{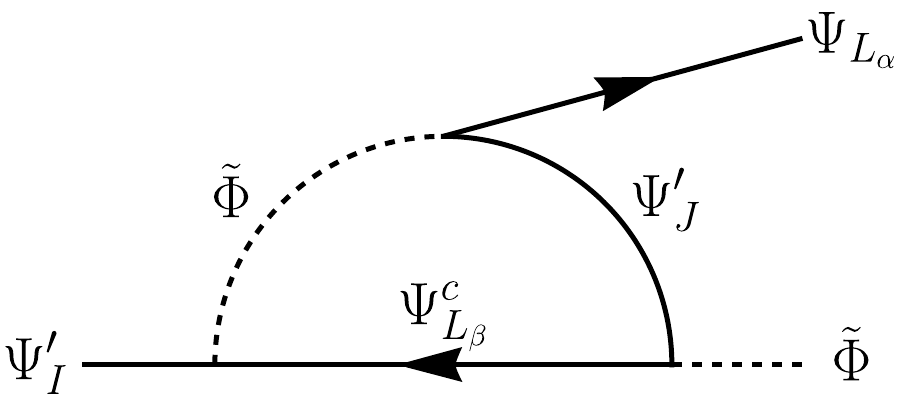}
        \\ {\footnotesize (b)}
    \end{minipage}
    \begin{minipage}[b]{0.3\linewidth}
        \centering
        \includegraphics[width=0.95\linewidth]{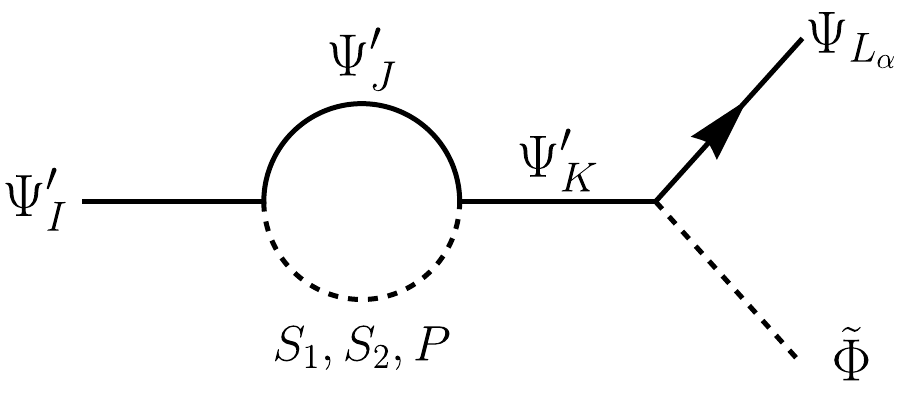}
        \\ {\footnotesize (c)}
    \end{minipage}
    \caption{\sl 
    One-loop diagrams for decays of right-handed neutrinos.
    }
    \label{fig:relevant diagrams}
\end{figure}

Next, we evaluate $\epsilon_I$.
To this end, we consider the tree and one-loop diagrams of the decay of the right-handed neutrinos into $\Psi_L^{(c)} + \tilde{\Phi}^{(\dagger)}$.
In our setup, three types of one-loop diagrams contribute to the asymmetry parameters.
In Fig.~\ref{fig:relevant diagrams}, we show the relevant diagrams: (a) wave-function ($\Psi_L \Phi$ loop), (b) vertex ($\Psi_L \Phi \Psi'$ loop), and (c) wave-function ($\Psi' X$ loop).
The contributions from other diagrams are negligible as discussed in the Appendix~\ref{App:asymmetry}.
Thus, $\epsilon_I$ is given by the sum of asymmetry parameters from the three diagrams as
\begin{align}
    \epsilon_I
    =
    \epsilon^{(a)}_I + \epsilon^{(b)}_I + \epsilon^{(c)}_I
    \ ,
\end{align}
where
\begin{align}
    \epsilon^{(a)}_I
    = &
    - \frac{1}{8 \pi}
    \sum_J
    \frac{ M_I M_J }{M_I^2 - M_J^2}
    \frac{(M_I^2 - M_J^2)^2}{(M_I^2 - M_J^2)^2 + M_I^2 \Gamma_{\mathrm{tree}, J}^2}
    \frac{\mathrm{Im}[ 
        [\lambda^{\prime \dagger}_\nu \lambda'_\nu]_{I J}^2
    ]}
    {[\lambda^{\prime \dagger}_\nu \lambda'_\nu]_{I I}}
    \ ,
    \\
    \epsilon^{(b)}_I
    = &
    - \frac{1}{8 \pi } 
    \sum_J
    \frac{
        \mathrm{Im} [ [\lambda^{\prime \dagger}_\nu \lambda^{\prime}_\nu]_{I J}^2 ]
    }{[\lambda^{\prime \dagger}_\nu \lambda'_\nu]_{I I}}
   r_J \left[ 
        1  - ( 1 + r_J^2 ) \log \left( \frac{1 + r_J^2}{ r_J^2 } \right)
    \right]
    \ ,
    \\
    \epsilon^{(c)}_I
    = &
    \sum_{J,K}
    \frac{1 - r_J^2}{128 \pi (1 - r_K^2) [\lambda^{\prime \dagger}_\nu \lambda'_\nu]_{I I}}
    \frac{(M_I^2 - M_K^2)^2}{(M_I^2 - M_K^2)^2 + M_I^2 \Gamma_{\mathrm{tree}, K}^2}
    \Theta(M_I - M_J)
    \nonumber \\
    & \times 
    \sum_{X = S_1, S_2, P}
    \left[
        - (1 + r_K)
        \mathrm{Im}[ \lambda_{\nu}^{\prime \dagger} \lambda'_{\nu}]_{I J} 
        \left\{
            (1 + r_J)^2 f^X_{I J} f^X_{J K}
         +  (1 - r_J)^2 g^X_{I J} g^X_{J K}  
        \right\}
    \right.
    \nonumber \\
    & \hspace{21mm}  +
    \left.
        (1 - r_K) 
        \mathrm{Re}[ \lambda_{\nu}^{\prime \dagger} \lambda'_{\nu}]_{I J}
        \left\{
            (1 + r_J)^2 f^X_{I J} g^X_{J K}
          - (1 - r_J)^2 g^X_{I J} f^X_{J K}           
        \right\}
    \right]
    \ .
\end{align}
The fractions with $\Gamma_{\mathrm{tree}, J}^2$ and $\Gamma_{\mathrm{tree}, K}^2$ come from the finite width of the right-handed neutrinos and regulates the asymmetry parameters in the resonant limit, $M_I \to M_{J,K}$~\cite{Pilaftsis:2003gt}.

From these formulae, we evaluate the asymmetry parameters in the normal and inverted ordering of active neutrino masses assuming the parameters obtained in Secs.\,\ref{subsec:degenerate in NO} and \ref{subsec:degenerate in IO}.
The results are summarized in Tab.~\ref{tab:asymmetry parameters}.
Here, we fix the expectation values of $\sigma_{1,2}$ in the vacuum and at the reheating for both the orderings as
\begin{gather}
    \langle \sigma_{1} \rangle_0 = 80\,\mathrm{GeV}
    \ , \quad 
    \langle \sigma_{2} \rangle_0 = 50\,\mathrm{GeV}
    \ , \quad  
    \langle \sigma_{1,2} \rangle_{T_R} = 10^5\,\mathrm{GeV}
    \ ,
\end{gather}
which can be realized by
\begin{gather}
    \lambda_{\sigma_1} = \lambda_{\sigma_2} = 3 \times 10^{-4}
    \ , \quad 
    \lambda_{\Phi \sigma_1} = \lambda_{\Phi \sigma_2} = - 3 \times 10^{-3}
    \ , \quad 
    T_R \simeq 3 \times 10^4\,\mathrm{GeV}
    \ .
\end{gather}
\setlength{\tabcolsep}{2.5mm}
{\renewcommand{\arraystretch}{1.2}
\begin{table}[t]
\begin{minipage}{\textwidth}
    \centering
    \caption{Asymmetry parameters in our setup.
    }
    \label{tab:asymmetry parameters}
    \begin{tabular}{ c | c | c c c | c }
        \hline
        \multicolumn{2}{c|}{} & \multicolumn{3}{c|}{Diagrams} & \multirow{2}{*}{Total $\epsilon$}
        \\
        \cline{3-5}
        \multicolumn{2}{c|}{} & (a) & (b) & (c) &
        \\
        \hline
        \multirow{3}{*}{Normal Ordering} & $\epsilon_1$ &
        $\phantom{-} 7 \times 10^{-8}$ & $\phantom{-} 3 \times 10^{-8}$ &
        $0$ & $\phantom{-} 1 \times 10^{-7}$
        \\
        & $\epsilon_2$ &
        $\phantom{-} 5 \times 10^{-7}$ & $- 4 \times 10^{-7}$ &
        $- 5 \times 10^{-6}$ & $- 5 \times 10^{-6}$
        \\
        & $\epsilon_3$ & 
        $- 5 \times 10^{-7}$ & $\phantom{-} 4 \times 10^{-7}$ &
        $- 5 \times 10^{-6}$ & $- 5 \times 10^{-6}$
        \\
        \hline 
        \multirow{3}{*}{Inverted Ordering} & $\epsilon_1$ &
        $- 5 \times 10^{-8}$ & $- 2 \times 10^{-8}$ &
        $0$ & $ - 6 \times 10^{-8}$
        \\
        & $\epsilon_2$ &
        $- 8 \times 10^{-8}$ & $\phantom{-} 7 \times 10^{-8}$ &
        $\phantom{-} 3 \times 10^{-6}$ & $\phantom{-} 3 \times 10^{-6}$
        \\
        & $\epsilon_3$ & 
        $\phantom{-} 6 \times 10^{-8}$ & $- 7 \times 10^{-8}$ &
        $\phantom{-} 3 \times 10^{-6}$ & $\phantom{-} 3 \times 10^{-6}$
        \\
        \hline
    \end{tabular}
\end{minipage}
\end{table}
}

In the normal ordering case, we also fix the parameters as\footnote{%
Here, we consider a large value of $h_{e \mu}$, whose effect was not discussed in the previous section.
Since we investigate non-thermal leptogenesis, the right-handed neutrinos are not in the thermal bath.
They decay promptly after being generated from the inflatons, and hence we expect that they do not alter the potential of $\sigma$.
Even if finite density effects of the right-handed neutrinos exist, such effects are negligible because their number density is much smaller than that of other particles in the thermal bath.
}
\begin{gather}
    |M_{e e}| = 5 \times 10^6\,\mathrm{GeV}
    \ , \quad 
    \lambda_\mu = 1
    \ , \quad 
    |h_{e \tau}| = 1
    \ .
\end{gather}
In this case, the mass eigenvalues of the right-handed neutrinos become
\begin{align}
\begin{gathered}
    M_1 \simeq 5.0 \times 10^6\,\mathrm{GeV}
    \ , \quad 
    M_2 \simeq M_3 \simeq 1.4 \times 10^7\,\mathrm{GeV}
    \ ,
    \\
    M_3 - M_2 \simeq 1.2 \times 10^3\,\mathrm{GeV}
    \ .
\end{gathered}
\end{align}
Here, $M_2$ and $M_3$ are highly degenerate because they are dominated by $M_{\mu \tau}$ and their degeneracy is slightly broken by $h_{\tau \tau} \langle \sigma_2 \rangle_{T_R}$, which is much smaller than $M_{\mu \tau}$.

On the other hand, in the inverted ordering case, we fix
\begin{gather}
    |M_{e e}| = 5 \times 10^6\,\mathrm{GeV}
    \ , \quad 
    \lambda_\tau = 1
    \ , \quad 
    |h_{e \mu}| = 1
    \ ,
\end{gather}
which leads to
\begin{align}
\begin{gathered}
    M_1 \simeq 5.0 \times 10^6\,\mathrm{GeV}
    \ , \quad 
    M_2 \simeq M_3 \simeq 1.4 \times 10^7\,\mathrm{GeV}
    \ ,
    \\
    M_3 - M_2 \simeq 5.2 \times 10^2\,\mathrm{GeV}
    \ ,
\end{gathered}
\end{align}
where $M_2$ and $M_3$ are highly degenerate as in the normal ordering case.

Finally, we evaluate the resultant yield of the lepton asymmetry.
Since the observed baryon asymmetry is $n_B/s \simeq 8.7 \times 10^{-11}$~\cite{Planck:2018vyg} and the sphaleron processes convert the lepton asymmetry generated at high temperatures into baryon asymmetry as $n_B \simeq (28/79) \times n_{B-L}$, the success of leptogenesis requires $Y_L \simeq - 2.5 \times 10^{-10}$.

We first consider the case where the inflaton mainly decays into $\Psi'_1$.
In this case, we adopt $f_1 = 2$ and $f_2 = f_3 = 0$ as a typical value when the inflaton decays into a pair of the right-handed neutrinos.
From Eq.\,\eqref{eq:lepton asymmetry}, the lepton asymmetry becomes
\begin{align}
    Y_L
    \simeq 
    \frac{2 T_R}{m_\mathrm{inf}} \epsilon_1
    \ .
\end{align}
Since the decay of the inflaton into a pair of $\Psi'_1$ requires $m_\mathrm{inf} > 2 M_1$, we obtain the maximum lepton asymmetry as
\begin{align}
 \label{eq:LAPsi1}
    Y_L
    \lesssim 
    \frac{T_R}{M_1} \epsilon_1
    \simeq 
    \left\{
    \begin{aligned}
        6 \times 10^{-10} 
        \quad &(\text{Normal Ordering})
        \\
        - 4 \times 10^{-10}
        \quad &(\text{Inverted Ordering})
    \end{aligned}
    \right.
    \ .
\end{align}
This value in the inverted ordering is compatible with the observed baryon asymmetry in the universe.
As we will see below, the washout effects reduce $Y_L$ by a factor of a few at most.

Next, we consider the case where the inflaton mainly decays into $\Psi'_2$ and $\Psi'_3$.
As a typical value, we adopt $f_2 = f_3 = 1$.
While $\Psi'_{2,3}$ mainly decays into the left-handed lepton and Higgs, a part of them generates $\Psi'_1$.
Thus, even if the inflaton does not directly decay into $\Psi'_1$, $f_1$ is effectively nonzero and estimated as%
\footnote{The branching ratio of $\Psi'_3$ into $\Psi'_2$ is negligible due to the degeneracy of the masses.}
\begin{align}
    f_1
    =
    \sum_{I = 2, 3} 
    f_I
    \left[
        1 - \mathrm{Br}(\Psi'_I \to \Psi_{L}^{(c)} + \tilde{\Phi}^{(\dagger)})
    \right]
    \simeq 
    0.22
    \ ,
\end{align}
for both the normal and inverted orderings.
From $m_\mathrm{inf} \gtrsim 2 M_3$, we obtain the maximum lepton asymmetry as
\begin{align}
 \label{eq:LAPsi23}
    Y_L
    \lesssim 
    \frac{T_R}{2 M_3} \sum_I 
    \tilde{\epsilon}_I f_I 
    \simeq 
    \left\{
    \begin{aligned}
        -1 \times 10^{-8} 
        \quad &(\text{Normal Ordering})
        \\
        6 \times 10^{-9}
        \quad &(\text{Inverted Ordering})
    \end{aligned}
    \right.
    \ .
\end{align}
This value in the normal ordering is compatible with the observed baryon asymmetry of the universe.

Note that the signs of asymmetry parameters of leptogenesis above are reversed when we adopt the CP phases opposite to those in Secs.~\ref{subsec:degenerate in NO} and \ref{subsec:degenerate in IO}.
However, such values of the Dirac phase $\delta$ ($\sim 90^\circ$) are disfavored in the current neutrino oscillation experiments.

\subsection{Wash-out effects in non-thermal leptogenesis}

In the presence of lepton number violating processes, lepton asymmetry in the thermal bath is washed out due to the unbalance between the rates of the processes including leptons and anti-leptons.
We evaluate the wash-out effects from the inverse decay of $\Psi'_I$ and lepton number violating scatterings with $\Delta L = 1$ and $2$ in our scenario.
It turns out that the effects are not significant with $T_R \lesssim 4 \times 10^4$\,GeV
in the case of the benchmark point in the previous subsection.

\subsubsection{Inverse decay}
The inverse decay of $\Psi'_I$ from the thermal bath violates lepton number by $\Delta L = 1$. 
Since non-thermal leptogenesis requires $M_{I} \gg T_R$, the inverse decay rate is suppressed.
We here estimate the condition of $z_I \equiv M_I/T_R$ for the wash-out effect of the inverse decay to be inefficient.
The inverse decay rate, $\Gamma_{\mathrm{ID}, I}$, is given by
\begin{align}
 \begin{split}
    \Gamma_{\mathrm{ID}, I}
    &=
    \Gamma_I^{\Psi_L \Phi}
    \times\frac{n_I^\mathrm{eq}}{n^{\mathrm{eq}}}, \\
    n_I^\mathrm{eq}
    &= g_N \left( \frac{M_{I} T}{2 \pi} \right)^{3/2} e^{-\frac{M_{I}}{T}}, \\
    n^\mathrm{eq}
    &= \frac{3}{4} \frac{\zeta (3)}{\pi^2} T^3,
 \end{split}
\end{align}
where $g_N=2$.%
\footnote{%
    In the limit of $\Delta n_L / n^\mathrm{eq} \ll 1$, the lepton asymmetry follows
    $\Delta \dot{n}_L + 3 H \Delta n_L
    \supset
    - \Gamma_{\mathrm{ID}, I} \Delta n_L$.
}
From the above equations, $\Gamma_{\mathrm{ID}, I}$ is obtained with respect to $z_I$ as
\begin{align}
    \Gamma_{\mathrm{ID}, I}(z_I)
    \simeq 
    \frac{1}{8\pi}\sum_\alpha |\lambda'_{\alpha I}|^2 M_{I} z_I^{3/2} e^{-z_I},
\end{align}
and
\begin{align}
    \frac{\Gamma_{\mathrm{ID}, I}(z_I)}{H(T_R)}
    \simeq 
    8\times10^9 \sum_\alpha |\lambda'_{\alpha I}|^2 \left( \frac{5 \times 10^6 \,\mathrm{GeV}}{M_{I}} \right) z_I^{7/2} e^{-z_I}.
\end{align}
From $\sum_\alpha |\lambda'_{\alpha I}|^2 \sim 1$ in our setup, it suggests that $\Gamma_{\mathrm{ID}, I}(z_I)/H(T_R) <1$ can be satisfied for all $I$'s if
\begin{align}
    T_R \lesssim 1.4 \times 10^5 \,\mathrm{GeV}
    \ ,
\end{align}
when $M_1 \simeq 5 \times 10^6$\,GeV and $M_{2,3} \simeq 1.4 \times 10^7$\,GeV.

\subsubsection{\texorpdfstring{$\Delta L = 1$}{} scatterings}
Scattering processes of $\Psi^\prime_{I}$ and left-handed leptons with $\Delta L = 1$ also contribute to the wash-out effects, $\eg$, $\Psi'_I + \Psi_{L_\alpha} \to f + \bar{f}$, where $f$ is a SM fermion. 
Here, the right-handed neutrinos do not appear in the final state since $M_I \gg T_R$ in non-thermal leptogenesis.
The Boltzmann equation of the lepton asymmetry in terms of $Y_L$ is given by,
\begin{align}
    \dot{Y}_L
    \simeq 
    \sum_I \left( 
        \tilde{\epsilon}_I \Gamma_{\mathrm{D},I}Y_I
        -
        \Gamma_{\Delta L = 1, I} Y_L Y_I
    \right) 
    \ .
\end{align}
where the second term expresses the wash-out effect from the $\Delta L = 1$ scatterings.
Here, $Y_L$ is at most $\sum_I \tilde{\epsilon}_I Y_I(T_R) = \sum_I \tilde{\epsilon}_I f_I T_R/m_\mathrm{inf}$.
Assuming that the inflaton mainly decays into one of $\Psi'_I$ with favorable $\tilde{\epsilon}_I$, 
it is sufficient to focus on such a $\Psi'_I$ in the Boltzmann equation;
\begin{align}
    \dot{Y}_L
    &\gtrsim 
    \left( 
        \Gamma_{\mathrm{D},I} - \Gamma_{\Delta L = 1, I} 
        f_I 
        \frac{T_R}{m_\mathrm{inf}}
    \right) \tilde{\epsilon}_I Y_I
    \ .
\end{align}
This shows that the wash-out effect due to the $\Delta L = 1$ scattering becomes negligible when $\Gamma_{\mathrm{D},I} \gg \Gamma_{\Delta L = 1, I}$ because $T_R/m_\mathrm{inf} \ll 1$ and $f_I = \order{1}$.
Actually, for $T \ll M_I$, $\Gamma_{\mathrm{D},I} \gg \Gamma_{\Delta L = 1, I}$ is satisfied 
for $\Psi_I'$'s interacting with the SM thermal bath as seen in, for example, Fig.\,5 of Ref.\,\cite{Giudice:2003jh}.
We have also checked that $\Gamma_{\mathrm{D},I} \gg \Gamma_{\Delta L = 1, I}$ is satisfied when we consider interactions involving the $\Umt$ sector particles.

\subsubsection{\texorpdfstring{$\Delta L = 2$}{} scatterings}
Finally, scattering processes between left-handed leptons and SM Higgs with a virtual $\Psi'_{I}$ can violate the lepton number by $\Delta L = 2$ like $\Psi_{L_\alpha} \Phi \to \Psi_{L_\beta}^c \Phi^\dagger$ and $\Psi_{L_\alpha} \Psi_{L_\beta} \to \Phi  \Phi$.

For $T\ll M_I$, the invariant amplitude squared is given by
\begin{align}
    |\mathcal{M}_{\Delta L = 2}|^2
    \simeq 
    \sum _I N_\mathrm{d} \left|
        \lambda'_{\beta I} M_I^{-1} \lambda'_{\alpha I}
    \right|^2 
    (p_{L_{\alpha}}\cdot p_{L_{\beta}}) 
    =
    N_\mathrm{d} \left|
        \lambda_{\beta} [M_{R, \mathrm{eff}}^{-1}]_{\beta \alpha} \lambda_{\alpha}
    \right|^2
    (p_{L_{\alpha}}\cdot p_{L_{\beta}}),
\end{align}
where $N_\mathrm{d} = 10$ is a numerical coefficient from relevant diagrams ($N_\mathrm{d}=8$ comes from 
$\Psi_{L_\alpha} \Phi \to \Psi^c_{L_\beta} \Phi^\dagger$ and $N_\mathrm{d} = 2$ comes from $\Psi_{L_\alpha}\Psi_{L_\beta} \to \Phi  \Phi$), and $p_{L_\alpha}$ is the four-momentum of left-handed lepton with $\alpha$ flavor.
We used the relations of Yukawa couplings and masses between the flavor and mass bases in the last equality (see Appendix \ref{App:mass basis}).
From this amplitude squared, the cross-section of the $\Delta L =2$ scatterings is obtained as
\begin{align}
     [\sigma v]_{\beta \alpha} 
     \sim
     \frac{N_\mathrm{d}}{32 \pi} 
     \left|
        \lambda_{\beta} [M_{R, \mathrm{eff}}^{-1}]_{\beta \alpha} \lambda_{\alpha}
    \right|^2,
\end{align}
where, as a typical value, we took the energies of SM particles as $T$.

For the neutrino oscillation parameters discussed in Sec.\,\ref{subsec:degenerate in NO} for the normal ordering, $h_{\mu \mu} = h_{e \mu} = 0$, which leads to
\begin{align}
    M_{R,\mathrm{eff}}^{-1} 
    =
    \begin{pmatrix}
        M_{e e}^{-1} & - \frac{h_{e \tau}\langle \sigma_1 \rangle}{M_{e e} M_{\mu \tau}} & 0
        \\
        - \frac{h_{e \tau} \langle \sigma_1 \rangle}{M_{e e}} & 
        - \frac{h_{e \tau}^2 \langle \sigma_1 \rangle^2 + h_{ \tau \tau} M_{e e} \langle \sigma_2 \rangle }{M_{e e} M_{\mu \tau}^2} & M_{\mu \tau}^{-1}
        \\
        0 & M_{\mu \tau}^{-1} & 0
    \end{pmatrix}
    \ .
\end{align}
For the benchmark point of non-thermal leptogenesis discussed in Sec.\,\ref{subsec: leptogenesis in broken},
\begin{align}
 \begin{gathered}
    \lambda_e \sim 10^{-4}
    \ , \quad 
    \lambda_\mu \sim 1
    \ , \quad 
    \lambda_\tau \sim 10^{-8}
    \ , \quad 
    h_{e \tau} \sim 1
    \ , \quad 
    h_{\tau \tau} \sim 10^{-3}
    \ ,
    \\
    M_{e e} \sim 5 \times 10^6\,\mathrm{GeV}
    \ , \quad
    M_{\mu \tau} \sim 10^7\,\mathrm{GeV}
    \ , \quad 
    \langle \sigma_i \rangle \sim T_R \sim 10^5\,\mathrm{GeV}
    \ ,
 \end{gathered}
\end{align}
the largest wash-out effect is for $\alpha=\beta=\mu$, and the scattering rate is given by
\begin{align}
    \Gamma_{\Delta L=2}|_{\mu \mu} \equiv [\sigma  v]_{\mu \mu} n^\mathrm{eq}
    \simeq \frac{N_\mathrm{d} \, \zeta (3)}{32 \pi^3} |\lambda_{\mu}|^4 |[M_{R, \mathrm{eff}}^{-1}]_{\mu \mu}|^2 T^3.
\end{align}
Then, $\Gamma_{\Delta L=2}|_{\mu \mu} < H$ requires
\begin{align}
    T_R 
    \lesssim 
    4 \times 10^4\, \mathrm{GeV}.
\end{align}
Since $T_R$ in our benchmark point is close to this bound, the generated lepton asymmetry estimated in the previous subsection may be reduced by about a factor of two.
Therefore, the result for inverted ordering in Eq.\,\eqref{eq:LAPsi1} is barely consistent with the observed baryon asymmetry.%
\footnote{In our analysis, we have not sought the optimal value of the lepton asymmetry for each ordering, and it is possible to achieve 
a larger lepton asymmetry by a factor of $\order{1}$.
}
On the other hand, that for normal ordering in Eq.\,\eqref{eq:LAPsi23} is sufficiently large to explain the observed value.

\section{Conclusion and Discussion}
\label{sec:conclusion}

An extension of the SM with the gauged $\Umt$ symmetry can explain the muon $g-2$ anomaly.
The gauged $\Umt$ symmetry is also consistent with the observed neutrino oscillations through the seesaw mechanism where three right-handed neutrinos $\bar{N}_e, \bar{N}_\mu, \bar{N}_\tau$ have the $\Umt$ charges $0, -1, +1$, respectively.
In this paper, we investigated if leptogenesis can work while explaining the neutrino masses and muon $g-2$ anomaly at the same time.%
\footnote{In Ref.\,\cite{Borah:2021mri}, leptogenesis 
has been discussed
in a setup
where the $Z'$ mass and the $\Umt$ breaking scale are highly separated by considering a hierarchical charge assignment between the $\Umt$ breaking fields and the SM leptons.
}

In our discussion, we sought the scenario where all the right-handed neutrinos are much heavier than the electroweak scale.
Such a spectrum is highly non-trivial because right-handed neutrino masses are typically tied to the $\Umt$ breaking scale, $10 \text{--} 100$\,GeV.
Nevertheless, we found that it is possible that all the right-handed neutrinos 
can be as heavy as $10^7$\,GeV only when the Yukawa couplings have specific structures as in Eqs.\,\eqref{eq:seesaw degenerate NO} and \eqref{eq:seesaw degenerate IO}.

We also found that leptogenesis requires $\Umt$ symmetry breaking in the early universe because the $\Umt$ symmetry prohibits the flavor mixing among the right-handed neutrinos.
As we have seen, however,
the gauge and self interactions of the breaking fields $\sigma$'s
tend to restore the $\Umt$ symmetry.
Thus, to achieve the $\Umt$ broken 
phase in the early universe, 
we assumed a sizable negative value of Higgs--$\sigma$ coupling $\lambda_{\Phi \sigma}$, which leads to $\langle \sigma \rangle \sim \sqrt{|\lambda_{\Phi \sigma}|/\lambda_{\sigma}} \, T$.

With these observations, we considered non-thermal leptogenesis and estimated the generated lepton asymmetry for both orderings of active neutrino masses.
Even taking into account the wash-out effects on the asymmetry, we found the parameter points to generate a sufficient lepton asymmetry compatible with the observed baryon asymmetry of the universe.
Therefore, we conclude that this model can explain the above three phenomena beyond the SM simultaneously.

In closing, we should mention that this scenario will be tested from various aspects in near future. 
Firstly, the extra neutral gauge boson $Z^\prime$ explaining the muon $g-2$ anomaly will be probed by COHERENT~\cite{Abdullah:2018ykz} and NA64$\mu$ at CERN~\cite{Gninenko:2014pea,Gninenko:2018tlp}.
Secondly, improvement of the upper bound on the sum of active neutrino masses from the CMB observations will probe this scenario.
To have the right-handed neutrino masses much larger than $\langle \sigma_{1,2} \rangle_0$, our scenario requires $\sum m_i \gtrsim 0.18$\,eV, which will be robustly tested in the future CMB observations such as CMB-S4~\cite{Abazajian:2019eic}.
Finally, theoretical and experimental progress on the $0 \nu \beta \beta$ decay will also be important for its test because relatively large values of the effective neutrino mass are suggested for this successful non-thermal leptogenesis as Eqs.\,\eqref{eq:mbetabetaNO} and \eqref{eq:mbetabetaIO}.

\section*{Acknowledgments}
The authors thank K.~Asai and T.~Shimomura for useful communications.
The authors also thank S.~Kobayashi for his useful Feynman diagram drawer.
This work was supported by JSPS KAKENHI Grant Nos. 18H05542, 21H04471, 22K03615 (M.I.), and 20J20248 (K.M.).
K.M. was supported by the Program of Excellence in Photon Science.

\appendix

\section{Symmetry breaking sector}
\label{sec:symmetrybreakingsector}
In this appendix, we summarize the symmetry breaking sector of the $\Umt$ model.
In our scenario, we considered
two SM singlet scalar bosons, $\sigma_{1,2}$, with the $\Umt$ charge $+1$ and $+2$, respectively.

\subsection{Model with a single scalar}
For the sake of brevity, 
let us first consider the model with a single scalar, $\sigma$.
The scalar potential of 
$\sigma$ and the Higgs doublet $\Phi$ is given by,
\begin{align}
V(\Phi,\sigma)
    =
    - \mu_\sigma^2 |\sigma|^2 
    - \mu_\Phi^2 \Phi^\dagger \Phi 
    + \lambda_\sigma |\sigma|^4 
    + \lambda_\Phi \left( \Phi^\dagger \Phi \right)^2
    + \lambda_{\Phi \sigma} |\sigma|^2 (\Phi^\dagger \Phi)\ ,
\end{align}
where $\mu_\sigma^2$, $\mu_\Phi^2$ express mass parameters for each scalar field, $\lambda_\sigma$, $\lambda_\Phi$ are quartic self-couplings, and $\lambda_{\Phi \sigma}$ is a Higgs--$\sigma$ coupling.
As we have discussed in Sec.\,\ref{subsec:breaking},
we assume $\lambda_{\Phi\sigma}<0$,
while 
$\lambda_{\sigma}$ and $\lambda_{\Phi}$ are positive.
Around the vacuum, we decompose $\Phi$ and $\sigma$ as,
\begin{align}
&\Phi = \left(
    \begin{array}{c}
         H^+ \\
         v_\mathrm{EW} +
         \displaystyle{\frac{1}{\sqrt{2}}}(
         \hat{H}+i a)
    \end{array}
    \right)\ , \quad
    \sigma = v + \frac{1}{\sqrt{2}}
    (
    \hat{S}+i A
    )\ ,
\end{align}
where $v_\mathrm{EW}$ and $v$ are VEVs .
The charged Higgs scalar $H^+$ and the CP-odd scalars $a$ and $A$ are would-be Goldstone modes, which are set to be zero in the unitary gauge.

From $\partial V/\partial \hat{H}=\partial V/\partial \hat{S} = 0$, we find 
\begin{align}
    \mu_{\Phi}^2 = 2\lambda_\Phi v_\mathrm{EW}^2 + \lambda_{\Phi \sigma} v^2 \ ,  \quad
    \mu_{\sigma}^2 = 2\lambda_\sigma v^2 + \lambda_{\Phi \sigma} v_\mathrm{EW}^2 \ .
\end{align}
The squared masses of the CP-even scalars are given by,
\begin{align}
M^2 =    \left(
    \begin{array}{cc}
      4\lambda_{\Phi} v_\mathrm{EW}^2 & 
      2 \lambda_{\Phi \sigma}v_\mathrm{EW} v 
      \\
      2 \lambda_{\Phi \sigma}v_\mathrm{EW} v  &
    4\lambda_{\sigma} v^2
    \end{array}
    \right)\ ,
\end{align}
in the $(\hat{H},\hat{S})$ basis.
The mass eigenstates are obtained by,
\begin{align}
    \left(
    \begin{array}{c}
      H \\
    S  
    \end{array}
    \right)
    =  \left(
    \begin{array}{cc}
      \cos\theta & 
      \sin\theta 
      \\
       -\sin\theta   &
    \cos\theta
    \end{array}
    \right) 
    \left(
    \begin{array}{c}
      \hat{H} \\
    \hat{S}  
    \end{array}
    \right)\ ,
\end{align}
where
\begin{align}
\label{eq:mixing}
    \tan 2\theta \simeq \frac{\lambda_{\Phi\sigma} v \, v_\mathrm{EW}}{\lambda_{\Phi}v_\mathrm{EW}^2 - \lambda_{\sigma} v^2}\ .
\end{align}
As we are interested in the parameter region where $\lambda_{\Phi} \gg |\lambda_{\Phi\sigma}|, \lambda_{\sigma}$
with $v \simeq 10 \text{\,--\,} 100$\,GeV, 
we approximate Eq.\,\eqref{eq:mixing} by,
\begin{align}
    \theta \simeq \frac{1}{2}
    \frac{\lambda_{\Phi\sigma}v \, v_\mathrm{EW}}{\lambda_{\Phi} v_\mathrm{EW}^2} \simeq \frac{2 \lambda_{\Phi\sigma}v \, v_\mathrm{EW}}{m_H^2}\ ,
\end{align}
where the mass eigenvalues are given by,
\begin{align}
    m_H^2 = 4 \lambda_\Phi v_\mathrm{EW}^2 (1 + \order{\lambda_{\Phi\sigma}^2})\ , \quad
     m_S^2 = 4 \lambda_\sigma v^2 (1 + \order{\lambda_{\Phi\sigma}^2})\ .
\end{align}
To reproduce the observed Higgs mass 
$m_H \simeq 125$\,GeV and $v_\mathrm{EW} \simeq 174$\,GeV, we take $\lambda_\Phi \simeq 0.13$.

To the leading order of $\lambda_{\Phi\sigma}$ and $\lambda_\sigma$, the 
scalar couplings relevant to $H$ and $S$ decays  are given by,
\begin{align}
    \mathcal{L} \supset 
-  \frac{1}{2} g_{HSS} H S^2
- \frac{1}{2} g_{HAA} H A^2 
- \frac{1}{2} g_{SAA} S A^2 
\ , 
\end{align}
with 
\begin{align}
    g_{HSS} \simeq  \frac{1}{\sqrt{2}}\frac{m_H^2}{v}\times \theta \ ,
    \quad 
    g_{HAA} \simeq \frac{1}{\sqrt{2}}\frac{m_H^2}{v} \times \theta \ , \quad
    g_{SAA} \simeq \frac{1}{\sqrt{2}}\frac{m_S^2}{v} \ .
\end{align}
Although the couplings to $A$ are vanishing in the unitary gauge,
they are useful to estimate the Higgs decay rate into $Z'$ through the 
Goldstone equivalence theorem.
The scalar couplings to $Z'$ in the unitary gauge are also obtained from the kinetic term, $|D_\mu \sigma|^2$, as
\begin{align}
    \mathcal{L} \supset \frac{1}{2}\frac{\sqrt{2}m_{Z'}^2}{v}(S-\theta H)Z_\mu'Z^{\prime \mu}\ ,
\end{align}
where
\begin{align}
    m_{Z'}^2 = 2 g_{Z'}^2 Q_\sigma^2 v^2\ ,
\end{align}
with $Q_\sigma$ being the $\Umt$ charge of $\sigma$.

\begin{figure}[t!]
	\centering{\includegraphics[width=0.5\textwidth]{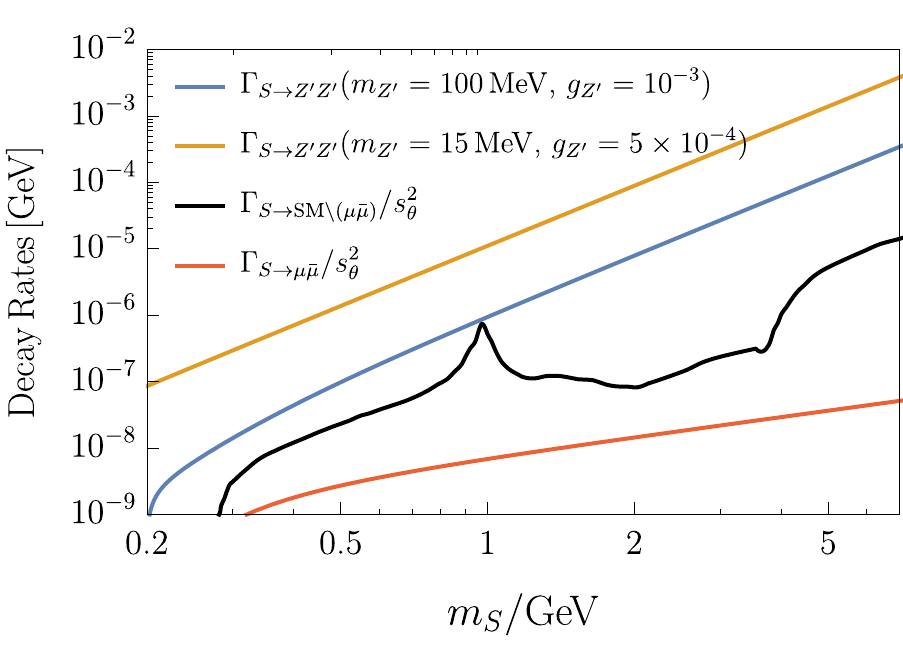}}
	\caption{\sl 
The decay rates of $S$.
The black line shows the total decay rate into the SM particles other than $\mu \bar{\mu}$ while the red line shows that into $\mu\bar{\mu}$, which are extracted from Ref.\,\cite{Winkler:2018qyg}.
They are normalized by the mixing angle $s_\theta^2 \equiv \sin^2 \theta$. 
The blue line shows the decay rate of $S$ into a pair of $Z'$'s for $m_{Z'} = 100$\,MeV and $g_{Z'}=10^{-3}$ ($\langle \sigma \rangle_0 = 100$\,GeV) for $\sin\theta \ll 1$.
The yellow line shows the
decay rate into a pair of $Z'$'s for $m_{Z'} = 15$\,MeV and $g_{Z'}=5\times 10^{-4}$ ($\langle \sigma \rangle_0 = 30$\,GeV) for $\sin\theta \ll 1$.
	 }
\label{fig:branching}
\end{figure}

Now, let us calculate the decay rates of $H$ and $S$.
The decay rates into a pair of $Z'$'s are given by,
\begin{align}
    &\Gamma_{H\to Z'Z'} 
    = \frac{1}{16\pi}\frac{M_Z'{}^4}{v^2m_H}\theta^2 
    \left(2 + \frac{m_H^4}{4m_{Z'}^4}
\left(1-\frac{2m_{Z'}^2}{m_H^2}\right)^2\right) \simeq \frac{1}{64\pi} \frac{m_H^3}{v^2}\theta^2\ , \\
    &\Gamma_{S\to Z'Z'} 
    = \frac{1}{16\pi}\frac{M_Z'{}^4}{v^2m_S}
    \left(2 + \frac{m_S^4}{4m_{Z'}^4}
\left(1-\frac{2m_{Z'}^2}{m_S^2}\right)^2\right) \simeq \frac{1}{64\pi} \frac{m_S^3}{v^2}\ .
\end{align}
These decay rates are in agreement with those into the Goldstone modes, $\Gamma_{H\to AA}$ and $\Gamma_{S\to AA}$, which demonstrates the Goldstone equivalence theorem.
The Higgs boson also decays into a pair of $S$'s with the decay rate
\begin{align}
    \Gamma_{H\to SS} = \frac{1}{32\pi}
    \frac{g_{HSS}^2}{m_H}\left(1-\frac{4m_S^2}{m_H^2}\right)^{1/2} 
    \simeq \frac{1}{64\pi}\frac{m_H^3}{ v^2}\theta^2\ ,
\end{align}
where we have assumed $m_H\gg m_S$ in the second equality.
In Fig.\,\ref{fig:branching}, we show
the partial decay rates of $S$ into a pair of $Z'$'s and those into the SM particles.
The figure shows that $S$ dominantly decays into $Z'$ for $\sin\theta\ll 1$.
As $Z'$ mainly decays into a pair of $\nu$'s, the decays of $S$ are virtually invisible.

Since decays of $S$ and $Z'$ are invisible, both  $\Gamma_{H\to Z'Z'}$ and $\Gamma_{H\to SS}$ contribute to the invisible decay mode of the Higgs boson as
\begin{align}
    \mathrm{Br}(H\to\mathrm{invisible}) 
    = \frac{\Gamma_{H\to Z'Z'}+\Gamma_{H\to SS}}{\Gamma_{\mathrm{SM}} +\Gamma_{H\to Z'Z'}+\Gamma_{H\to SS} } \ ,
\end{align}
where $\Gamma_\mathrm{SM}$ is the predicted value of the Higgs decay width into the SM particles, $ \Gamma_{\mathrm{SM}}\simeq 4.1\,$MeV~\cite{LHCHiggsCrossSectionWorkingGroup:2011wcg}.
For $m_H \gg m_{Z'}$, $m_S$,
we obtain
\begin{align}
    \Gamma_{H\to Z'Z'}+\Gamma_{H\to SS}\simeq \frac{\lambda_{\Phi\sigma}^2v_\mathrm{EW}^2}{8\pi m_H}.
\end{align}
Thus, from the upper limit on the branching fraction of Higgs invisible decay mode, $\mathrm{Br}(H\to \mathrm{invisible}) < 0.11$~\cite{Albert:2021wjs},
we find the upper limit on $|\lambda_{\Phi\sigma}|$,
\begin{align}
\label{eq:invisible1scalar}
|\lambda_{\Phi\sigma}|  \simeq  \left(\frac{8\pi m_H \Gamma_{\mathrm{SM}}} {v_\mathrm{EW}^2}\times\mathrm{Br}(H\to\mathrm{invisible})\right)^{1/2} 
\lesssim 7\times 10^{-3}\ .
\end{align}

\subsection{Model with two scalars}
\label{subapp: two scalar}
Next, let us summarize the scalar sector with two $\Umt$ charged fields 
$\sigma_1$ and $\sigma_2$.
In this case, the scalar potential is extended to,
\begin{align}
    V(\Phi,\sigma_1,\sigma_2)
    =&
    - \mu_\Phi^2 \Phi^\dagger \Phi 
    - \mu_{\sigma_1}^2 |\sigma_1|^2 
    - \mu_{\sigma_2}^2 |\sigma_2|^2 
    \nonumber \\
    &
    + \lambda_\Phi \left( \Phi^\dagger \Phi \right)^2
    + \lambda_{\sigma_1} |\sigma_1|^4 
    + \lambda_{\sigma_2} |\sigma_2|^4 
    \nonumber \\
    &
    + \lambda_{\Phi \sigma_1} (\Phi^\dagger \Phi)|\sigma_1|^2
    + \lambda_{\Phi \sigma_2} (\Phi^\dagger \Phi)|\sigma_2|^2
    \nonumber \\
    &   
    -c \sigma_2^* \sigma_1^2 - c \sigma_2 \sigma_{1}^{*2}
    \ ,
\end{align}
where $c$ is a real-valued coupling constant with mass dimension one.
We have omitted a possible $|\sigma_1|^2|\sigma_2|^2$ term for simplicity.
We assume $\lambda_{\Phi \sigma_i}<0$ ($i=1,2$) and  $\lambda_{\Phi,\sigma_1,\sigma_2}>0$.

Around the vacuum, the scalar fields are decomposed as
\begin{align}
&\Phi = \left(
    \begin{array}{c}
         H^+ \\
         v_\mathrm{EW} +
         \displaystyle{\frac{1}{\sqrt{2}}}(
         \hat{H}+i a)
    \end{array}
    \right)\ , \quad
    \sigma_i = v_i + \frac{1}{\sqrt{2}}
    (
    \hat{S}_i+i \hat{A}_i
    )\ , \quad(i = 1,2)\ .
\end{align}
In this case, $H^+$, $a$, and a linear combination of $\hat{A}_{1,2}$ are would-be Goldstone modes, which are set to zero in the unitary gauge.

From $\partial V/\partial \hat{H}=\partial V/\partial \hat{S}_i = 0$, we find 
\begin{align}
    \mu_{\Phi}^2 &= 2\lambda_H v_\mathrm{EW}^2 + \lambda_{\Phi \sigma_1} v_1^2+\lambda_{\Phi \sigma_2} v_2^2 \ , \\    
    \mu_{\sigma_1}^2 &= 2\lambda_{\sigma_1} v_1^2 + \lambda_{\Phi \sigma_1} v_\mathrm{EW}^2 -2c v_2 \ ,
    \\ 
    \mu_{\sigma_2}^2 &= 2\lambda_{\sigma_2} v_2^2 + \lambda_{\Phi \sigma_2} v_\mathrm{EW}^2 - c \frac{v_1^2}{v_2} \ . 
\end{align}
The squared mass matrix of the CP-even scalars are given by,
\begin{align}
M^2 =  \left(
\begin{array}{ccc}
 4 \lambda_\Phi  {v_\mathrm{EW}}^2 & 2 \lambda_{ \Phi \sigma_1} v_1 v_\mathrm{EW} & 2
   \lambda_{\Phi 
   \sigma_2} v_2 v_\mathrm{EW} \\
 2 \lambda_{\Phi\sigma_1}  v_1 v_\mathrm{EW} & 4 \lambda_{\sigma_1}v_1^2 &
   -2 c v_1 \\
 2 \lambda_{\Phi\sigma_2} v_2 
v_\mathrm{EW} & -2 c v_1 &
\displaystyle{\frac{c
   v_1^2}{v_2}}+4 {\lambda_{\sigma_2}} v_2^2 \\
\end{array}
\right)\ ,
\end{align}
in the $(\hat{H},\hat{S}_1,\hat{S}_2)$ basis.
As we are interested in the case where
$\lambda_{\Phi} \gg |\lambda_{\Phi \sigma_{i}}|$, $\lambda_{\sigma_{i}}$ ($i=1,2$), we can approximately diagonalize the 
matrix by,
\begin{align}
    &\hat{H} = H - \theta_{1} \tilde{S}_1 - \theta_{2} \tilde{S}_2 \ ,\\
    & \hat{S}_{1} = \tilde{S}_1 + \theta_1 H\ ,\\
     & \hat{S}_{2} = \tilde{S}_2 + \theta_2 H\ ,
\end{align}
where the small mixing angles are given by,
\begin{align}
 \theta_i \simeq \frac{1}{2}
    \frac{\lambda_{\Phi\sigma_i}v_i v_\mathrm{EW}}{\lambda_H v_\mathrm{EW}^2 - \lambda_{\sigma_i}v_i^2} 
     \simeq \frac{2\lambda_{\Phi\sigma_i}v_i v_\mathrm{EW}}{m_H^2} \ ,
\end{align}
for $i=1,2$ with small multiplicative correction factors of $\order{\lambda_{\Phi\sigma_i},cv_i/m_H, \lambda_{\sigma_i}}$.
To the leading order of $\lambda_{\Phi\sigma_i},cv_i/m_H, \lambda_{\sigma_i}$, the squared mass matrix of $(\tilde{S}_1,\tilde{S}_2)$ is given by,
\begin{align}
    M_S^2 \simeq  \left(
\begin{array}{cc}
 4 \lambda_{\sigma_1}v_1^2 &
   -2 c v_1 \\
   -2 c v_1 &
\displaystyle{\frac{c
   v_1^2}{v_2}}+4 {\lambda_{\sigma_2}} v_2^2 \\
\end{array}
\right)\ .
\end{align}
We define the mass eigenstates of $M_S^2$ as $(S_1,S_2)$ whose
masses are denoted by $m_{S_1}$ and $m_{S_2}$.

The squared mass matrix of the CP-odd scalars is given by,
\begin{align}
M_A^2 =    \left(
    \begin{array}{cc}
      4c v_2 & 
     -2cv_1 
      \\
     -2 cv_1  &
c \displaystyle{\frac{v_1^2}{v_2}}
    \end{array}
    \right)\ ,
\end{align}
in the $(\hat{A}_1,\hat{A}_2)$ basis.
The matrix can be diagonalized by
\begin{align}
    \left(
    \begin{array}{c}
      A \\
    P
    \end{array}
    \right)
    =  \left(
    \begin{array}{cc}
      \cos\alpha & 
      \sin\alpha 
      \\
       -\sin\alpha   &
    \cos\alpha
    \end{array}
    \right) 
    \left(
    \begin{array}{c}
      \hat{A}_1 \\
    \hat{A}_2  
    \end{array}
    \right)\ ,
\end{align}
where 
\begin{align}
    \sin\alpha= \frac{2 v_2}{\sqrt{v_1^2 + 4 v_2^2}}\ , \quad 
    \cos\alpha= \frac{v_1}{\sqrt{v_1^2 + 4 v_2^2}}\ .
\end{align}
Here, $A$ corresponds to the massless would-be Goldstone mode, while the mass of $P$ is given by,
\begin{align}
    m_P^2 = \frac{c(v_1^2 + 4 v_2^2)}{v_2}\ .
\end{align}

After eliminating the mixings to $\hat{S}_i$'s, 
the Higgs couplings to the scalar sector are given by,
\begin{align}
    \mathcal{L} 
    =
    \sum_{i=1,2}
    \left( 
        - \frac{1}{2} g_{H\tilde{S}_i\tilde{S}_i}H \tilde{S}_i^2 
        - \frac{1}{2} g_{H\hat{A}_i\hat{A}_i} H \hat{A}_i^2
    \right)
    \ ,
\end{align}
where
\begin{align}
    g_{H\tilde{S}_i\tilde{S}_i} = g_{H\hat{A}_i\hat{A}_i} = \frac{1}{\sqrt{2}}\frac{m_H^2}{v_i}\times \theta_i \ . 
\end{align}
The absence of $H\tilde{S}_1\tilde{S}_2$
and $H\hat{A}_1\hat{A}_2$ is due to our simplification that we have omitted the $|\sigma_1|^2 |\sigma_2|^2$ term.
The Higgs couplings to $Z'$'s and $P$ are, on the other hand, given by 
\begin{align}
   \mathcal{L} =& 
-    \sqrt{2}g_{Z'}^2 (v_1 \theta_1 + 4v_2 \theta_2)
    H Z'_\mu Z^{\prime \mu}
- \frac{m_H^2}{2\sqrt{2}}
\left(
s_\alpha^2\frac{\theta_1}{v_1}
+c_\alpha^2\frac{\theta_2}{v_2}
\right)
HP^2
\cr 
&    
-g_{Z'}(s_\alpha \theta_1 - 2c_\alpha \theta_2) 
(P \partial_\mu H Z^{\prime \mu}
- H  \partial_\mu P Z^{\prime \mu})
    \ , \\
    =& 
    -\frac{1}{2}\frac{\sqrt{2} m_{Z'}^2}{v_1^2 + 4 v_2^2} 
    \left( 
    v_1{\theta_1} +4 v_2{\theta_2}\right)
    H Z'_\mu Z^{\prime \mu}
    -\frac{1}{2\sqrt{2}} \frac{m_H^2}{v_1+4v_2^2}
\left(
4v_2^2\frac{\theta_1}{v_1}
+v_1^2\frac{\theta_2}{v_2}
\right)
HP^2\cr
    & 
    -\frac{\sqrt{2}m_{Z'}v_1 v_2}{v_1^2+4v_2^2}
    \left(
    \frac{\theta_1}{v_1} - \frac{\theta_2}{v_2}
    \right)(P\partial_\mu H Z^{\prime \mu}-H\partial_\mu P Z^{\prime \mu})\ ,
\end{align}
where $s_\alpha \equiv \sin\alpha$ and $c_\alpha \equiv \cos\alpha$.

As in the case of the single scalar model, $S_{1,2}$ dominantly decay into a pair of $Z'$'s.
As we will see below, $P$ can decay into a CP-even scalar and $Z'$.
Thus, the branching ratio of the invisible Higgs decay is given by,
\begin{align}
    \mathrm{Br}(H\to\mathrm{invisible}) 
    = \frac{\Gamma_{H\to Z'Z'}+\Gamma_{H\to SS}+\Gamma_{H\to PP}+
    \Gamma_{H\to Z'P}}{\Gamma_{\mathrm{SM}} +\Gamma_{H\to Z'Z'}+\Gamma_{H\to SS}+\Gamma_{H\to PP}+
    \Gamma_{H\to Z'P} } \ ,
\end{align}
where $\Gamma_{H\to SS}$ 
denotes the sum of the decay rates into the CP-even scalars $(S_1,S_2)$.
For $m_H\gg m_{Z'}, m_{S_1}, m_{S_2}, m_P$, we find%
\footnote{Note that  
\begin{align}
   \Gamma_{H\to Z'Z'}+\Gamma_{H\to PP}+
    \Gamma_{H\to Z'P} = 
    \Gamma_{H\to \hat{A}_1+\hat{A}_1}+ \Gamma_{H\to \hat{A}_2+\hat{A}_2}\ ,
\end{align}
in the limit of $m_H \gg m_P$,
which is in agreement with the Goldstone equivalence theorem.}
\begin{align}
    \Gamma_{H\to Z'Z'}+\Gamma_{H\to SS}+\Gamma_{H\to PP}+
    \Gamma_{H\to Z'P}
    \simeq \frac{(|\lambda_{\Phi\sigma_1}|^2+|\lambda_{\Phi\sigma_2}|^2)v_\mathrm{EW}^2}{8\pi m_H}\ .
\end{align}
As a result,
 the upper limit on the branching fraction of Higgs invisible decay mode, $\mathrm{Br}(H\to \mathrm{invisible}) < 0.11$~\cite{Albert:2021wjs}, results in a constraint,
\begin{align}
\label{eq:invisible2scalar in app}
(|\lambda_{\Phi\sigma_1}|^2 + 
|\lambda_{\Phi\sigma_2}|^2)^{1/2}
< 7\times 10^{-3}\ ,
\end{align}
for the two scalar models (see Eq.\,\eqref{eq:invisible1scalar}).

Finally, let  us comment on the fate of the CP-odd scalar $P$.
Through the interactions,
\begin{align}
    \mathcal{L} 
    =
      g_{Z'}s_\alpha P \partial_\mu \tilde{S}_1 Z^{\prime \mu} 
    - g_{Z'}s_\alpha  \tilde{S}_1 \partial_\mu P Z^{\prime \mu} -2g_{Z'}c_\alpha P\partial_\mu \tilde{S}_2 Z^{\prime \mu}
    + 2g_{Z'}c_\alpha \tilde{S}_2 \partial_\mu P Z^{\prime \mu}
    \ ,
\end{align}
$P$ decays into $Z'$ and a CP-even scalar when kinematically allowed.
In such a case, the decay rate of $P$ is comparable to those of the CP-even scalars 
in Fig.\,\ref{fig:branching} for $m_P = \order{m_{S_i}}$.
For example, for 
\begin{align}
    &v_1 = 80\,\mathrm{GeV}\ , 
    \quad
    v_2 = 50\,\mathrm{GeV}\ ,
    \quad
     c = 0.03\,\mathrm{GeV}\ ,
    \\
    &\lambda_{\sigma_1} = \lambda_{\sigma_2} = 3 \times 10^{-4}\ , 
    \quad
    \lambda_{\Phi\sigma_1} = \lambda_{\Phi\sigma_2} = 3 \times 10^{-3}\ , 
\end{align}
we obtain
\begin{align}
    m_{S_1} = 3.5\,\mathrm{GeV}
    \ , \quad
    m_{S_2} = 1.4\, \mathrm{GeV}
    \ , \quad
    m_P = 3.1\, \mathrm{GeV}
    \ ,
\end{align}
which allows $P\to Z' + S_2$.
Hence, $P$ does not cause any cosmological problems for $m_P = \order{1}$\,GeV.

When $P$ is lighter than both the CP-even scalars, $P$ decays into $Z'Z'Z'$ through 
a virtual CP-even scalar.
In this case, the decay rate of $P$ is expected to be roughly suppressed by $\lambda_{\sigma_i}^2 m_P^4/(8\pi m_{S_i}^4)$ compared to the two-body decay rate in Fig.\,\ref{fig:branching}.
Even in such a case, $P$ does not cause cosmological problems as long as its decay temperature, $T_D \simeq \sqrt{\Gamma_P M_P}$, is much higher than $\order{10}$\,MeV.

\section{Finite density effect}
\label{sec:density}
In this appendix, we derive an effective mass of a scalar field that couples to the particles with finite densities.
For simplicity, let us consider a model where a complex scalar field $\sigma$ couples to a real scalar field $\chi$ that has finite density.
The Lagrangian density of them is assumed to be
\begin{align}
    &\mathcal{L} = |\partial_\mu \sigma|^2 +\frac{1}{2}(\partial_\mu \chi)^2 - V\ , \\
    &V =\lambda_\sigma(|\sigma|^2-v^2)^2 +
    \frac{1}{2}m_{\chi}^2 
    \chi^2 + \frac{\kappa}{2}{\chi^2|\sigma|^2} \ ,
\end{align}
where $\lambda_\sigma$ and $\kappa$
are dimensionless coupling constants, and $v$ and $m_{\chi}$
are the parameters with mass dimension one.
We assume $\lambda_\sigma > 0$ so that 
the scalar potential of $\sigma$ is bounded from below.
In general, the sign of $\kappa$ can be positive or negative.
In this model, $\sigma$ obtains a non-vanishing expectation value at the vacuum, which breaks the global U(1) symmetry of $\sigma$.

Let us consider a system with a finite volume $V_\mathrm{3D}= L^3$, where $L$ is a large length scale.
The mode expansion of $\chi$ is then given by,
\begin{align}
    \hat{\chi}(x) = 
    \frac{1}{L^{3/2}}
    \sum_{\mathbf{n}}
    \frac{1}{\sqrt{2p^0}} 
    \left(
    \hat{a}_{\mathbf{n}}e^{-ipx} + \hat{a}_{\mathbf{n}}^{\dagger}e^{ipx}
    \right)\ ,
\end{align}
where $\mathbf{n}=(n_1,n_2,n_3)$ is a set of three integers, with which the 3D momentum of $\chi$ is given by,
\begin{align}
    \mathbf{p}= \frac{2\pi}{L}\mathbf{n}\ .
\end{align}
The creation and annihilation operators satisfy 
\begin{align}
    [\hat{a}_{\mathbf{n}},
    \hat{a}_{\mathbf{n'}}^\dagger] = \delta_{\mathbf{n}\mathbf{n}'}
    \ .
\end{align}

Let us assume that the number density of $\chi$ is $n_{\chi}\neq 0$, which is given by
\begin{align}
    n_\chi = \frac{1}{V_\mathrm{3D}} \sum_{\mathbf{n}} N_{\mathbf{n}}\ ,
\end{align}
where $N_{\mathbf{n}}$ is the particle number for each mode in the 3D volume $V_\mathrm{3D}$.
The corresponding particle state of $\hat\chi$ is given by,
\begin{align}
    |n_{\chi}\rangle = 
    \prod_{\mathbf{n}}\left[\frac{1}{\sqrt{N_{\mathbf{n}}!}} (\hat{a}_{\mathbf{n}}^{\dagger})^{N_{\mathbf{n}}}\right] |0\rangle \ ,
\end{align}
where
\begin{align}
    &\langle 0|0\rangle = 1 \ , \\
    &\langle n_{\chi}|n_\chi \rangle  = 1\ . 
\end{align}

In this state, the expectation value of $\chi^2(x)$ is given by
\begin{align}
    \langle n_{\chi}|
    \left[\hat{\chi}^2(x)\right]_\mathrm{R}
    |n_{\chi}\rangle 
    &= 
    \sum_{\mathbf{n},\mathbf{n}'}
    \frac{2}{L^3 (2p_0 2p_0')^{1/2}}
    \langle n_{\chi}|
    a_{\mathbf{n}}^\dagger
    a_{\mathbf{n}'} e^{-i(p-p')x}
    |n_{\chi}\rangle\ ,\\
     &= 
    \sum_{\mathbf{n}}
    \frac{N_{\mathbf{n}}}{L^3 p_0} \ ,\\ 
    &= n_{\chi}
    \langle p_0^{-1} \rangle\ .
\end{align}
Note that we define the 
renormalized operator $\left[\hat{\chi}^2(x)\right]_\mathrm{R}$ by 
\begin{align}
       \left[\hat{\chi}^2(x)\right]_\mathrm{R}
       \equiv \hat{\chi}^2(x) - 
       \langle 0 |\chi^2(x)|0\rangle\ .
\end{align}

Now consider the effective mass of $\sigma$ around its origin,
\begin{align}
    m_{\mathrm{eff}}^2 
    =
    \frac{\partial^2 V}{\partial \sigma^* \partial\sigma}\Bigg|_{\sigma = 0} 
    &=
    -2{\lambda_\sigma} v^2 
    + \frac{1}{2}\kappa \langle \left[\chi^2\right]_\mathrm{R}\rangle 
    \\
    &=
    -2{\lambda_\sigma}v^2 + \frac{1}{2}\kappa n_{\chi} \langle p_0^{-1} \rangle\ .
\end{align}
Thus, for $\kappa >0$, the effective mass of $\sigma$ becomes positive for 
\begin{align}
    n_{\chi} 
    >
    \frac{4\lambda_\sigma}{\kappa} \frac{v^2}{\langle p_0^{-1}\rangle}\ ,
\end{align}
and hence, the U(1) symmetry of $\sigma$ is restored due to the finite density effect.
When $\chi$ is thermalized, for example, 
the above mass term reproduces the thermal mass
\begin{align}
    m_\mathrm{th}^2|_\chi = 
    \frac{\kappa}{24}T^2 \ .
\end{align}

Now, let us apply this result to the $\Umt$ model.
In this case, $\sigma$ couples to the $\Umt$  gauge boson, which is expected to have a finite density in the early universe. 
The relevant coupling is 
\begin{align}
    \mathcal{L} =|D_\mu \sigma|^2 \supset g^2 A_{\mu}A^\mu |\sigma|^2 \to - g^2 A_{i}A_i |\sigma|^2\ ,
\end{align}
where we take the unitary gauge.
Thus, $\sigma$ couples to the finite density particle with $\kappa = 2 g^2 > 0 $, and hence, we find that the U(1) symmetry of $\sigma$ can be restored by the finite density of the gauge boson.

Let us also comment that the 
finite density of $\sigma$ itself can also contribute to the effective mass of $\sigma$.
Since the quartic coupling constant $\lambda_\sigma$ is required to be positive to avoid the potential unbounded from below.
Thus, the finite density of $\sigma$ also can restore the symmetry.
In the case of the thermalized $\sigma$,
the finite density effects on the mass reduce to the thermal mass,
\begin{align}
    m_{\mathrm{th}}^2|_\sigma = \frac{\lambda_\sigma}{3} T^2 \ .
\end{align}

\section{Mass basis of right-handed neutrinos}
\label{App:mass basis}

Here, we introduce the mass basis of the right-handed neutrinos in the $\Umt$ broken phase and summarize the Lagrangian for four-component fermions in the mass basis.
This Lagrangian will be used to evaluate the asymmetry parameters in the Appendix~\ref{App:asymmetry}.
On the notations of fermions, we follow Ref.~\cite{Dreiner:2008tw}.

First, we consider the mass basis in the $\Umt$ broken phase.
In terms of two-component fermions, the Lagrangian related to right-handed neutrinos is given by
\begin{align}
    \mathcal{L}
    =&
    i \bar{N}_{\alpha}^\dagger \overline{\sigma}^{\mu} D_\mu \bar{N}_\alpha
    +
    \left(
        - \frac{M_{R \alpha \beta}}{2} \bar{N}_\alpha \bar{N}_\beta
        - \lambda_{\nu \alpha \beta} L_\alpha \tilde{\Phi} \bar{N}_{\beta} 
        -h_{e \mu} \sigma_1 \bar{N}_{e} \bar{N}_{\mu}
        -h_{e \tau} \sigma_1^* \bar{N}_{e} \bar{N}_{\tau} 
    \right.
    \nonumber\\
    &
    \left.
        - \frac{1}{2} h_{\mu \mu} \sigma_2 \bar{N}_{\mu} \bar{N}_{\mu} 
        - \frac{1}{2} h_{\tau \tau} \sigma_2^* \bar{N}_{\tau} \bar{N}_{\tau} 
        + \mathrm{h.c.} 
    \right)
    \ .
    \label{eq: Lagrangian in Weyl}
\end{align}
When the $\Umt$ symmetry is spontaneously broken due to nonzero $\langle \sigma_{1,2} \rangle$, the Majorana mass term receives additional contributions from the Majorana Yukawa terms as
\begin{align}
    M_{R,\mathrm{eff}} =
    \begin{pmatrix}
         M_{ee} & h_{e \mu} \langle \sigma_1 \rangle & h_{e \tau} \langle \sigma_1 \rangle \\
         h_{e \mu} \langle \sigma_1 \rangle & h_{\mu \mu} \langle \sigma_2 \rangle & M_{\mu \tau} \\
         h_{e \tau} \langle \sigma_1 \rangle & M_{\mu \tau} & h_{\tau \tau} \langle \sigma_2 \rangle
    \end{pmatrix}
    \ .     
\end{align}
Thus, either of $\bar{N}_\alpha$ is not the mass eigenstates in general.
Since $M_{R, \mathrm{eff}}$ is a complex symmetric matrix, it is diagonalized by a unitary matrix $\Omega$ using the Autonne-Takagi factorization as
\begin{align}
\label{eq:Takagi factorization}
    M_{R, \mathrm{eff}}'
    \equiv 
    \begin{pmatrix}
        M_1 &   0 &   0 \\
          0 & M_2 &   0 \\
          0 &   0 & M_3 
    \end{pmatrix}
    =
    \Omega^\mathrm{T} M_{R, \mathrm{eff}} \Omega
    \ .
\end{align}
Here, $M_I$ is the mass eigenvalue with $I = 1, 2, 3$ and we choose $\Omega$ so that $M_I$ is real and $M_1 \leq M_2 \leq M_3$ without loss of generality.
Thus, the Majorana mass term is diagonalized as
\begin{align}
    \bar{N}^\mathrm{T} \frac{M_{R, \mathrm{eff}}}{2} \bar{N}
    =
    \bar{N}' \frac{M'_{R, \mathrm{eff}}}{2} \bar{N}'
    \ ,
\end{align}
where $\bar{N}'_I \equiv \Omega^\dagger_{I \alpha} \bar{N}_\alpha$ represents the mass eigenstates.

Using this mass basis, we rewrite the Lagrangian in Eq.\,\eqref{eq: Lagrangian in Weyl} as
\begin{align}
    \mathcal{L}'
    &=
    i \bar{N}_I^{\prime \dagger} \overline{\sigma}^{\mu} D'_{\mu I J} \bar{N}'_J
    - 
    \left(
         \frac{M^{\prime}_{R, \mathrm{eff} I J}}{2} \bar{N'}_I \bar{N}'_J
        + \lambda'_{\nu \alpha I} L_\alpha \tilde{\Phi} \bar{N}'_I
        + \frac{1}{2} H^{e \mu \prime}_{I J} \delta \sigma_1 \bar{N}'_{I} \bar{N}'_J
        + \frac{1}{2} H^{e \tau \prime}_{I J} \delta \sigma_1^* \bar{N}'_{I} \bar{N}'_J 
    \right.
    \nonumber\\
    & \hspace{40mm}
    \left.
        + \frac{1}{2} H^{\mu \mu \prime}_{I J} \delta \sigma_2 \bar{N}'_{I} \bar{N}'_J 
        + \frac{1}{2} H^{\tau \tau \prime}_{I J} \delta \sigma_2^* \bar{N}'_{I} \bar{N}'_J
        + \mathrm{h.c.} 
    \right)
    \ .
\end{align}
Here, 
\begin{align}
    \delta \sigma_i
    \equiv 
    \sigma_i - \langle \sigma_i \rangle 
    =
    \frac{1}{\sqrt{2}} \left( \hat{S}_i + i \hat{A}_i \right)
    \ ,
\end{align}
and the covariant derivative is given by,
\begin{align}
    \label{eq:covariant D in mass basis}
    D'_{\mu I J}
    &\equiv 
    \partial_\mu \delta_{I J}
    - i g_{Z'} Z'_\mu \Omega^\dagger_{I \alpha} Q^\mathrm{f}_{\alpha \beta} \Omega_{\beta J}
    \nonumber \\
    &\equiv 
    \partial_\mu \delta_{I J}
    - i g_{Z'} Z'_\mu Q'_{I J}
    \ ,
\end{align}
with
\begin{align}
    Q^\mathrm{f}
    \equiv 
    \begin{pmatrix}
        0 & 0 & 0
        \\
        0 & 1 & 0
        \\
        0 & 0 & -1
    \end{pmatrix}
    \ .
\end{align}
The Dirac Yukawa couplings are given by
\begin{align}
    \lambda'_{\nu \alpha I}
    =
    \lambda_{\nu \alpha \beta} \Omega_{\beta I}
    \ ,
\end{align}
and the Majorana Yukawa couplings are represented by the matrices:
\begin{align}
    H^{\alpha \beta \prime}
    =
    \Omega^\mathrm{T} H^{\alpha \beta} \Omega
    \ ,
\end{align}
with
\begin{gather}
    H^{e \mu}
    \equiv 
    \begin{pmatrix}
        0 & h_{e \mu} & 0
        \\
        h_{e \mu} & 0 & 0 
        \\
        0 & 0 & 0
    \end{pmatrix}
    \ , \quad 
    H^{e \tau}
    \equiv 
    \begin{pmatrix}
        0 & 0 & h_{e \tau}
        \\
        0 & 0 & 0 
        \\
        h_{e \tau} & 0 & 0
    \end{pmatrix}
    \ ,
    \\
    H^{\mu \mu}
    \equiv 
    \begin{pmatrix}
        0 & 0 & 0
        \\
        0 & h_{\mu \mu} & 0 
        \\
        0 & 0 & 0
    \end{pmatrix}
    \ , \quad 
    H^{\tau \tau}
    \equiv 
    \begin{pmatrix}
        0 & 0 & 0
        \\
        0 & 0 & 0 
        \\
        0 & 0 & h_{\tau \tau}
    \end{pmatrix}
    \ .
\end{gather}

Next, we rewrite the Lagrangian in terms of four-component fermions:
\begin{align}
    \Psi_\alpha
    \equiv 
    \begin{pmatrix}
        \bar{N}_\alpha \\ \bar{N}^{\dagger}_\alpha
    \end{pmatrix}
    \ , \quad 
    \Psi_{L_\alpha} 
    \equiv 
    \begin{pmatrix}
        L_\alpha \\ 0
    \end{pmatrix}
    \ .
\end{align}
The mass eigenstates of right-handed neutrinos as four-component fermions are given by
\begin{align}
\label{eq: Diagonal Majorana}
    \Psi'_I
    \equiv 
    \Omega^\dagger_{I \alpha} \Psi_\alpha
    \ .
\end{align}
Then, the Lagrangian becomes
\begin{align}
    \mathcal{L}'
    =
    \mathcal{L}'_\mathrm{kin}
    +
    \mathcal{L}'_\mathrm{mass}
    +
    \mathcal{L}'_\mathrm{DY}
    +
    \mathcal{L}'_\mathrm{MY}
    \ ,
\end{align}
with
\begin{align}
    \mathcal{L}'_\mathrm{kin}
    &=
    \frac{i}{2} \overline{\Psi'}_I \slashed{D}_{I J} \Psi'_J 
    \nonumber \\
    &=
    \frac{i}{2} \overline{\Psi'}_I 
    \gamma^\mu
    \left( 
        \partial_\mu \delta_{I J}
        -
        i g_{Z'} \gamma^5 Z'_\mu Q'_{I J}
    \right) 
    \Psi'_J 
    \ ,
\\
    \mathcal{L}'_\mathrm{mass}
    &=
    - \frac{1}{2} \overline{\Psi'}_I 
    \left(
        \mathrm{Re}[M'_{R, \mathrm{eff}}] 
        - i \gamma^5 \mathrm{Im} [M'_{R, \mathrm{eff}}]
    \right)_{I J}
    \Psi'_J 
    \ ,
\\
    \mathcal{L}'_\mathrm{DY}
    &=
    - \lambda'_{\nu \alpha I}
    \overline{\Psi'}_I \tilde{\Phi} P_L \Psi_{L_\alpha}
    - \lambda^{\prime *}_{\nu \alpha I}
    \overline{\Psi}_{L_\alpha} P_R \tilde{\Phi}^\dagger \Psi'_I
    \ ,
\\
    \mathcal{L}'_\mathrm{MY}
    &=
    - \sum_{X = \hat{S}_1, \hat{A}_1, \hat{S}_2, \hat{A}_2} \overline{\Psi'}_I F_{I J}^{X} X \Psi'_J
    \ .
\end{align}
Here, 
\begin{align}
    F_{I J}^{\hat{S}_1}
    & \equiv 
    \frac{
        \left( 
            \mathrm{Re}[ H^{e \mu \prime} + H^{e \tau  \prime} ]
            - i \gamma^5 \mathrm{Im}[ H^{e \mu \prime} + H^{e \tau \prime} ]
        \right)_{I J}
    }{2 \sqrt{2}} 
    \ ,
\\
    F_{I J}^{\hat{A}_1}
    & \equiv 
    \frac{
        \left( 
            \mathrm{Re}[ i( H^{e \mu \prime} - H^{e \tau \prime} ) ]
            - i \gamma^5 \mathrm{Im}[ i( H^{e \mu \prime} - H^{e \tau \prime} ) ]
        \right)_{I J}
    }{2 \sqrt{2}} 
    \ ,
\\
    F_{I J}^{\hat{S}_2}
    & \equiv 
    \frac{
        \left( 
            \mathrm{Re}[ H^{\mu \mu \prime} + H^{\tau \tau \prime} ]
            - i \gamma^5 \mathrm{Im}[ H^{\mu \mu \prime} + H^{\tau \tau \prime} ]
        \right)_{I J}
    }{2 \sqrt{2}} 
    \ ,
\\
    F_{I J}^{\hat{A}_2}
    & \equiv 
    \frac{
        \left( 
            \mathrm{Re}[ i(H^{\mu \mu \prime} - H^{\tau \tau \prime}) ]
            - i \gamma^5 \mathrm{Im}[ i(H^{\mu \mu \prime} - H^{\tau \tau \prime}) ]
        \right)_{I J}
    }{2 \sqrt{2}} 
    \ .
\end{align}
For later convenience, we also define
\begin{align}
    F_{I J}^X
    \equiv &
    \frac{1}{2\sqrt{2}}
    \left( 
        f^X_{I J} - i \gamma^5 g^X_{I J}
    \right)
    \ .
\end{align}
\section{Asymmetry parameters}
\label{App:asymmetry}

\begin{figure}[t]
    \begin{minipage}[b]{0.3\linewidth}
        \centering
        \includegraphics[width=0.9\linewidth]{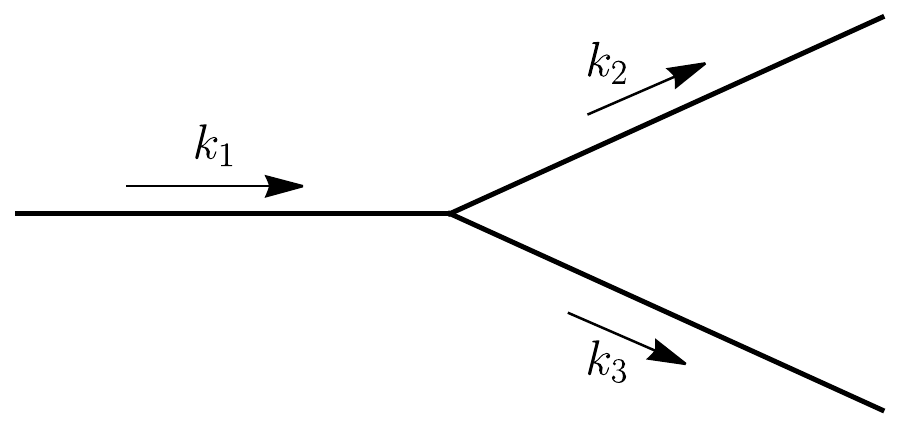}
    \end{minipage}
    \begin{minipage}[b]{0.3\linewidth}
        \centering
        \includegraphics[width=0.9\linewidth]{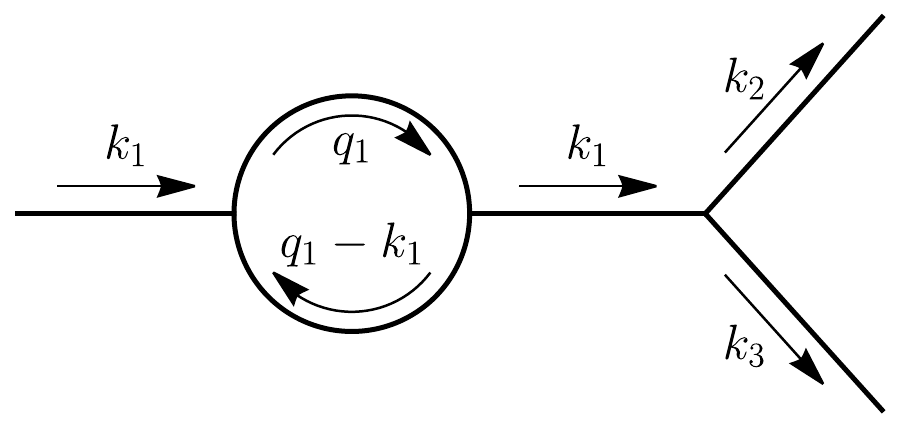}
    \end{minipage}
    \begin{minipage}[b]{0.3\linewidth}
        \centering
        \includegraphics[width=0.9\linewidth]{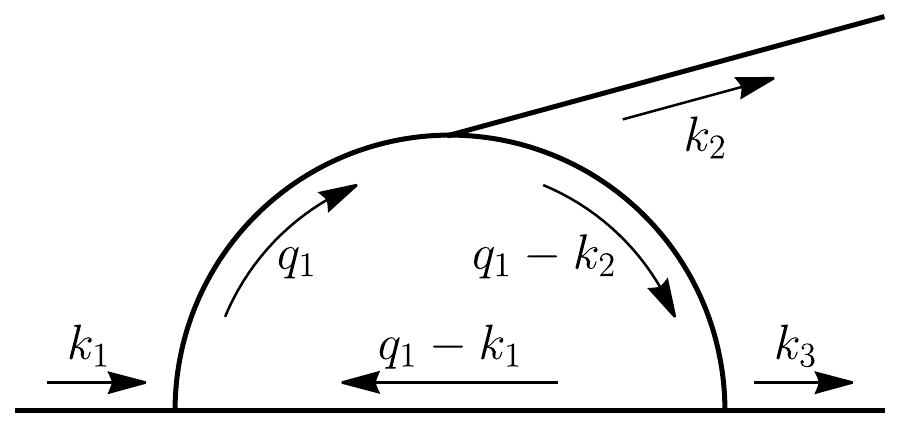}
    \end{minipage}
    \caption{\sl 
    Momentum assignment in tree, wave-function, and vertex diagrams.}
    \label{fig:momentum}
\end{figure}

\begin{figure}[t]
    \begin{minipage}[b]{0.45\linewidth}
        \centering
        \includegraphics[width=0.9\linewidth]{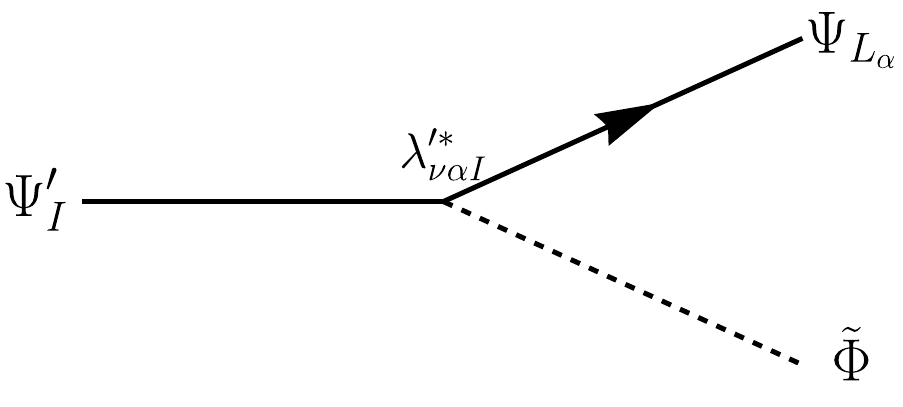}
    \end{minipage}
    \begin{minipage}[b]{0.45\linewidth}
        \centering
        \includegraphics[width=0.9\linewidth]{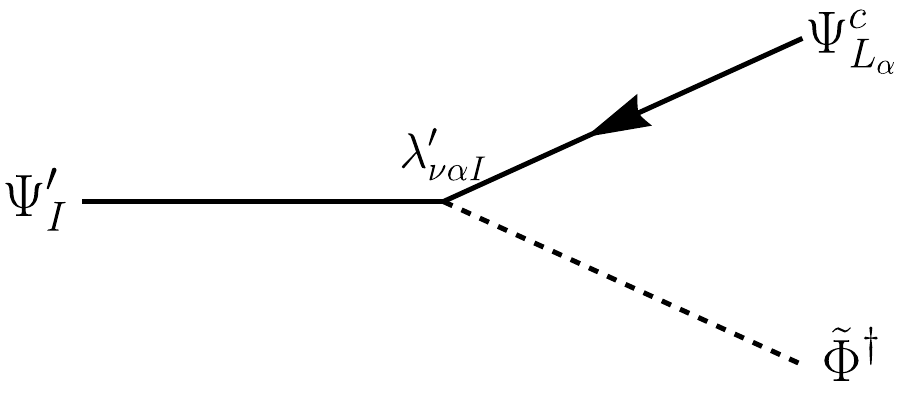}
    \end{minipage}
    \caption{\sl 
    Tree diagrams for decays of right-handed neutrinos.
    Plain solid lines, solid lines with arrows, and dashed lines correspond to right-handed neutrinos, left-handed leptons, and scalar bosons, respectively.}
    \label{fig:tree}
\end{figure}

Using the Lagrangian in the Appendix~\ref{App:mass basis}, we evaluate the asymmetry parameters in the decays of the right-handed neutrinos.
In particular, we consider the decays of 
\begin{align}
    \Psi'_I &\to \Psi_{L_\alpha} \tilde{\Phi}
    \ ,
    \\
    \Psi'_I &\to \Psi_{L_\alpha}^c \tilde{\Phi}^\dagger
    \ .
\end{align}
To evaluate the asymmetry parameters, we consider tree (Fig.~\ref{fig:tree}) and one-loop (Figs.~\ref{fig:standard wave}\,--\,\ref{fig:Z'-loop}) diagrams.
The one-loop diagrams are classified into wave-function and vertex diagrams.
The assignment of the momenta in the following calculations is shown in Fig.~\ref{fig:momentum}.
Hereafter, we approximate that the particles other than the right-handed neutrinos are massless.
We also take the leading order of $\lambda_{\Phi \sigma_i}$ and $c v_i/ m_H$.
In other words, we ignore the mixings among the Higgs and $\hat{S}_{1, 2}$ and identify $\hat{S}_{1,2}$ with the mass eigenstates $S_{1,2}$.
The CP-odd scalars $\hat{A}_1$ and $\hat{A}_2$ are related to the physical degree of freedom $P$ by
\begin{align}
    \hat{A}_1
    =
    - P \sin \alpha
    \ , \quad 
    \hat{A}_2
    =
    P \cos \alpha
    \ ,
\end{align}
where we neglect the would-be Goldstone mode, $A$ (see also the Appendix~\ref{sec:symmetrybreakingsector}).

First, we consider the tree diagrams in Fig.~\ref{fig:tree}.
The amplitudes of these processes are given by
\begin{align}
    i \mathcal{M}_\mathrm{tree}
    &=
    \bar{u} (- i \lambda^{\prime *}_{\nu \alpha I} P_R ) u
    \ ,
    \\
    i \overline{\mathcal{M}}_\mathrm{tree}
    &=
    \bar{u} (- i \lambda^{\prime}_{\nu \alpha I} P_L ) u
    \ .
\end{align}
The decay rate of right-handed neutrinos through these processes is proportional to
\begin{align}
    |\mathcal{M}_\mathrm{tree}|^2 + |\overline{\mathcal{M}}_\mathrm{tree}|^2
    =
    2 [\lambda^{\prime \dagger}_\nu \lambda'_\nu]_{I I}
    \bar{u} P_L u \bar{u} P_R u
\end{align}

\subsection{Wave-function diagram with \texorpdfstring{$\Psi_L \Phi$ loop}{}}

\begin{figure}[t]
    \begin{minipage}[b]{0.45\linewidth}
        \centering
        \includegraphics[width=0.9\linewidth]{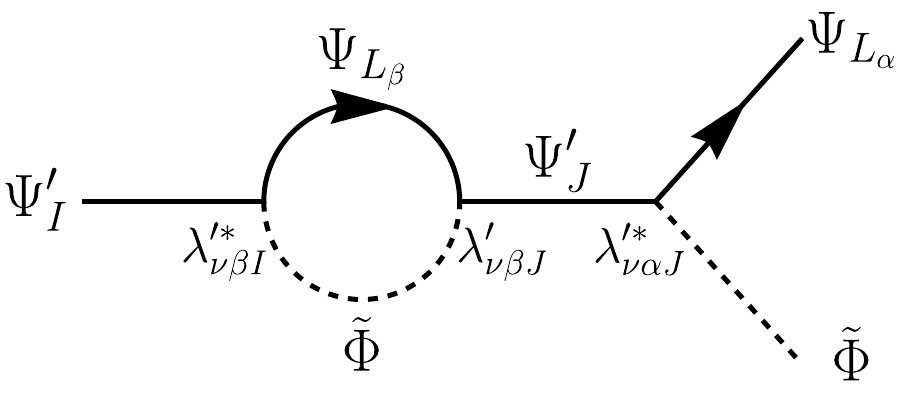}
    \end{minipage}
    \begin{minipage}[b]{0.45\linewidth}
        \centering
        \includegraphics[width=0.9\linewidth]{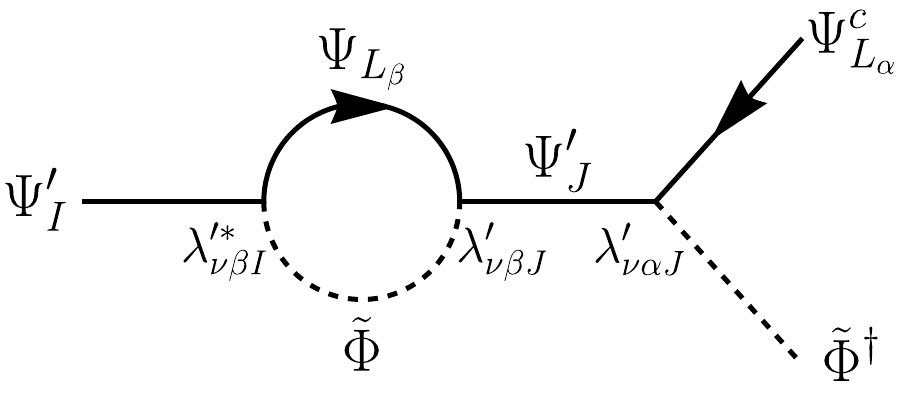}
    \end{minipage}
    \\
    \vspace{5mm}
    \begin{minipage}[b]{0.45\linewidth}
        \centering
        \includegraphics[width=0.9\linewidth]{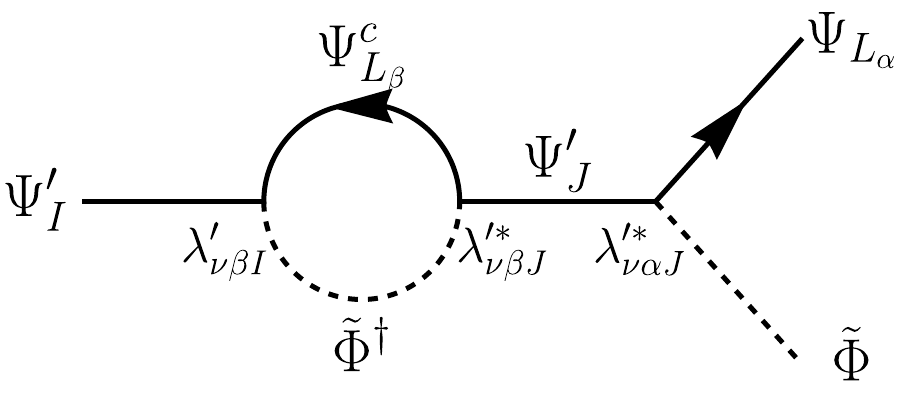}
    \end{minipage}
    \begin{minipage}[b]{0.45\linewidth}
        \centering
        \includegraphics[width=0.9\linewidth]{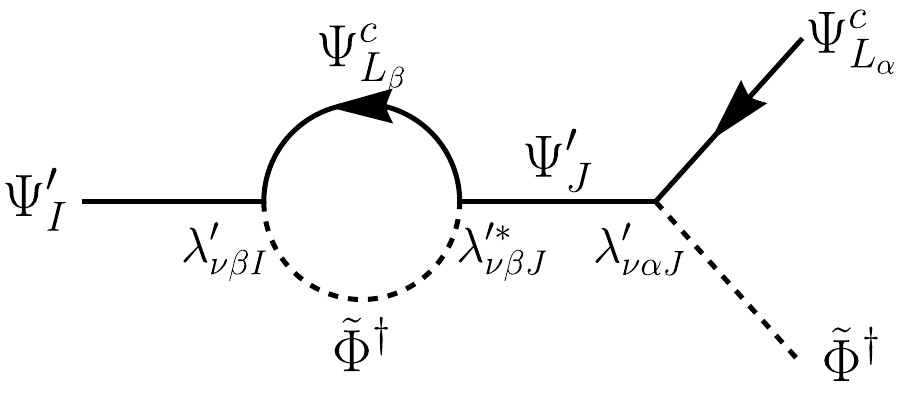}
    \end{minipage}
    \caption{\sl 
    Wave-function diagrams with $\Psi_L \Phi$ loop for decays of right-handed neutrinos.}
    \label{fig:standard wave}
\end{figure}

The diagrams with $\Psi_{L_\beta}$ loop in the upper side of Fig.~\ref{fig:standard wave} give
\begin{align}
    i \mathcal{M}_a^L
    &=
    \bar{u} 
    (- i \lambda^{\prime *}_{\nu \alpha J} P_R )
    \frac{i(\slashed{k}_1 + M_J)}{k_1^2 - M_J^2}
    (- i \lambda^{\prime}_{\nu \beta J} P_L )
    n_w \int \frac{\mathrm{d}^4 q_1}{(2 \pi)^4}
    \frac{i \slashed{q}_1}{q_1^2}
    \frac{i}{(q_1 - k_1)^2}
    (- i \lambda^{\prime *}_{\nu \beta I} P_R )
    u
    \ ,
    \\
    i \overline{\mathcal{M}}_a^L
    &=
    \bar{u} 
    (- i \lambda^{\prime}_{\nu \alpha J} P_L )
    \frac{i(\slashed{k}_1 + M_J)}{k_1^2 - M_J^2}
    (- i \lambda^{\prime}_{\nu \beta J} P_L )
    n_w \int \frac{\mathrm{d}^4 q_1}{(2 \pi)^4}
    \frac{i \slashed{q}_1}{q_1^2}
    \frac{i}{(q_1 - k_1)^2}
    (- i \lambda^{\prime *}_{\nu \beta I} P_R )
    u
    \ ,
\end{align}
where $n_w = 2$ represents the degree of freedom of the $SU(2)$ doublet in the loop.

The diagrams with $\Psi_{L_\beta}^c$ loop in the lower side of Fig.~\ref{fig:standard wave} give
\begin{align}
    i \mathcal{M}_a^{L^c}
    &=
    \bar{u} 
    (- i \lambda^{\prime *}_{\nu \alpha J} P_R )
    \frac{i(\slashed{k}_1 + M_J)}{k_1^2 - M_J^2}
    (- i \lambda^{\prime *}_{\nu \beta J} P_R )
    n_w \int \frac{\mathrm{d}^4 q_1}{(2 \pi)^4}
    \frac{i \slashed{q}_1}{q_1^2}
    \frac{i}{(q_1 - k_1)^2}
    (- i \lambda^{\prime}_{\nu \beta I} P_L )
    u
    \ ,
    \\
    i \overline{\mathcal{M}}_a^{L^c}
    &=
    \bar{u} 
    (- i \lambda^{\prime}_{\nu \alpha J} P_L )
    \frac{i(\slashed{k}_1 + M_J)}{k_1^2 - M_J^2}
    (- i \lambda^{\prime *}_{\nu \beta J} P_R )
    n_w \int \frac{\mathrm{d}^4 q_1}{(2 \pi)^4}
    \frac{i \slashed{q}_1}{q_1^2}
    \frac{i}{(q_1 - k_1)^2}
    (- i \lambda^{\prime}_{\nu \beta I} P_L )
    u
    \ .
\end{align}

As discussed in Sec.~\ref{subsec:failure}, the asymmetry parameters are proportional to the imaginary part coming from loop integrals.
Thus, we pick up the imaginary part in the loop integrals and obtain
\begin{align}
    \mathcal{M}_a^L
    &\supset
    \frac{i M_I^2 \lambda^{\prime *}_{\nu \alpha J} [\lambda^{\prime \mathrm{T}}_\nu \lambda^{\prime *}_\nu]_{J I} }{16 \pi (M_I^2 - M_J^2)}
    \bar{u} P_R u
    \ ,
    \\
    \overline{\mathcal{M}}_a^L
    &\supset
    \frac{i M_I M_J \lambda^{\prime}_{\nu \alpha J} [\lambda^{\prime \mathrm{T}}_\nu \lambda^{\prime *}_\nu]_{J I} }{16 \pi (M_I^2 - M_J^2)}
    \bar{u} P_L u
    \ ,
    \\
    \mathcal{M}_a^{L^c}
    &\supset
    \frac{i M_I M_J \lambda^{\prime *}_{\nu \alpha J} [\lambda^{\prime \dagger}_\nu \lambda^{\prime}_\nu]_{J I} }{16 \pi (M_I^2 - M_J^2)}
    \bar{u} P_R u
    \ ,
    \\
    \overline{\mathcal{M}}_a^{L^c}
    &\supset
    \frac{i M_I^2 \lambda^{\prime}_{\nu \alpha J} [\lambda^{\prime \dagger}_\nu \lambda^{\prime}_\nu]_{J I} }{16 \pi (M_I^2 - M_J^2)}
    \bar{u} P_L u
    \ .
\end{align}
Then, the difference between $|\mathcal{M}|^2$ and $|\overline{\mathcal{M}}|^2$ from diagram (a) is 
\begin{align}
    (|\mathcal{M}|^2 - |\overline{\mathcal{M}}|^2)_a
    &=
    - \mathrm{Im}[ 
        [\lambda^{\prime \dagger}_\nu \lambda'_\nu]_{I J}^2
    ]
    \frac{ M_I M_J \bar{u} P_L u \bar{u} P_R u}{4 \pi (M_I^2 - M_J^2)}
    \ .
\end{align}
As a result, we obtain the asymmetric parameter as
\begin{align}
    \epsilon^{(a)}_I
    =
    \frac{(|\mathcal{M}|^2 - |\overline{\mathcal{M}}|^2)_a}{|\mathcal{M}_\mathrm{tree}|^2 + |\overline{\mathcal{M}}_\mathrm{tree}|^2}
    = 
    - \frac{1}{8 \pi}
    \sum_J
    \frac{ M_I M_J }{M_I^2 - M_J^2}
    \frac{(M_I^2 - M_J^2)^2}{(M_I^2 - M_J^2)^2 + M_I^2 \Gamma_{\mathrm{tree}, J}^2}
    \frac{\mathrm{Im}[ 
        [\lambda^{\prime \dagger}_\nu \lambda'_\nu]_{I J}^2
    ]}
    {[\lambda^{\prime \dagger}_\nu \lambda'_\nu]_{I I}}
    \ ,
\end{align}
where we introduce the regulator with $\Gamma_{\mathrm{tree}, J}^2$ reflecting the finite width of the right-handed neutrinos~\cite{Pilaftsis:2003gt}.

\subsection{Vertex diagram with \texorpdfstring{$\Psi_L \Phi \Psi'$}{} loop}

\begin{figure}[t]
    \begin{minipage}[b]{0.45\linewidth}
        \centering
        \includegraphics[width=0.9\linewidth]{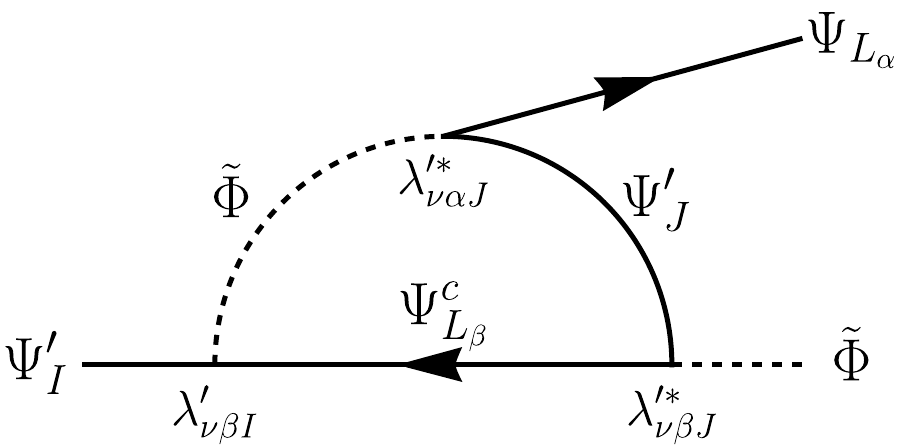}
    \end{minipage}
    \begin{minipage}[b]{0.45\linewidth}
        \centering
        \includegraphics[width=0.9\linewidth]{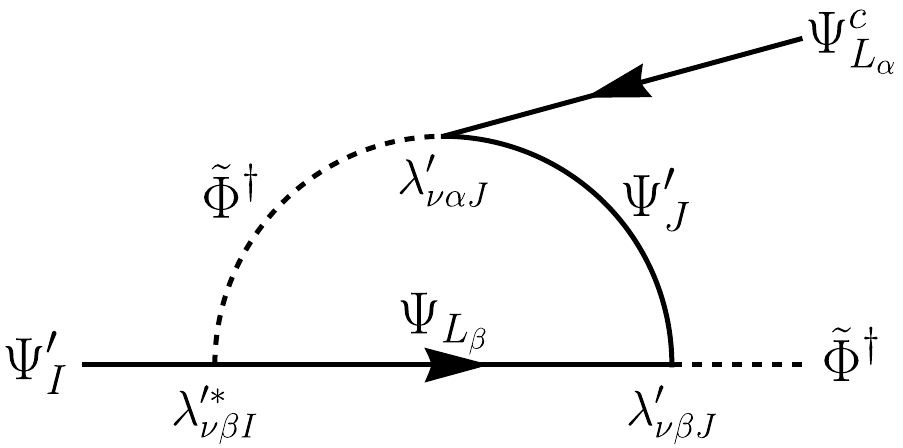}
    \end{minipage}
    \caption{\sl 
    Vertex diagrams with $\Psi_L \Phi \Psi'$ loop for decays of right-handed neutrinos.}
    \label{fig:standard vertex}
\end{figure}

The vertex diagrams with $\Psi_L \Phi \Psi'$ loop in Fig.~\ref{fig:standard vertex} give
\begin{align}
    i \mathcal{M}_b
    &=
    \bar{u} 
    (- i \lambda^{\prime *}_{\nu \alpha J} P_R )
    \int \frac{\mathrm{d}^4 q_1}{(2 \pi)^4}
    \frac{i(\slashed{k}_2 - \slashed{q}_1 + M_J)}{(q_1 - k_2)^2 - M_J^2}
    (- i \lambda^{\prime *}_{\nu \beta J} P_R )
    \frac{i (\slashed{k}_1 - \slashed{q}_1) }{(q_1 - k_1)^2}
    (- i \lambda^{\prime}_{\nu \beta I} P_L )
    \frac{i}{q_1^2}
    u
    \ ,
    \\
    i \overline{\mathcal{M}}_b
    &=
    \bar{u} 
    (- i \lambda^{\prime}_{\nu \alpha J} P_L )
    \int \frac{\mathrm{d}^4 q_1}{(2 \pi)^4}
    \frac{i(\slashed{k}_2 - \slashed{q}_1 + M_J)}{(q_1 - k_2)^2 - M_J^2}
    (- i \lambda^{\prime}_{\nu \beta J} P_L )
    \frac{i (\slashed{k}_1 - \slashed{q}_1) }{(q_1 - k_1)^2}
    (- i \lambda^{\prime *}_{\nu \beta I} P_R )
    \frac{i}{q_1^2}
    u
    \ .
\end{align}
Note that $\mathcal{M}_b$ does not include $n_w$.

As in the case of diagram (a), we pick the imaginary part in the loop integrals and obtain
\begin{align}
    \mathcal{M}_b
    &\supset
    \frac{i r_J}{16 \pi} \lambda^{\prime *}_{\nu \alpha J} [\lambda^{\prime \dagger}_\nu \lambda^{\prime}_\nu]_{J I}
    \left( 
        1  + (1 + r_J^2) \log r_J^2 - ( 1 + r_J^2 ) \log(1 + r_J^2)
    \right)
    \bar{u} P_R u
    \ ,
    \\
    \overline{\mathcal{M}}_b
    &\supset
    \frac{i r_J}{16 \pi} \lambda^{\prime}_{\nu \alpha J} [\lambda^{\prime \mathrm{T}}_\nu \lambda^{\prime *}_\nu]_{J I}
    \left( 
        1  + (1 + r_J^2) \log r_J^2 - ( 1 +r_J^2 ) \log(1 + r_J^2)
    \right)
    \bar{u} P_R u
    \ .
\end{align}
As a result, the asymmetry parameter from diagram (b) is given by
\begin{align}
    \epsilon^{(b)}_I
    = &
    - \frac{1}{8 \pi } 
    \sum_J
    \frac{
        \mathrm{Im} [ [\lambda^{\prime \dagger}_\nu \lambda^{\prime}_\nu]_{I J}^2 ]
    }{[\lambda^{\prime \dagger}_\nu \lambda'_\nu]_{I I}}
   r_J \left[ 
        1  - ( 1 + r_J^2 ) \log \left( \frac{1 + r_J^2}{ r_J^2 } \right)
    \right]
    \ ,
\end{align}
where $r_J \equiv M_J/M_I$.

\subsection{Wave-function diagram with \texorpdfstring{$\Psi' X$}{} loop}

\begin{figure}[t]
    \begin{minipage}[b]{0.45\linewidth}
        \centering
        \includegraphics[width=0.9\linewidth]{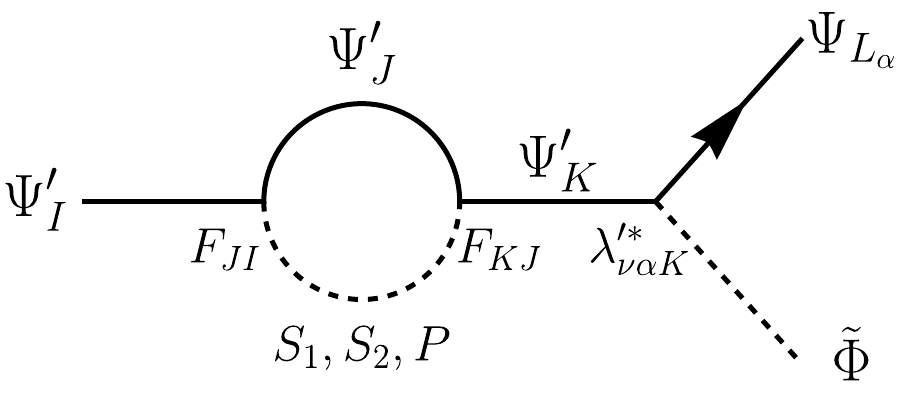}
    \end{minipage}
    \begin{minipage}[b]{0.45\linewidth}
        \centering
        \includegraphics[width=0.9\linewidth]{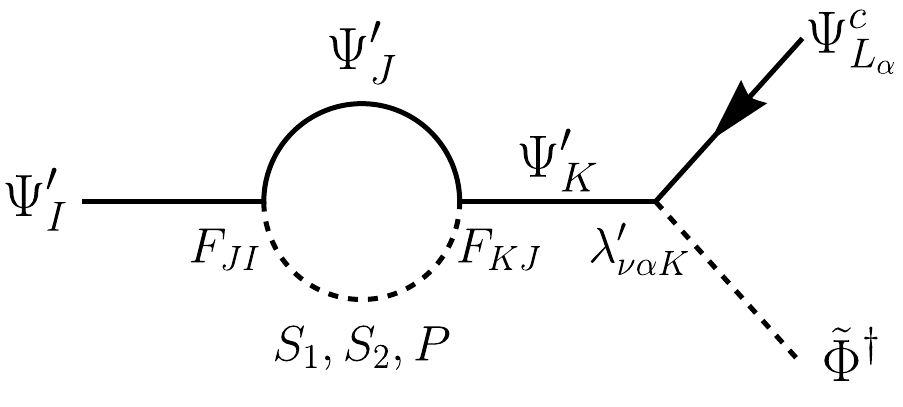}
    \end{minipage}
    \caption{\sl 
    Wave-function diagrams with $\Psi' X$ loop for decays of right-handed neutrinos.}
    \label{fig:nonstandard wave}
\end{figure}

The wave-function diagrams with $\Psi' X$ loop in Fig.~\ref{fig:nonstandard wave} give
\begin{align}
    i \mathcal{M}_c^{X}
    &=
    \bar{u} 
    (- i \lambda^{\prime *}_{\nu \alpha K} P_R )
    \frac{i(\slashed{k}_1 + M_K)}{k_1^2 - M_K^2}
    ( - i F_{K J}^X )
    \int \frac{\mathrm{d}^4 q_1}{(2 \pi)^4}
    \frac{i (\slashed{q}_1 + M_J)}{q_1^2 - M_J^2}
    \frac{i}{(q_1 - k_1)^2 - M_X^2}
    ( - i F_{J I}^X )
    u
    \ ,
    \\
    i \overline{\mathcal{M}}_c^X
    &=
    \bar{u} 
    (- i \lambda^{\prime}_{\nu \alpha K} P_L )
    \frac{i(\slashed{k}_1 + M_K)}{k_1^2 - M_K^2}
    ( - i F_{K J}^X )
    \int \frac{\mathrm{d}^4 q_1}{(2 \pi)^4}
    \frac{i (\slashed{q}_1 + M_J)}{q_1^2 - M_J^2}
    \frac{i}{(q_1 - k_1)^2 - M_X^2}
    ( - i F_{J I}^X )
    u
    \ ,
\end{align}
where $X = S_1, S_2, P$ is the scalar particle in the loop.

The loop integral in $\mathcal{M}_c$ and $\overline{\mathcal{M}}_c$ has a nonzero imaginary part only when $M_I > M_J$ as expected from the optical theorem.
For $M_I > M_J$, we pick the imaginary part in the loop integrals and obtain 
\begin{align}
    \mathcal{M}_c^{X}
    &\supset 
    \frac{i \lambda^{\prime *}_{\nu \alpha K}(M_I^2 - M_J^2)}{256 \pi M_I^3 (M_I^2 - M_K^2)}
    \left[ 
        (M_I - M_J)^2 g^X_{I J} \{ (M_I + M_K) g^X_{J K} - i f^X_{J K} (M_I - M_K) \}
    \right. 
    \nonumber \\
    & \hspace{45mm}
    \left.
        +
        (M_I + M_J)^2 f^X_{I J} \{ (M_I + M_K) f^X_{J K} + i g^X_{J K} (M_I - M_K) \}
    \right]
    \bar{u} P_R u
    \ ,
    \\
    \overline{\mathcal{M}}_c^{X}
    &\supset 
    \frac{i \lambda'_{\nu \alpha K}(M_I^2 - M_J^2)}{256 \pi M_I^3 (M_I^2 - M_K^2)}
    \left[ 
        (M_I - M_J)^2 g^X_{I J} \{ (M_I + M_K) g^X_{J K} + i f^X_{J K} (M_I - M_K) \}
    \right. 
    \nonumber \\
    & \hspace{45mm}
    \left.
        +
        (M_I + M_J)^2 f^X_{I J} \{ (M_I + M_K) f^X_{J K} - i g^X_{J K} (M_I - M_K) \}
    \right]
    \bar{u} P_L u
    \ .
\end{align}
As a result, the difference between $|\mathcal{M}|^2$ and $|\overline{\mathcal{M}}|^2$ from diagram (c) is given by,
\begin{align}
    &(|\mathcal{M}|^2 - |\overline{\mathcal{M}}|^2)_c
    \nonumber \\
    &=
    \sum_{X = S_1, S_2, P}
    \frac{1 - r_J^2}{64 \pi (1 - r_K^2)}
    \nonumber \\
    & \phantom{=} \times 
    \left[
        (1 + r_K)
        \left\{
            \mathrm{Im} \lambda'_{\nu \alpha I} \mathrm{Re}\lambda'_{\nu \alpha K}
          - \mathrm{Re} \lambda'_{\nu \alpha I} \mathrm{Im}\lambda'_{\nu \alpha K}
        \right\}
        \left\{
            (1 + r_J)^2 f^X_{I J} f^X_{J K}
         +  (1 - r_J)^2 g^X_{I J} g^X_{J K}  
        \right\}
    \right.
    \nonumber \\
    & \phantom{=} \  + 
    \left.
        (1 - r_K) 
        \left\{
            \mathrm{Re} \lambda'_{\nu \alpha I} \mathrm{Re}\lambda'_{\nu \alpha K}
          + \mathrm{Im} \lambda'_{\nu \alpha I} \mathrm{Im}\lambda'_{\nu \alpha K}
        \right\}
        \left\{
            (1 + r_J)^2 f^X_{I J} g^X_{J K}
          - (1 - r_J)^2 g^X_{I J} f^X_{J K}           
        \right\}
    \right]
    \nonumber \\
    & \phantom{=} \times 
    \bar{u} P_L u \bar{u} P_R u
    \ .
\end{align}
The corresponding asymmetry parameter is given by,
\begin{align}
    \epsilon^{(c)}_I
    = &
    \sum_{J,K}
    \frac{1 - r_J^2}{128 \pi (1 - r_K^2) [\lambda^{\prime \dagger}_\nu \lambda'_\nu]_{I I}}
    \frac{(M_I^2 - M_K^2)^2}{(M_I^2 - M_K^2)^2 + M_I^2 \Gamma_{\mathrm{tree}, K}^2}
    \Theta(M_I - M_J)
    \nonumber \\
    & \times 
    \sum_{X = S_1, S_2, P}
    \left[
        - (1 + r_K)
        \mathrm{Im}[ \lambda_{\nu}^{\prime \dagger} \lambda'_{\nu}]_{I J} 
        \left\{
            (1 + r_J)^2 f^X_{I J} f^X_{J K}
         +  (1 - r_J)^2 g^X_{I J} g^X_{J K}  
        \right\}
    \right.
    \nonumber \\
    & \hspace{21mm}  +
    \left.
        (1 - r_K) 
        \mathrm{Re}[ \lambda_{\nu}^{\prime \dagger} \lambda'_{\nu}]_{I J}
        \left\{
            (1 + r_J)^2 f^X_{I J} g^X_{J K}
          - (1 - r_J)^2 g^X_{I J} f^X_{J K}           
        \right\}
    \right]
    \ ,
\end{align}
where we introduced the regulator again.

\subsection{Vertex diagram with \texorpdfstring{$\Phi \Psi' X$}{} loop}

\begin{figure}[t]
    \begin{minipage}[b]{0.45\linewidth}
        \centering
        \includegraphics[width=0.9\linewidth]{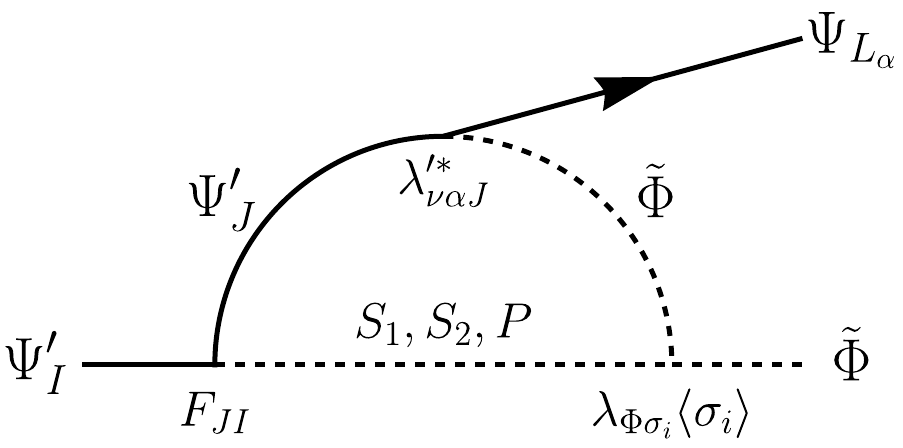}
    \end{minipage}
    \begin{minipage}[b]{0.45\linewidth}
        \centering
        \includegraphics[width=0.9\linewidth]{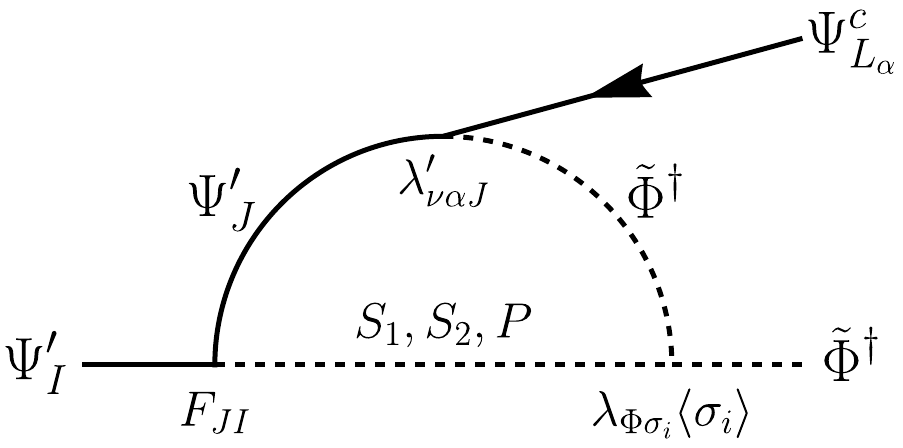}
    \end{minipage}
    \caption{\sl 
    Vertex diagrams with $\Phi \Psi' X$ loop for decays of right-handed neutrinos.}
    \label{fig:nonstandard vertex}
\end{figure}

The vertex diagrams with $\Phi \Psi' X$ loop in Fig.~\ref{fig:nonstandard vertex} include the scalar coupling proportional to $\lambda_{\Phi \sigma} \langle \sigma_i \rangle$.
From the dimensional analysis, the contribution from this diagram is suppressed by $\langle \sigma_i \rangle / M_I$ compared with the contributions above.
Moreover, $\lambda_{\Phi \sigma_i}$ is bounded from above as $\lambda_{\Phi \sigma_i} \lesssim 7 \times 10^{-3}$ as shown in the Appendix~\ref{sec:symmetrybreakingsector} while $\lambda'_{\nu \alpha I}$ and $F^X_{I J}$ are $\order{1}$ at most.
Thus, the contribution of these diagrams to the total asymmetric parameter can be safely neglected.

\subsection{Vertex diagram with \texorpdfstring{$\Psi_L \Psi' Z'$}{} loop}

\begin{figure}[t]
    \begin{minipage}[b]{0.45\linewidth}
        \centering
        \includegraphics[width=0.9\linewidth]{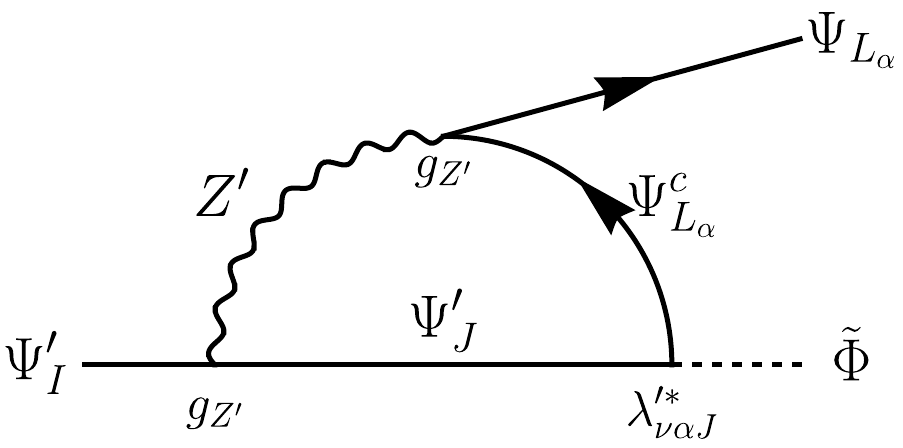}
    \end{minipage}
    \begin{minipage}[b]{0.45\linewidth}
        \centering
        \includegraphics[width=0.9\linewidth]{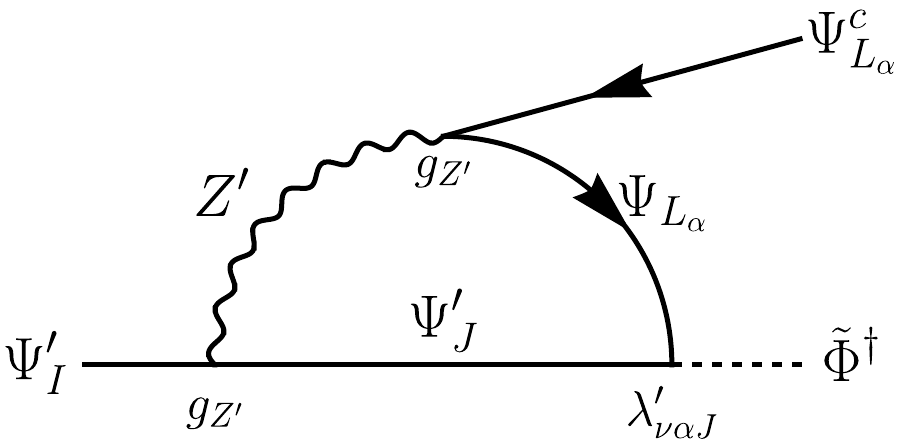}
    \end{minipage}
    \caption{\sl 
    Vertex diagrams with $\Psi_L \Psi' Z'$ loop for decays of right-handed neutrinos.
    Wavy lines correspond to $Z'$.}
    \label{fig:Z'-loop}
\end{figure}

The vertex diagrams with $\Psi_L \Psi' Z'$ loop in Fig.~\ref{fig:Z'-loop} include two of the gauge couplings $g_{Z'}$.
Thus, the asymmetric parameter from these diagrams is suppressed by $g_{Z'}^2 \lesssim 10^{-6}$ and can be safely neglected.

\bibliography{Refs}

\begin{thebibliography}{77}%
\makeatletter
\providecommand \@ifxundefined [1]{%
 \@ifx{#1\undefined}
}%
\providecommand \@ifnum [1]{%
 \ifnum #1\expandafter \@firstoftwo
 \else \expandafter \@secondoftwo
 \fi
}%
\providecommand \@ifx [1]{%
 \ifx #1\expandafter \@firstoftwo
 \else \expandafter \@secondoftwo
 \fi
}%
\providecommand \natexlab [1]{#1}%
\providecommand \enquote  [1]{``#1''}%
\providecommand \bibnamefont  [1]{#1}%
\providecommand \bibfnamefont [1]{#1}%
\providecommand \citenamefont [1]{#1}%
\providecommand \href@noop [0]{\@secondoftwo}%
\providecommand \href [0]{\begingroup \@sanitize@url \@href}%
\providecommand \@href[1]{\@@startlink{#1}\@@href}%
\providecommand \@@href[1]{\endgroup#1\@@endlink}%
\providecommand \@sanitize@url [0]{\catcode `\\12\catcode `\$12\catcode
  `\&12\catcode `\#12\catcode `\^12\catcode `\_12\catcode `\%12\relax}%
\providecommand \@@startlink[1]{}%
\providecommand \@@endlink[0]{}%
\providecommand \url  [0]{\begingroup\@sanitize@url \@url }%
\providecommand \@url [1]{\endgroup\@href {#1}{\urlprefix }}%
\providecommand \urlprefix  [0]{URL }%
\providecommand \Eprint [0]{\href }%
\providecommand \doibase [0]{http://dx.doi.org/}%
\providecommand \selectlanguage [0]{\@gobble}%
\providecommand \bibinfo  [0]{\@secondoftwo}%
\providecommand \bibfield  [0]{\@secondoftwo}%
\providecommand \translation [1]{[#1]}%
\providecommand \BibitemOpen [0]{}%
\providecommand \bibitemStop [0]{}%
\providecommand \bibitemNoStop [0]{.\EOS\space}%
\providecommand \EOS [0]{\spacefactor3000\relax}%
\providecommand \BibitemShut  [1]{\csname bibitem#1\endcsname}%
\let\auto@bib@innerbib\@empty
\bibitem [{\citenamefont {Bennett}\ \emph {et~al.}(2006)\citenamefont {Bennett}
  \emph {et~al.}}]{Muong-2:2006rrc}%
  \BibitemOpen
  \bibfield  {author} {\bibinfo {author} {\bibfnamefont {G.~W.}\ \bibnamefont
  {Bennett}} \emph {et~al.} (\bibinfo {collaboration} {Muon g-2}),\ }\href
  {\doibase 10.1103/PhysRevD.73.072003} {\bibfield  {journal} {\bibinfo
  {journal} {Phys. Rev. D}\ }\textbf {\bibinfo {volume} {73}},\ \bibinfo
  {pages} {072003} (\bibinfo {year} {2006})},\ \Eprint
  {http://arxiv.org/abs/hep-ex/0602035} {arXiv:hep-ex/0602035} \BibitemShut
  {NoStop}%
\bibitem [{\citenamefont {Abi}\ \emph {et~al.}(2021)\citenamefont {Abi} \emph
  {et~al.}}]{Muong-2:2021ojo}%
  \BibitemOpen
  \bibfield  {author} {\bibinfo {author} {\bibfnamefont {B.}~\bibnamefont
  {Abi}} \emph {et~al.} (\bibinfo {collaboration} {Muon g-2}),\ }\href
  {\doibase 10.1103/PhysRevLett.126.141801} {\bibfield  {journal} {\bibinfo
  {journal} {Phys. Rev. Lett.}\ }\textbf {\bibinfo {volume} {126}},\ \bibinfo
  {pages} {141801} (\bibinfo {year} {2021})},\ \Eprint
  {http://arxiv.org/abs/2104.03281} {arXiv:2104.03281 [hep-ex]} \BibitemShut
  {NoStop}%
\bibitem [{\citenamefont {Albahri}\ \emph {et~al.}(2021)\citenamefont {Albahri}
  \emph {et~al.}}]{Muong-2:2021vma}%
  \BibitemOpen
  \bibfield  {author} {\bibinfo {author} {\bibfnamefont {T.}~\bibnamefont
  {Albahri}} \emph {et~al.} (\bibinfo {collaboration} {Muon g-2}),\ }\href
  {\doibase 10.1103/PhysRevD.103.072002} {\bibfield  {journal} {\bibinfo
  {journal} {Phys. Rev. D}\ }\textbf {\bibinfo {volume} {103}},\ \bibinfo
  {pages} {072002} (\bibinfo {year} {2021})},\ \Eprint
  {http://arxiv.org/abs/2104.03247} {arXiv:2104.03247 [hep-ex]} \BibitemShut
  {NoStop}%
\bibitem [{\citenamefont {Aoyama}\ \emph {et~al.}(2020)\citenamefont {Aoyama}
  \emph {et~al.}}]{Aoyama:2020ynm}%
  \BibitemOpen
  \bibfield  {author} {\bibinfo {author} {\bibfnamefont {T.}~\bibnamefont
  {Aoyama}} \emph {et~al.},\ }\href {\doibase 10.1016/j.physrep.2020.07.006}
  {\bibfield  {journal} {\bibinfo  {journal} {Phys. Rept.}\ }\textbf {\bibinfo
  {volume} {887}},\ \bibinfo {pages} {1} (\bibinfo {year} {2020})},\ \Eprint
  {http://arxiv.org/abs/2006.04822} {arXiv:2006.04822 [hep-ph]} \BibitemShut
  {NoStop}%
\bibitem [{\citenamefont {Aoyama}\ \emph {et~al.}(2012)\citenamefont {Aoyama},
  \citenamefont {Hayakawa}, \citenamefont {Kinoshita},\ and\ \citenamefont
  {Nio}}]{Aoyama:2012wk}%
  \BibitemOpen
  \bibfield  {author} {\bibinfo {author} {\bibfnamefont {T.}~\bibnamefont
  {Aoyama}}, \bibinfo {author} {\bibfnamefont {M.}~\bibnamefont {Hayakawa}},
  \bibinfo {author} {\bibfnamefont {T.}~\bibnamefont {Kinoshita}}, \ and\
  \bibinfo {author} {\bibfnamefont {M.}~\bibnamefont {Nio}},\ }\href {\doibase
  10.1103/PhysRevLett.109.111808} {\bibfield  {journal} {\bibinfo  {journal}
  {Phys. Rev. Lett.}\ }\textbf {\bibinfo {volume} {109}},\ \bibinfo {pages}
  {111808} (\bibinfo {year} {2012})},\ \Eprint {http://arxiv.org/abs/1205.5370}
  {arXiv:1205.5370 [hep-ph]} \BibitemShut {NoStop}%
\bibitem [{\citenamefont {Aoyama}\ \emph {et~al.}(2019)\citenamefont {Aoyama},
  \citenamefont {Kinoshita},\ and\ \citenamefont {Nio}}]{Aoyama:2019ryr}%
  \BibitemOpen
  \bibfield  {author} {\bibinfo {author} {\bibfnamefont {T.}~\bibnamefont
  {Aoyama}}, \bibinfo {author} {\bibfnamefont {T.}~\bibnamefont {Kinoshita}}, \
  and\ \bibinfo {author} {\bibfnamefont {M.}~\bibnamefont {Nio}},\ }\href
  {\doibase 10.3390/atoms7010028} {\bibfield  {journal} {\bibinfo  {journal}
  {Atoms}\ }\textbf {\bibinfo {volume} {7}},\ \bibinfo {pages} {28} (\bibinfo
  {year} {2019})}\BibitemShut {NoStop}%
\bibitem [{\citenamefont {Czarnecki}\ \emph {et~al.}(2003)\citenamefont
  {Czarnecki}, \citenamefont {Marciano},\ and\ \citenamefont
  {Vainshtein}}]{Czarnecki:2002nt}%
  \BibitemOpen
  \bibfield  {author} {\bibinfo {author} {\bibfnamefont {A.}~\bibnamefont
  {Czarnecki}}, \bibinfo {author} {\bibfnamefont {W.~J.}\ \bibnamefont
  {Marciano}}, \ and\ \bibinfo {author} {\bibfnamefont {A.}~\bibnamefont
  {Vainshtein}},\ }\href {\doibase 10.1103/PhysRevD.67.073006} {\bibfield
  {journal} {\bibinfo  {journal} {Phys. Rev. D}\ }\textbf {\bibinfo {volume}
  {67}},\ \bibinfo {pages} {073006} (\bibinfo {year} {2003})},\ \bibinfo {note}
  {[Erratum: Phys.Rev.D 73, 119901 (2006)]},\ \Eprint
  {http://arxiv.org/abs/hep-ph/0212229} {arXiv:hep-ph/0212229} \BibitemShut
  {NoStop}%
\bibitem [{\citenamefont {Gnendiger}\ \emph {et~al.}(2013)\citenamefont
  {Gnendiger}, \citenamefont {St\"ockinger},\ and\ \citenamefont
  {St\"ockinger-Kim}}]{Gnendiger:2013pva}%
  \BibitemOpen
  \bibfield  {author} {\bibinfo {author} {\bibfnamefont {C.}~\bibnamefont
  {Gnendiger}}, \bibinfo {author} {\bibfnamefont {D.}~\bibnamefont
  {St\"ockinger}}, \ and\ \bibinfo {author} {\bibfnamefont {H.}~\bibnamefont
  {St\"ockinger-Kim}},\ }\href {\doibase 10.1103/PhysRevD.88.053005} {\bibfield
   {journal} {\bibinfo  {journal} {Phys. Rev. D}\ }\textbf {\bibinfo {volume}
  {88}},\ \bibinfo {pages} {053005} (\bibinfo {year} {2013})},\ \Eprint
  {http://arxiv.org/abs/1306.5546} {arXiv:1306.5546 [hep-ph]} \BibitemShut
  {NoStop}%
\bibitem [{\citenamefont {Davier}\ \emph {et~al.}(2017)\citenamefont {Davier},
  \citenamefont {Hoecker}, \citenamefont {Malaescu},\ and\ \citenamefont
  {Zhang}}]{Davier:2017zfy}%
  \BibitemOpen
  \bibfield  {author} {\bibinfo {author} {\bibfnamefont {M.}~\bibnamefont
  {Davier}}, \bibinfo {author} {\bibfnamefont {A.}~\bibnamefont {Hoecker}},
  \bibinfo {author} {\bibfnamefont {B.}~\bibnamefont {Malaescu}}, \ and\
  \bibinfo {author} {\bibfnamefont {Z.}~\bibnamefont {Zhang}},\ }\href
  {\doibase 10.1140/epjc/s10052-017-5161-6} {\bibfield  {journal} {\bibinfo
  {journal} {Eur. Phys. J. C}\ }\textbf {\bibinfo {volume} {77}},\ \bibinfo
  {pages} {827} (\bibinfo {year} {2017})},\ \Eprint
  {http://arxiv.org/abs/1706.09436} {arXiv:1706.09436 [hep-ph]} \BibitemShut
  {NoStop}%
\bibitem [{\citenamefont {Keshavarzi}\ \emph {et~al.}(2018)\citenamefont
  {Keshavarzi}, \citenamefont {Nomura},\ and\ \citenamefont
  {Teubner}}]{Keshavarzi:2018mgv}%
  \BibitemOpen
  \bibfield  {author} {\bibinfo {author} {\bibfnamefont {A.}~\bibnamefont
  {Keshavarzi}}, \bibinfo {author} {\bibfnamefont {D.}~\bibnamefont {Nomura}},
  \ and\ \bibinfo {author} {\bibfnamefont {T.}~\bibnamefont {Teubner}},\ }\href
  {\doibase 10.1103/PhysRevD.97.114025} {\bibfield  {journal} {\bibinfo
  {journal} {Phys. Rev. D}\ }\textbf {\bibinfo {volume} {97}},\ \bibinfo
  {pages} {114025} (\bibinfo {year} {2018})},\ \Eprint
  {http://arxiv.org/abs/1802.02995} {arXiv:1802.02995 [hep-ph]} \BibitemShut
  {NoStop}%
\bibitem [{\citenamefont {Colangelo}\ \emph {et~al.}(2019)\citenamefont
  {Colangelo}, \citenamefont {Hoferichter},\ and\ \citenamefont
  {Stoffer}}]{Colangelo:2018mtw}%
  \BibitemOpen
  \bibfield  {author} {\bibinfo {author} {\bibfnamefont {G.}~\bibnamefont
  {Colangelo}}, \bibinfo {author} {\bibfnamefont {M.}~\bibnamefont
  {Hoferichter}}, \ and\ \bibinfo {author} {\bibfnamefont {P.}~\bibnamefont
  {Stoffer}},\ }\href {\doibase 10.1007/JHEP02(2019)006} {\bibfield  {journal}
  {\bibinfo  {journal} {JHEP}\ }\textbf {\bibinfo {volume} {02}},\ \bibinfo
  {pages} {006} (\bibinfo {year} {2019})},\ \Eprint
  {http://arxiv.org/abs/1810.00007} {arXiv:1810.00007 [hep-ph]} \BibitemShut
  {NoStop}%
\bibitem [{\citenamefont {Hoferichter}\ \emph {et~al.}(2019)\citenamefont
  {Hoferichter}, \citenamefont {Hoid},\ and\ \citenamefont
  {Kubis}}]{Hoferichter:2019mqg}%
  \BibitemOpen
  \bibfield  {author} {\bibinfo {author} {\bibfnamefont {M.}~\bibnamefont
  {Hoferichter}}, \bibinfo {author} {\bibfnamefont {B.-L.}\ \bibnamefont
  {Hoid}}, \ and\ \bibinfo {author} {\bibfnamefont {B.}~\bibnamefont {Kubis}},\
  }\href {\doibase 10.1007/JHEP08(2019)137} {\bibfield  {journal} {\bibinfo
  {journal} {JHEP}\ }\textbf {\bibinfo {volume} {08}},\ \bibinfo {pages} {137}
  (\bibinfo {year} {2019})},\ \Eprint {http://arxiv.org/abs/1907.01556}
  {arXiv:1907.01556 [hep-ph]} \BibitemShut {NoStop}%
\bibitem [{\citenamefont {Davier}\ \emph {et~al.}(2020)\citenamefont {Davier},
  \citenamefont {Hoecker}, \citenamefont {Malaescu},\ and\ \citenamefont
  {Zhang}}]{Davier:2019can}%
  \BibitemOpen
  \bibfield  {author} {\bibinfo {author} {\bibfnamefont {M.}~\bibnamefont
  {Davier}}, \bibinfo {author} {\bibfnamefont {A.}~\bibnamefont {Hoecker}},
  \bibinfo {author} {\bibfnamefont {B.}~\bibnamefont {Malaescu}}, \ and\
  \bibinfo {author} {\bibfnamefont {Z.}~\bibnamefont {Zhang}},\ }\href
  {\doibase 10.1140/epjc/s10052-020-7792-2} {\bibfield  {journal} {\bibinfo
  {journal} {Eur. Phys. J. C}\ }\textbf {\bibinfo {volume} {80}},\ \bibinfo
  {pages} {241} (\bibinfo {year} {2020})},\ \bibinfo {note} {[Erratum:
  Eur.Phys.J.C 80, 410 (2020)]},\ \Eprint {http://arxiv.org/abs/1908.00921}
  {arXiv:1908.00921 [hep-ph]} \BibitemShut {NoStop}%
\bibitem [{\citenamefont {Keshavarzi}\ \emph {et~al.}(2020)\citenamefont
  {Keshavarzi}, \citenamefont {Nomura},\ and\ \citenamefont
  {Teubner}}]{Keshavarzi:2019abf}%
  \BibitemOpen
  \bibfield  {author} {\bibinfo {author} {\bibfnamefont {A.}~\bibnamefont
  {Keshavarzi}}, \bibinfo {author} {\bibfnamefont {D.}~\bibnamefont {Nomura}},
  \ and\ \bibinfo {author} {\bibfnamefont {T.}~\bibnamefont {Teubner}},\ }\href
  {\doibase 10.1103/PhysRevD.101.014029} {\bibfield  {journal} {\bibinfo
  {journal} {Phys. Rev. D}\ }\textbf {\bibinfo {volume} {101}},\ \bibinfo
  {pages} {014029} (\bibinfo {year} {2020})},\ \Eprint
  {http://arxiv.org/abs/1911.00367} {arXiv:1911.00367 [hep-ph]} \BibitemShut
  {NoStop}%
\bibitem [{\citenamefont {Kurz}\ \emph {et~al.}(2014)\citenamefont {Kurz},
  \citenamefont {Liu}, \citenamefont {Marquard},\ and\ \citenamefont
  {Steinhauser}}]{Kurz:2014wya}%
  \BibitemOpen
  \bibfield  {author} {\bibinfo {author} {\bibfnamefont {A.}~\bibnamefont
  {Kurz}}, \bibinfo {author} {\bibfnamefont {T.}~\bibnamefont {Liu}}, \bibinfo
  {author} {\bibfnamefont {P.}~\bibnamefont {Marquard}}, \ and\ \bibinfo
  {author} {\bibfnamefont {M.}~\bibnamefont {Steinhauser}},\ }\href {\doibase
  10.1016/j.physletb.2014.05.043} {\bibfield  {journal} {\bibinfo  {journal}
  {Phys. Lett. B}\ }\textbf {\bibinfo {volume} {734}},\ \bibinfo {pages} {144}
  (\bibinfo {year} {2014})},\ \Eprint {http://arxiv.org/abs/1403.6400}
  {arXiv:1403.6400 [hep-ph]} \BibitemShut {NoStop}%
\bibitem [{\citenamefont {Melnikov}\ and\ \citenamefont
  {Vainshtein}(2004)}]{Melnikov:2003xd}%
  \BibitemOpen
  \bibfield  {author} {\bibinfo {author} {\bibfnamefont {K.}~\bibnamefont
  {Melnikov}}\ and\ \bibinfo {author} {\bibfnamefont {A.}~\bibnamefont
  {Vainshtein}},\ }\href {\doibase 10.1103/PhysRevD.70.113006} {\bibfield
  {journal} {\bibinfo  {journal} {Phys. Rev. D}\ }\textbf {\bibinfo {volume}
  {70}},\ \bibinfo {pages} {113006} (\bibinfo {year} {2004})},\ \Eprint
  {http://arxiv.org/abs/hep-ph/0312226} {arXiv:hep-ph/0312226} \BibitemShut
  {NoStop}%
\bibitem [{\citenamefont {Masjuan}\ and\ \citenamefont
  {Sanchez-Puertas}(2017)}]{Masjuan:2017tvw}%
  \BibitemOpen
  \bibfield  {author} {\bibinfo {author} {\bibfnamefont {P.}~\bibnamefont
  {Masjuan}}\ and\ \bibinfo {author} {\bibfnamefont {P.}~\bibnamefont
  {Sanchez-Puertas}},\ }\href {\doibase 10.1103/PhysRevD.95.054026} {\bibfield
  {journal} {\bibinfo  {journal} {Phys. Rev. D}\ }\textbf {\bibinfo {volume}
  {95}},\ \bibinfo {pages} {054026} (\bibinfo {year} {2017})},\ \Eprint
  {http://arxiv.org/abs/1701.05829} {arXiv:1701.05829 [hep-ph]} \BibitemShut
  {NoStop}%
\bibitem [{\citenamefont {Colangelo}\ \emph {et~al.}(2017)\citenamefont
  {Colangelo}, \citenamefont {Hoferichter}, \citenamefont {Procura},\ and\
  \citenamefont {Stoffer}}]{Colangelo:2017fiz}%
  \BibitemOpen
  \bibfield  {author} {\bibinfo {author} {\bibfnamefont {G.}~\bibnamefont
  {Colangelo}}, \bibinfo {author} {\bibfnamefont {M.}~\bibnamefont
  {Hoferichter}}, \bibinfo {author} {\bibfnamefont {M.}~\bibnamefont
  {Procura}}, \ and\ \bibinfo {author} {\bibfnamefont {P.}~\bibnamefont
  {Stoffer}},\ }\href {\doibase 10.1007/JHEP04(2017)161} {\bibfield  {journal}
  {\bibinfo  {journal} {JHEP}\ }\textbf {\bibinfo {volume} {04}},\ \bibinfo
  {pages} {161} (\bibinfo {year} {2017})},\ \Eprint
  {http://arxiv.org/abs/1702.07347} {arXiv:1702.07347 [hep-ph]} \BibitemShut
  {NoStop}%
\bibitem [{\citenamefont {Hoferichter}\ \emph {et~al.}(2018)\citenamefont
  {Hoferichter}, \citenamefont {Hoid}, \citenamefont {Kubis}, \citenamefont
  {Leupold},\ and\ \citenamefont {Schneider}}]{Hoferichter:2018kwz}%
  \BibitemOpen
  \bibfield  {author} {\bibinfo {author} {\bibfnamefont {M.}~\bibnamefont
  {Hoferichter}}, \bibinfo {author} {\bibfnamefont {B.-L.}\ \bibnamefont
  {Hoid}}, \bibinfo {author} {\bibfnamefont {B.}~\bibnamefont {Kubis}},
  \bibinfo {author} {\bibfnamefont {S.}~\bibnamefont {Leupold}}, \ and\
  \bibinfo {author} {\bibfnamefont {S.~P.}\ \bibnamefont {Schneider}},\ }\href
  {\doibase 10.1007/JHEP10(2018)141} {\bibfield  {journal} {\bibinfo  {journal}
  {JHEP}\ }\textbf {\bibinfo {volume} {10}},\ \bibinfo {pages} {141} (\bibinfo
  {year} {2018})},\ \Eprint {http://arxiv.org/abs/1808.04823} {arXiv:1808.04823
  [hep-ph]} \BibitemShut {NoStop}%
\bibitem [{\citenamefont {G\'erardin}\ \emph {et~al.}(2019)\citenamefont
  {G\'erardin}, \citenamefont {Meyer},\ and\ \citenamefont
  {Nyffeler}}]{Gerardin:2019vio}%
  \BibitemOpen
  \bibfield  {author} {\bibinfo {author} {\bibfnamefont {A.}~\bibnamefont
  {G\'erardin}}, \bibinfo {author} {\bibfnamefont {H.~B.}\ \bibnamefont
  {Meyer}}, \ and\ \bibinfo {author} {\bibfnamefont {A.}~\bibnamefont
  {Nyffeler}},\ }\href {\doibase 10.1103/PhysRevD.100.034520} {\bibfield
  {journal} {\bibinfo  {journal} {Phys. Rev. D}\ }\textbf {\bibinfo {volume}
  {100}},\ \bibinfo {pages} {034520} (\bibinfo {year} {2019})},\ \Eprint
  {http://arxiv.org/abs/1903.09471} {arXiv:1903.09471 [hep-lat]} \BibitemShut
  {NoStop}%
\bibitem [{\citenamefont {Bijnens}\ \emph {et~al.}(2019)\citenamefont
  {Bijnens}, \citenamefont {Hermansson-Truedsson},\ and\ \citenamefont
  {Rodr\'\i{}guez-S\'anchez}}]{Bijnens:2019ghy}%
  \BibitemOpen
  \bibfield  {author} {\bibinfo {author} {\bibfnamefont {J.}~\bibnamefont
  {Bijnens}}, \bibinfo {author} {\bibfnamefont {N.}~\bibnamefont
  {Hermansson-Truedsson}}, \ and\ \bibinfo {author} {\bibfnamefont
  {A.}~\bibnamefont {Rodr\'\i{}guez-S\'anchez}},\ }\href {\doibase
  10.1016/j.physletb.2019.134994} {\bibfield  {journal} {\bibinfo  {journal}
  {Phys. Lett. B}\ }\textbf {\bibinfo {volume} {798}},\ \bibinfo {pages}
  {134994} (\bibinfo {year} {2019})},\ \Eprint
  {http://arxiv.org/abs/1908.03331} {arXiv:1908.03331 [hep-ph]} \BibitemShut
  {NoStop}%
\bibitem [{\citenamefont {Colangelo}\ \emph {et~al.}(2020)\citenamefont
  {Colangelo}, \citenamefont {Hagelstein}, \citenamefont {Hoferichter},
  \citenamefont {Laub},\ and\ \citenamefont {Stoffer}}]{Colangelo:2019uex}%
  \BibitemOpen
  \bibfield  {author} {\bibinfo {author} {\bibfnamefont {G.}~\bibnamefont
  {Colangelo}}, \bibinfo {author} {\bibfnamefont {F.}~\bibnamefont
  {Hagelstein}}, \bibinfo {author} {\bibfnamefont {M.}~\bibnamefont
  {Hoferichter}}, \bibinfo {author} {\bibfnamefont {L.}~\bibnamefont {Laub}}, \
  and\ \bibinfo {author} {\bibfnamefont {P.}~\bibnamefont {Stoffer}},\ }\href
  {\doibase 10.1007/JHEP03(2020)101} {\bibfield  {journal} {\bibinfo  {journal}
  {JHEP}\ }\textbf {\bibinfo {volume} {03}},\ \bibinfo {pages} {101} (\bibinfo
  {year} {2020})},\ \Eprint {http://arxiv.org/abs/1910.13432} {arXiv:1910.13432
  [hep-ph]} \BibitemShut {NoStop}%
\bibitem [{\citenamefont {Blum}\ \emph {et~al.}(2020)\citenamefont {Blum},
  \citenamefont {Christ}, \citenamefont {Hayakawa}, \citenamefont {Izubuchi},
  \citenamefont {Jin}, \citenamefont {Jung},\ and\ \citenamefont
  {Lehner}}]{Blum:2019ugy}%
  \BibitemOpen
  \bibfield  {author} {\bibinfo {author} {\bibfnamefont {T.}~\bibnamefont
  {Blum}}, \bibinfo {author} {\bibfnamefont {N.}~\bibnamefont {Christ}},
  \bibinfo {author} {\bibfnamefont {M.}~\bibnamefont {Hayakawa}}, \bibinfo
  {author} {\bibfnamefont {T.}~\bibnamefont {Izubuchi}}, \bibinfo {author}
  {\bibfnamefont {L.}~\bibnamefont {Jin}}, \bibinfo {author} {\bibfnamefont
  {C.}~\bibnamefont {Jung}}, \ and\ \bibinfo {author} {\bibfnamefont
  {C.}~\bibnamefont {Lehner}},\ }\href {\doibase
  10.1103/PhysRevLett.124.132002} {\bibfield  {journal} {\bibinfo  {journal}
  {Phys. Rev. Lett.}\ }\textbf {\bibinfo {volume} {124}},\ \bibinfo {pages}
  {132002} (\bibinfo {year} {2020})},\ \Eprint
  {http://arxiv.org/abs/1911.08123} {arXiv:1911.08123 [hep-lat]} \BibitemShut
  {NoStop}%
\bibitem [{\citenamefont {Colangelo}\ \emph {et~al.}(2014)\citenamefont
  {Colangelo}, \citenamefont {Hoferichter}, \citenamefont {Nyffeler},
  \citenamefont {Passera},\ and\ \citenamefont {Stoffer}}]{Colangelo:2014qya}%
  \BibitemOpen
  \bibfield  {author} {\bibinfo {author} {\bibfnamefont {G.}~\bibnamefont
  {Colangelo}}, \bibinfo {author} {\bibfnamefont {M.}~\bibnamefont
  {Hoferichter}}, \bibinfo {author} {\bibfnamefont {A.}~\bibnamefont
  {Nyffeler}}, \bibinfo {author} {\bibfnamefont {M.}~\bibnamefont {Passera}}, \
  and\ \bibinfo {author} {\bibfnamefont {P.}~\bibnamefont {Stoffer}},\ }\href
  {\doibase 10.1016/j.physletb.2014.06.012} {\bibfield  {journal} {\bibinfo
  {journal} {Phys. Lett. B}\ }\textbf {\bibinfo {volume} {735}},\ \bibinfo
  {pages} {90} (\bibinfo {year} {2014})},\ \Eprint
  {http://arxiv.org/abs/1403.7512} {arXiv:1403.7512 [hep-ph]} \BibitemShut
  {NoStop}%
\bibitem [{\citenamefont {Workman}\ \emph {et~al.}(2022)\citenamefont {Workman}
  \emph {et~al.}}]{ParticleDataGroup:2022pth}%
  \BibitemOpen
  \bibfield  {author} {\bibinfo {author} {\bibfnamefont {R.~L.}\ \bibnamefont
  {Workman}} \emph {et~al.} (\bibinfo {collaboration} {Particle Data Group}),\
  }\href {\doibase 10.1093/ptep/ptac097} {\bibfield  {journal} {\bibinfo
  {journal} {PTEP}\ }\textbf {\bibinfo {volume} {2022}},\ \bibinfo {pages}
  {083C01} (\bibinfo {year} {2022})}\BibitemShut {NoStop}%
\bibitem [{\citenamefont {Ignatov}\ \emph {et~al.}(2023)\citenamefont {Ignatov}
  \emph {et~al.}}]{CMD-3:2023alj}%
  \BibitemOpen
  \bibfield  {author} {\bibinfo {author} {\bibfnamefont {F.~V.}\ \bibnamefont
  {Ignatov}} \emph {et~al.} (\bibinfo {collaboration} {CMD-3}),\ }\href@noop {}
  {\bibfield  {journal} {\bibinfo  {journal} {arXiv preprint}\ } (\bibinfo
  {year} {2023})},\ \Eprint {http://arxiv.org/abs/2302.08834} {arXiv:2302.08834
  [hep-ex]} \BibitemShut {NoStop}%
\bibitem [{\citenamefont {Foot}(1991)}]{Foot:1990mn}%
  \BibitemOpen
  \bibfield  {author} {\bibinfo {author} {\bibfnamefont {R.}~\bibnamefont
  {Foot}},\ }\href {\doibase 10.1142/S0217732391000543} {\bibfield  {journal}
  {\bibinfo  {journal} {Mod. Phys. Lett. A}\ }\textbf {\bibinfo {volume} {6}},\
  \bibinfo {pages} {527} (\bibinfo {year} {1991})}\BibitemShut {NoStop}%
\bibitem [{\citenamefont {He}\ \emph {et~al.}(1991)\citenamefont {He},
  \citenamefont {Joshi}, \citenamefont {Lew},\ and\ \citenamefont
  {Volkas}}]{He:1991qd}%
  \BibitemOpen
  \bibfield  {author} {\bibinfo {author} {\bibfnamefont {X.-G.}\ \bibnamefont
  {He}}, \bibinfo {author} {\bibfnamefont {G.~C.}\ \bibnamefont {Joshi}},
  \bibinfo {author} {\bibfnamefont {H.}~\bibnamefont {Lew}}, \ and\ \bibinfo
  {author} {\bibfnamefont {R.~R.}\ \bibnamefont {Volkas}},\ }\href {\doibase
  10.1103/PhysRevD.44.2118} {\bibfield  {journal} {\bibinfo  {journal} {Phys.
  Rev. D}\ }\textbf {\bibinfo {volume} {44}},\ \bibinfo {pages} {2118}
  (\bibinfo {year} {1991})}\BibitemShut {NoStop}%
\bibitem [{\citenamefont {Foot}\ \emph {et~al.}(1994)\citenamefont {Foot},
  \citenamefont {He}, \citenamefont {Lew},\ and\ \citenamefont
  {Volkas}}]{Foot:1994vd}%
  \BibitemOpen
  \bibfield  {author} {\bibinfo {author} {\bibfnamefont {R.}~\bibnamefont
  {Foot}}, \bibinfo {author} {\bibfnamefont {X.~G.}\ \bibnamefont {He}},
  \bibinfo {author} {\bibfnamefont {H.}~\bibnamefont {Lew}}, \ and\ \bibinfo
  {author} {\bibfnamefont {R.~R.}\ \bibnamefont {Volkas}},\ }\href {\doibase
  10.1103/PhysRevD.50.4571} {\bibfield  {journal} {\bibinfo  {journal} {Phys.
  Rev. D}\ }\textbf {\bibinfo {volume} {50}},\ \bibinfo {pages} {4571}
  (\bibinfo {year} {1994})},\ \Eprint {http://arxiv.org/abs/hep-ph/9401250}
  {arXiv:hep-ph/9401250} \BibitemShut {NoStop}%
\bibitem [{\citenamefont {Gninenko}\ and\ \citenamefont
  {Krasnikov}(2001)}]{Gninenko:2001hx}%
  \BibitemOpen
  \bibfield  {author} {\bibinfo {author} {\bibfnamefont {S.~N.}\ \bibnamefont
  {Gninenko}}\ and\ \bibinfo {author} {\bibfnamefont {N.~V.}\ \bibnamefont
  {Krasnikov}},\ }\href {\doibase 10.1016/S0370-2693(01)00693-1} {\bibfield
  {journal} {\bibinfo  {journal} {Phys. Lett. B}\ }\textbf {\bibinfo {volume}
  {513}},\ \bibinfo {pages} {119} (\bibinfo {year} {2001})},\ \Eprint
  {http://arxiv.org/abs/hep-ph/0102222} {arXiv:hep-ph/0102222} \BibitemShut
  {NoStop}%
\bibitem [{\citenamefont {Baek}\ \emph {et~al.}(2001)\citenamefont {Baek},
  \citenamefont {Deshpande}, \citenamefont {He},\ and\ \citenamefont
  {Ko}}]{Baek:2001kca}%
  \BibitemOpen
  \bibfield  {author} {\bibinfo {author} {\bibfnamefont {S.}~\bibnamefont
  {Baek}}, \bibinfo {author} {\bibfnamefont {N.~G.}\ \bibnamefont {Deshpande}},
  \bibinfo {author} {\bibfnamefont {X.~G.}\ \bibnamefont {He}}, \ and\ \bibinfo
  {author} {\bibfnamefont {P.}~\bibnamefont {Ko}},\ }\href {\doibase
  10.1103/PhysRevD.64.055006} {\bibfield  {journal} {\bibinfo  {journal} {Phys.
  Rev. D}\ }\textbf {\bibinfo {volume} {64}},\ \bibinfo {pages} {055006}
  (\bibinfo {year} {2001})},\ \Eprint {http://arxiv.org/abs/hep-ph/0104141}
  {arXiv:hep-ph/0104141} \BibitemShut {NoStop}%
\bibitem [{\citenamefont {Murakami}(2002)}]{Murakami:2001cs}%
  \BibitemOpen
  \bibfield  {author} {\bibinfo {author} {\bibfnamefont {B.}~\bibnamefont
  {Murakami}},\ }\href {\doibase 10.1103/PhysRevD.65.055003} {\bibfield
  {journal} {\bibinfo  {journal} {Phys. Rev. D}\ }\textbf {\bibinfo {volume}
  {65}},\ \bibinfo {pages} {055003} (\bibinfo {year} {2002})},\ \Eprint
  {http://arxiv.org/abs/hep-ph/0110095} {arXiv:hep-ph/0110095} \BibitemShut
  {NoStop}%
\bibitem [{\citenamefont {Ma}\ \emph {et~al.}(2002)\citenamefont {Ma},
  \citenamefont {Roy},\ and\ \citenamefont {Roy}}]{Ma:2001md}%
  \BibitemOpen
  \bibfield  {author} {\bibinfo {author} {\bibfnamefont {E.}~\bibnamefont
  {Ma}}, \bibinfo {author} {\bibfnamefont {D.~P.}\ \bibnamefont {Roy}}, \ and\
  \bibinfo {author} {\bibfnamefont {S.}~\bibnamefont {Roy}},\ }\href {\doibase
  10.1016/S0370-2693(01)01428-9} {\bibfield  {journal} {\bibinfo  {journal}
  {Phys. Lett. B}\ }\textbf {\bibinfo {volume} {525}},\ \bibinfo {pages} {101}
  (\bibinfo {year} {2002})},\ \Eprint {http://arxiv.org/abs/hep-ph/0110146}
  {arXiv:hep-ph/0110146} \BibitemShut {NoStop}%
\bibitem [{\citenamefont {Bauer}\ \emph {et~al.}(2018)\citenamefont {Bauer},
  \citenamefont {Foldenauer},\ and\ \citenamefont {Jaeckel}}]{Bauer:2018onh}%
  \BibitemOpen
  \bibfield  {author} {\bibinfo {author} {\bibfnamefont {M.}~\bibnamefont
  {Bauer}}, \bibinfo {author} {\bibfnamefont {P.}~\bibnamefont {Foldenauer}}, \
  and\ \bibinfo {author} {\bibfnamefont {J.}~\bibnamefont {Jaeckel}},\ }\href
  {\doibase 10.1007/JHEP07(2018)094} {\bibfield  {journal} {\bibinfo  {journal}
  {JHEP}\ }\textbf {\bibinfo {volume} {07}},\ \bibinfo {pages} {094} (\bibinfo
  {year} {2018})},\ \Eprint {http://arxiv.org/abs/1803.05466} {arXiv:1803.05466
  [hep-ph]} \BibitemShut {NoStop}%
\bibitem [{\citenamefont {Dreiner}\ \emph {et~al.}(2010)\citenamefont
  {Dreiner}, \citenamefont {Haber},\ and\ \citenamefont
  {Martin}}]{Dreiner:2008tw}%
  \BibitemOpen
  \bibfield  {author} {\bibinfo {author} {\bibfnamefont {H.~K.}\ \bibnamefont
  {Dreiner}}, \bibinfo {author} {\bibfnamefont {H.~E.}\ \bibnamefont {Haber}},
  \ and\ \bibinfo {author} {\bibfnamefont {S.~P.}\ \bibnamefont {Martin}},\
  }\href {\doibase 10.1016/j.physrep.2010.05.002} {\bibfield  {journal}
  {\bibinfo  {journal} {Phys. Rept.}\ }\textbf {\bibinfo {volume} {494}},\
  \bibinfo {pages} {1} (\bibinfo {year} {2010})},\ \Eprint
  {http://arxiv.org/abs/0812.1594} {arXiv:0812.1594 [hep-ph]} \BibitemShut
  {NoStop}%
\bibitem [{\citenamefont {Harigaya}\ \emph {et~al.}(2014)\citenamefont
  {Harigaya}, \citenamefont {Igari}, \citenamefont {Nojiri}, \citenamefont
  {Takeuchi},\ and\ \citenamefont {Tobe}}]{Harigaya:2013twa}%
  \BibitemOpen
  \bibfield  {author} {\bibinfo {author} {\bibfnamefont {K.}~\bibnamefont
  {Harigaya}}, \bibinfo {author} {\bibfnamefont {T.}~\bibnamefont {Igari}},
  \bibinfo {author} {\bibfnamefont {M.~M.}\ \bibnamefont {Nojiri}}, \bibinfo
  {author} {\bibfnamefont {M.}~\bibnamefont {Takeuchi}}, \ and\ \bibinfo
  {author} {\bibfnamefont {K.}~\bibnamefont {Tobe}},\ }\href {\doibase
  10.1007/JHEP03(2014)105} {\bibfield  {journal} {\bibinfo  {journal} {JHEP}\
  }\textbf {\bibinfo {volume} {03}},\ \bibinfo {pages} {105} (\bibinfo {year}
  {2014})},\ \Eprint {http://arxiv.org/abs/1311.0870} {arXiv:1311.0870
  [hep-ph]} \BibitemShut {NoStop}%
\bibitem [{\citenamefont {Asai}\ \emph {et~al.}(2017)\citenamefont {Asai},
  \citenamefont {Hamaguchi},\ and\ \citenamefont {Nagata}}]{Asai:2017ryy}%
  \BibitemOpen
  \bibfield  {author} {\bibinfo {author} {\bibfnamefont {K.}~\bibnamefont
  {Asai}}, \bibinfo {author} {\bibfnamefont {K.}~\bibnamefont {Hamaguchi}}, \
  and\ \bibinfo {author} {\bibfnamefont {N.}~\bibnamefont {Nagata}},\ }\href
  {\doibase 10.1140/epjc/s10052-017-5348-x} {\bibfield  {journal} {\bibinfo
  {journal} {Eur. Phys. J. C}\ }\textbf {\bibinfo {volume} {77}},\ \bibinfo
  {pages} {763} (\bibinfo {year} {2017})},\ \Eprint
  {http://arxiv.org/abs/1705.00419} {arXiv:1705.00419 [hep-ph]} \BibitemShut
  {NoStop}%
\bibitem [{\citenamefont {Asai}\ \emph {et~al.}(2019)\citenamefont {Asai},
  \citenamefont {Hamaguchi}, \citenamefont {Nagata}, \citenamefont {Tseng},\
  and\ \citenamefont {Tsumura}}]{Asai:2018ocx}%
  \BibitemOpen
  \bibfield  {author} {\bibinfo {author} {\bibfnamefont {K.}~\bibnamefont
  {Asai}}, \bibinfo {author} {\bibfnamefont {K.}~\bibnamefont {Hamaguchi}},
  \bibinfo {author} {\bibfnamefont {N.}~\bibnamefont {Nagata}}, \bibinfo
  {author} {\bibfnamefont {S.-Y.}\ \bibnamefont {Tseng}}, \ and\ \bibinfo
  {author} {\bibfnamefont {K.}~\bibnamefont {Tsumura}},\ }\href {\doibase
  10.1103/PhysRevD.99.055029} {\bibfield  {journal} {\bibinfo  {journal} {Phys.
  Rev. D}\ }\textbf {\bibinfo {volume} {99}},\ \bibinfo {pages} {055029}
  (\bibinfo {year} {2019})},\ \Eprint {http://arxiv.org/abs/1811.07571}
  {arXiv:1811.07571 [hep-ph]} \BibitemShut {NoStop}%
\bibitem [{\citenamefont {Heeck}\ and\ \citenamefont
  {Rodejohann}(2011)}]{Heeck:2011wj}%
  \BibitemOpen
  \bibfield  {author} {\bibinfo {author} {\bibfnamefont {J.}~\bibnamefont
  {Heeck}}\ and\ \bibinfo {author} {\bibfnamefont {W.}~\bibnamefont
  {Rodejohann}},\ }\href {\doibase 10.1103/PhysRevD.84.075007} {\bibfield
  {journal} {\bibinfo  {journal} {Phys. Rev. D}\ }\textbf {\bibinfo {volume}
  {84}},\ \bibinfo {pages} {075007} (\bibinfo {year} {2011})},\ \Eprint
  {http://arxiv.org/abs/1107.5238} {arXiv:1107.5238 [hep-ph]} \BibitemShut
  {NoStop}%
\bibitem [{\citenamefont {Araki}\ \emph {et~al.}(2012)\citenamefont {Araki},
  \citenamefont {Heeck},\ and\ \citenamefont {Kubo}}]{Araki:2012ip}%
  \BibitemOpen
  \bibfield  {author} {\bibinfo {author} {\bibfnamefont {T.}~\bibnamefont
  {Araki}}, \bibinfo {author} {\bibfnamefont {J.}~\bibnamefont {Heeck}}, \ and\
  \bibinfo {author} {\bibfnamefont {J.}~\bibnamefont {Kubo}},\ }\href {\doibase
  10.1007/JHEP07(2012)083} {\bibfield  {journal} {\bibinfo  {journal} {JHEP}\
  }\textbf {\bibinfo {volume} {07}},\ \bibinfo {pages} {083} (\bibinfo {year}
  {2012})},\ \Eprint {http://arxiv.org/abs/1203.4951} {arXiv:1203.4951
  [hep-ph]} \BibitemShut {NoStop}%
\bibitem [{\citenamefont {Minkowski}(1977)}]{Minkowski:1977sc}%
  \BibitemOpen
  \bibfield  {author} {\bibinfo {author} {\bibfnamefont {P.}~\bibnamefont
  {Minkowski}},\ }\href {\doibase 10.1016/0370-2693(77)90435-X} {\bibfield
  {journal} {\bibinfo  {journal} {Phys. Lett.}\ }\textbf {\bibinfo {volume}
  {67B}},\ \bibinfo {pages} {421} (\bibinfo {year} {1977})}\BibitemShut
  {NoStop}%
\bibitem [{\citenamefont {Yanagida}(1979)}]{Yanagida:1979as}%
  \BibitemOpen
  \bibfield  {author} {\bibinfo {author} {\bibfnamefont {T.}~\bibnamefont
  {Yanagida}},\ }in\ \href@noop {} {\emph {\bibinfo {booktitle} {Horizontal
  gauge symmetry and masses of neutrinos}}},\ Vol.\ \bibinfo {volume}
  {7902131}\ (\bibinfo {year} {1979})\ pp.\ \bibinfo {pages}
  {95--99}\BibitemShut {NoStop}%
\bibitem [{\citenamefont {Glashow}(1980)}]{Glashow:1979nm}%
  \BibitemOpen
  \bibfield  {author} {\bibinfo {author} {\bibfnamefont {S.~L.}\ \bibnamefont
  {Glashow}},\ }\href {\doibase 10.1007/978-1-4684-7197-7_15} {\bibfield
  {journal} {\bibinfo  {journal} {NATO Sci. Ser. B}\ }\textbf {\bibinfo
  {volume} {61}},\ \bibinfo {pages} {687} (\bibinfo {year} {1980})}\BibitemShut
  {NoStop}%
\bibitem [{\citenamefont {Mohapatra}\ and\ \citenamefont
  {Senjanovic}(1980)}]{Mohapatra:1979ia}%
  \BibitemOpen
  \bibfield  {author} {\bibinfo {author} {\bibfnamefont {R.~N.}\ \bibnamefont
  {Mohapatra}}\ and\ \bibinfo {author} {\bibfnamefont {G.}~\bibnamefont
  {Senjanovic}},\ }\href {\doibase 10.1103/PhysRevLett.44.912} {\bibfield
  {journal} {\bibinfo  {journal} {Phys. Rev. Lett.}\ }\textbf {\bibinfo
  {volume} {44}},\ \bibinfo {pages} {912} (\bibinfo {year} {1980})},\ \bibinfo
  {note} {[,231(1979)]}\BibitemShut {NoStop}%
\bibitem [{\citenamefont {Gell-Mann}\ \emph {et~al.}(1979)\citenamefont
  {Gell-Mann}, \citenamefont {Ramond},\ and\ \citenamefont
  {Slansky}}]{GellMann:1980vs}%
  \BibitemOpen
  \bibfield  {author} {\bibinfo {author} {\bibfnamefont {M.}~\bibnamefont
  {Gell-Mann}}, \bibinfo {author} {\bibfnamefont {P.}~\bibnamefont {Ramond}}, \
  and\ \bibinfo {author} {\bibfnamefont {R.}~\bibnamefont {Slansky}},\
  }\bibfield  {booktitle} {\emph {\bibinfo {booktitle} {{Supergravity Workshop
  Stony Brook, New York, September 27-28, 1979}}},\ }\href@noop {} {\bibfield
  {journal} {\bibinfo  {journal} {Conf. Proc.}\ }\textbf {\bibinfo {volume}
  {C790927}},\ \bibinfo {pages} {315} (\bibinfo {year} {1979})},\ \Eprint
  {http://arxiv.org/abs/1306.4669} {arXiv:1306.4669 [hep-th]} \BibitemShut
  {NoStop}%
\bibitem [{\citenamefont {Schechter}\ and\ \citenamefont
  {Valle}(1980)}]{Schechter:1980gr}%
  \BibitemOpen
  \bibfield  {author} {\bibinfo {author} {\bibfnamefont {J.}~\bibnamefont
  {Schechter}}\ and\ \bibinfo {author} {\bibfnamefont {J.~W.~F.}\ \bibnamefont
  {Valle}},\ }\href {\doibase 10.1103/PhysRevD.22.2227} {\bibfield  {journal}
  {\bibinfo  {journal} {Phys. Rev.}\ }\textbf {\bibinfo {volume} {D22}},\
  \bibinfo {pages} {2227} (\bibinfo {year} {1980})}\BibitemShut {NoStop}%
\bibitem [{\citenamefont {Fukugita}\ and\ \citenamefont
  {Yanagida}(1986)}]{Fukugita:1986hr}%
  \BibitemOpen
  \bibfield  {author} {\bibinfo {author} {\bibfnamefont {M.}~\bibnamefont
  {Fukugita}}\ and\ \bibinfo {author} {\bibfnamefont {T.}~\bibnamefont
  {Yanagida}},\ }\href {\doibase 10.1016/0370-2693(86)91126-3} {\bibfield
  {journal} {\bibinfo  {journal} {Phys. Lett. B}\ }\textbf {\bibinfo {volume}
  {174}},\ \bibinfo {pages} {45} (\bibinfo {year} {1986})}\BibitemShut
  {NoStop}%
\bibitem [{\citenamefont {Kumekawa}\ \emph {et~al.}(1994)\citenamefont
  {Kumekawa}, \citenamefont {Moroi},\ and\ \citenamefont
  {Yanagida}}]{Kumekawa:1994gx}%
  \BibitemOpen
  \bibfield  {author} {\bibinfo {author} {\bibfnamefont {K.}~\bibnamefont
  {Kumekawa}}, \bibinfo {author} {\bibfnamefont {T.}~\bibnamefont {Moroi}}, \
  and\ \bibinfo {author} {\bibfnamefont {T.}~\bibnamefont {Yanagida}},\ }\href
  {\doibase 10.1143/PTP.92.437} {\bibfield  {journal} {\bibinfo  {journal}
  {Prog. Theor. Phys.}\ }\textbf {\bibinfo {volume} {92}},\ \bibinfo {pages}
  {437} (\bibinfo {year} {1994})},\ \Eprint
  {http://arxiv.org/abs/hep-ph/9405337} {arXiv:hep-ph/9405337} \BibitemShut
  {NoStop}%
\bibitem [{\citenamefont {Asaka}\ \emph {et~al.}(2000)\citenamefont {Asaka},
  \citenamefont {Hamaguchi}, \citenamefont {Kawasaki},\ and\ \citenamefont
  {Yanagida}}]{Asaka:1999jb}%
  \BibitemOpen
  \bibfield  {author} {\bibinfo {author} {\bibfnamefont {T.}~\bibnamefont
  {Asaka}}, \bibinfo {author} {\bibfnamefont {K.}~\bibnamefont {Hamaguchi}},
  \bibinfo {author} {\bibfnamefont {M.}~\bibnamefont {Kawasaki}}, \ and\
  \bibinfo {author} {\bibfnamefont {T.}~\bibnamefont {Yanagida}},\ }\href
  {\doibase 10.1103/PhysRevD.61.083512} {\bibfield  {journal} {\bibinfo
  {journal} {Phys. Rev. D}\ }\textbf {\bibinfo {volume} {61}},\ \bibinfo
  {pages} {083512} (\bibinfo {year} {2000})},\ \Eprint
  {http://arxiv.org/abs/hep-ph/9907559} {arXiv:hep-ph/9907559} \BibitemShut
  {NoStop}%
\bibitem [{\citenamefont {Lavoura}(2005)}]{Lavoura:2004tu}%
  \BibitemOpen
  \bibfield  {author} {\bibinfo {author} {\bibfnamefont {L.}~\bibnamefont
  {Lavoura}},\ }\href {\doibase 10.1016/j.physletb.2005.01.047} {\bibfield
  {journal} {\bibinfo  {journal} {Phys. Lett. B}\ }\textbf {\bibinfo {volume}
  {609}},\ \bibinfo {pages} {317} (\bibinfo {year} {2005})},\ \Eprint
  {http://arxiv.org/abs/hep-ph/0411232} {arXiv:hep-ph/0411232} \BibitemShut
  {NoStop}%
\bibitem [{\citenamefont {Lashin}\ and\ \citenamefont
  {Chamoun}(2008)}]{Lashin:2007dm}%
  \BibitemOpen
  \bibfield  {author} {\bibinfo {author} {\bibfnamefont {E.~I.}\ \bibnamefont
  {Lashin}}\ and\ \bibinfo {author} {\bibfnamefont {N.}~\bibnamefont
  {Chamoun}},\ }\href {\doibase 10.1103/PhysRevD.78.073002} {\bibfield
  {journal} {\bibinfo  {journal} {Phys. Rev. D}\ }\textbf {\bibinfo {volume}
  {78}},\ \bibinfo {pages} {073002} (\bibinfo {year} {2008})},\ \Eprint
  {http://arxiv.org/abs/0708.2423} {arXiv:0708.2423 [hep-ph]} \BibitemShut
  {NoStop}%
\bibitem [{\citenamefont {Altmannshofer}\ \emph {et~al.}(2014)\citenamefont
  {Altmannshofer}, \citenamefont {Gori}, \citenamefont {Pospelov},\ and\
  \citenamefont {Yavin}}]{Altmannshofer:2014pba}%
  \BibitemOpen
  \bibfield  {author} {\bibinfo {author} {\bibfnamefont {W.}~\bibnamefont
  {Altmannshofer}}, \bibinfo {author} {\bibfnamefont {S.}~\bibnamefont {Gori}},
  \bibinfo {author} {\bibfnamefont {M.}~\bibnamefont {Pospelov}}, \ and\
  \bibinfo {author} {\bibfnamefont {I.}~\bibnamefont {Yavin}},\ }\href
  {\doibase 10.1103/PhysRevLett.113.091801} {\bibfield  {journal} {\bibinfo
  {journal} {Phys. Rev. Lett.}\ }\textbf {\bibinfo {volume} {113}},\ \bibinfo
  {pages} {091801} (\bibinfo {year} {2014})},\ \Eprint
  {http://arxiv.org/abs/1406.2332} {arXiv:1406.2332 [hep-ph]} \BibitemShut
  {NoStop}%
\bibitem [{\citenamefont {Harnik}\ \emph {et~al.}(2012)\citenamefont {Harnik},
  \citenamefont {Kopp},\ and\ \citenamefont {Machado}}]{Harnik:2012ni}%
  \BibitemOpen
  \bibfield  {author} {\bibinfo {author} {\bibfnamefont {R.}~\bibnamefont
  {Harnik}}, \bibinfo {author} {\bibfnamefont {J.}~\bibnamefont {Kopp}}, \ and\
  \bibinfo {author} {\bibfnamefont {P.~A.~N.}\ \bibnamefont {Machado}},\ }\href
  {\doibase 10.1088/1475-7516/2012/07/026} {\bibfield  {journal} {\bibinfo
  {journal} {JCAP}\ }\textbf {\bibinfo {volume} {07}},\ \bibinfo {pages} {026}
  (\bibinfo {year} {2012})},\ \Eprint {http://arxiv.org/abs/1202.6073}
  {arXiv:1202.6073 [hep-ph]} \BibitemShut {NoStop}%
\bibitem [{\citenamefont {Bilmis}\ \emph {et~al.}(2015)\citenamefont {Bilmis},
  \citenamefont {Turan}, \citenamefont {Aliev}, \citenamefont {Deniz},
  \citenamefont {Singh},\ and\ \citenamefont {Wong}}]{Bilmis:2015lja}%
  \BibitemOpen
  \bibfield  {author} {\bibinfo {author} {\bibfnamefont {S.}~\bibnamefont
  {Bilmis}}, \bibinfo {author} {\bibfnamefont {I.}~\bibnamefont {Turan}},
  \bibinfo {author} {\bibfnamefont {T.~M.}\ \bibnamefont {Aliev}}, \bibinfo
  {author} {\bibfnamefont {M.}~\bibnamefont {Deniz}}, \bibinfo {author}
  {\bibfnamefont {L.}~\bibnamefont {Singh}}, \ and\ \bibinfo {author}
  {\bibfnamefont {H.~T.}\ \bibnamefont {Wong}},\ }\href {\doibase
  10.1103/PhysRevD.92.033009} {\bibfield  {journal} {\bibinfo  {journal} {Phys.
  Rev. D}\ }\textbf {\bibinfo {volume} {92}},\ \bibinfo {pages} {033009}
  (\bibinfo {year} {2015})},\ \Eprint {http://arxiv.org/abs/1502.07763}
  {arXiv:1502.07763 [hep-ph]} \BibitemShut {NoStop}%
\bibitem [{\citenamefont {Lees}\ \emph {et~al.}(2016)\citenamefont {Lees} \emph
  {et~al.}}]{BaBar:2016sci}%
  \BibitemOpen
  \bibfield  {author} {\bibinfo {author} {\bibfnamefont {J.~P.}\ \bibnamefont
  {Lees}} \emph {et~al.} (\bibinfo {collaboration} {BaBar}),\ }\href {\doibase
  10.1103/PhysRevD.94.011102} {\bibfield  {journal} {\bibinfo  {journal} {Phys.
  Rev. D}\ }\textbf {\bibinfo {volume} {94}},\ \bibinfo {pages} {011102}
  (\bibinfo {year} {2016})},\ \Eprint {http://arxiv.org/abs/1606.03501}
  {arXiv:1606.03501 [hep-ex]} \BibitemShut {NoStop}%
\bibitem [{\citenamefont {Patrignani}\ \emph {et~al.}(2016)\citenamefont
  {Patrignani} \emph {et~al.}}]{ParticleDataGroup:2016lqr}%
  \BibitemOpen
  \bibfield  {author} {\bibinfo {author} {\bibfnamefont {C.}~\bibnamefont
  {Patrignani}} \emph {et~al.} (\bibinfo {collaboration} {Particle Data
  Group}),\ }\href {\doibase 10.1088/1674-1137/40/10/100001} {\bibfield
  {journal} {\bibinfo  {journal} {Chin. Phys. C}\ }\textbf {\bibinfo {volume}
  {40}},\ \bibinfo {pages} {100001} (\bibinfo {year} {2016})}\BibitemShut
  {NoStop}%
\bibitem [{\citenamefont {Esteban}\ \emph {et~al.}(2020)\citenamefont
  {Esteban}, \citenamefont {Gonzalez-Garcia}, \citenamefont {Maltoni},
  \citenamefont {Schwetz},\ and\ \citenamefont {Zhou}}]{Esteban:2020cvm}%
  \BibitemOpen
  \bibfield  {author} {\bibinfo {author} {\bibfnamefont {I.}~\bibnamefont
  {Esteban}}, \bibinfo {author} {\bibfnamefont {M.~C.}\ \bibnamefont
  {Gonzalez-Garcia}}, \bibinfo {author} {\bibfnamefont {M.}~\bibnamefont
  {Maltoni}}, \bibinfo {author} {\bibfnamefont {T.}~\bibnamefont {Schwetz}}, \
  and\ \bibinfo {author} {\bibfnamefont {A.}~\bibnamefont {Zhou}},\ }\href
  {\doibase 10.1007/JHEP09(2020)178} {\bibfield  {journal} {\bibinfo  {journal}
  {JHEP}\ }\textbf {\bibinfo {volume} {09}},\ \bibinfo {pages} {178} (\bibinfo
  {year} {2020})},\ \Eprint {http://arxiv.org/abs/2007.14792} {arXiv:2007.14792
  [hep-ph]} \BibitemShut {NoStop}%
\bibitem [{\citenamefont {Aghanim}\ \emph {et~al.}(2020)\citenamefont {Aghanim}
  \emph {et~al.}}]{Planck:2018vyg}%
  \BibitemOpen
  \bibfield  {author} {\bibinfo {author} {\bibfnamefont {N.}~\bibnamefont
  {Aghanim}} \emph {et~al.} (\bibinfo {collaboration} {Planck}),\ }\href
  {\doibase 10.1051/0004-6361/201833910} {\bibfield  {journal} {\bibinfo
  {journal} {Astron. Astrophys.}\ }\textbf {\bibinfo {volume} {641}},\ \bibinfo
  {pages} {A6} (\bibinfo {year} {2020})},\ \bibinfo {note} {[Erratum:
  Astron.Astrophys. 652, C4 (2021)]},\ \Eprint
  {http://arxiv.org/abs/1807.06209} {arXiv:1807.06209 [astro-ph.CO]}
  \BibitemShut {NoStop}%
\bibitem [{\citenamefont {Baldini}\ \emph {et~al.}(2016)\citenamefont {Baldini}
  \emph {et~al.}}]{MEG:2016leq}%
  \BibitemOpen
  \bibfield  {author} {\bibinfo {author} {\bibfnamefont {A.~M.}\ \bibnamefont
  {Baldini}} \emph {et~al.} (\bibinfo {collaboration} {MEG}),\ }\href {\doibase
  10.1140/epjc/s10052-016-4271-x} {\bibfield  {journal} {\bibinfo  {journal}
  {Eur. Phys. J. C}\ }\textbf {\bibinfo {volume} {76}},\ \bibinfo {pages} {434}
  (\bibinfo {year} {2016})},\ \Eprint {http://arxiv.org/abs/1605.05081}
  {arXiv:1605.05081 [hep-ex]} \BibitemShut {NoStop}%
\bibitem [{\citenamefont {Aubert}\ \emph {et~al.}(2010)\citenamefont {Aubert}
  \emph {et~al.}}]{BaBar:2009hkt}%
  \BibitemOpen
  \bibfield  {author} {\bibinfo {author} {\bibfnamefont {B.}~\bibnamefont
  {Aubert}} \emph {et~al.} (\bibinfo {collaboration} {BaBar}),\ }\href
  {\doibase 10.1103/PhysRevLett.104.021802} {\bibfield  {journal} {\bibinfo
  {journal} {Phys. Rev. Lett.}\ }\textbf {\bibinfo {volume} {104}},\ \bibinfo
  {pages} {021802} (\bibinfo {year} {2010})},\ \Eprint
  {http://arxiv.org/abs/0908.2381} {arXiv:0908.2381 [hep-ex]} \BibitemShut
  {NoStop}%
\bibitem [{\citenamefont {Abe}\ \emph {et~al.}(2022)\citenamefont {Abe} \emph
  {et~al.}}]{KamLAND-Zen:2022tow}%
  \BibitemOpen
  \bibfield  {author} {\bibinfo {author} {\bibfnamefont {S.}~\bibnamefont
  {Abe}} \emph {et~al.} (\bibinfo {collaboration} {KamLAND-Zen}),\ }\href@noop
  {} {\bibfield  {journal} {\bibinfo  {journal} {arXiv preprint}\ } (\bibinfo
  {year} {2022})},\ \Eprint {http://arxiv.org/abs/2203.02139} {arXiv:2203.02139
  [hep-ex]} \BibitemShut {NoStop}%
\bibitem [{\citenamefont {Agostini}\ \emph {et~al.}(2020)\citenamefont
  {Agostini} \emph {et~al.}}]{GERDA:2020xhi}%
  \BibitemOpen
  \bibfield  {author} {\bibinfo {author} {\bibfnamefont {M.}~\bibnamefont
  {Agostini}} \emph {et~al.} (\bibinfo {collaboration} {GERDA}),\ }\href
  {\doibase 10.1103/PhysRevLett.125.252502} {\bibfield  {journal} {\bibinfo
  {journal} {Phys. Rev. Lett.}\ }\textbf {\bibinfo {volume} {125}},\ \bibinfo
  {pages} {252502} (\bibinfo {year} {2020})},\ \Eprint
  {http://arxiv.org/abs/2009.06079} {arXiv:2009.06079 [nucl-ex]} \BibitemShut
  {NoStop}%
\bibitem [{\citenamefont {Escudero}\ \emph {et~al.}(2019)\citenamefont
  {Escudero}, \citenamefont {Hooper}, \citenamefont {Krnjaic},\ and\
  \citenamefont {Pierre}}]{Escudero:2019gzq}%
  \BibitemOpen
  \bibfield  {author} {\bibinfo {author} {\bibfnamefont {M.}~\bibnamefont
  {Escudero}}, \bibinfo {author} {\bibfnamefont {D.}~\bibnamefont {Hooper}},
  \bibinfo {author} {\bibfnamefont {G.}~\bibnamefont {Krnjaic}}, \ and\
  \bibinfo {author} {\bibfnamefont {M.}~\bibnamefont {Pierre}},\ }\href
  {\doibase 10.1007/JHEP03(2019)071} {\bibfield  {journal} {\bibinfo  {journal}
  {JHEP}\ }\textbf {\bibinfo {volume} {03}},\ \bibinfo {pages} {071} (\bibinfo
  {year} {2019})},\ \Eprint {http://arxiv.org/abs/1901.02010} {arXiv:1901.02010
  [hep-ph]} \BibitemShut {NoStop}%
\bibitem [{\citenamefont {Albert}(2021)}]{Albert:2021wjs}%
  \BibitemOpen
  \bibfield  {author} {\bibinfo {author} {\bibfnamefont {A.}~\bibnamefont
  {Albert}} (\bibinfo {collaboration} {ATLAS, CMS}),\ }\href {\doibase
  10.22323/1.397.0076} {\bibfield  {journal} {\bibinfo  {journal} {PoS}\
  }\textbf {\bibinfo {volume} {LHCP2021}},\ \bibinfo {pages} {076} (\bibinfo
  {year} {2021})}\BibitemShut {NoStop}%
\bibitem [{\citenamefont {Nomura}\ and\ \citenamefont
  {Shimomura}(2019)}]{Nomura:2018yej}%
  \BibitemOpen
  \bibfield  {author} {\bibinfo {author} {\bibfnamefont {T.}~\bibnamefont
  {Nomura}}\ and\ \bibinfo {author} {\bibfnamefont {T.}~\bibnamefont
  {Shimomura}},\ }\href {\doibase 10.1140/epjc/s10052-019-7094-8} {\bibfield
  {journal} {\bibinfo  {journal} {Eur. Phys. J. C}\ }\textbf {\bibinfo {volume}
  {79}},\ \bibinfo {pages} {594} (\bibinfo {year} {2019})},\ \Eprint
  {http://arxiv.org/abs/1803.00842} {arXiv:1803.00842 [hep-ph]} \BibitemShut
  {NoStop}%
\bibitem [{\citenamefont {Adhikary}(2006)}]{Adhikary:2006rf}%
  \BibitemOpen
  \bibfield  {author} {\bibinfo {author} {\bibfnamefont {B.}~\bibnamefont
  {Adhikary}},\ }\href {\doibase 10.1103/PhysRevD.74.033002} {\bibfield
  {journal} {\bibinfo  {journal} {Phys. Rev. D}\ }\textbf {\bibinfo {volume}
  {74}},\ \bibinfo {pages} {033002} (\bibinfo {year} {2006})},\ \Eprint
  {http://arxiv.org/abs/hep-ph/0604009} {arXiv:hep-ph/0604009} \BibitemShut
  {NoStop}%
\bibitem [{\citenamefont {Akhmedov}\ \emph {et~al.}(1998)\citenamefont
  {Akhmedov}, \citenamefont {Rubakov},\ and\ \citenamefont
  {Smirnov}}]{Akhmedov:1998qx}%
  \BibitemOpen
  \bibfield  {author} {\bibinfo {author} {\bibfnamefont {E.~K.}\ \bibnamefont
  {Akhmedov}}, \bibinfo {author} {\bibfnamefont {V.~A.}\ \bibnamefont
  {Rubakov}}, \ and\ \bibinfo {author} {\bibfnamefont {A.~{\relax Yu}.}\
  \bibnamefont {Smirnov}},\ }\href {\doibase 10.1103/PhysRevLett.81.1359}
  {\bibfield  {journal} {\bibinfo  {journal} {Phys. Rev. Lett.}\ }\textbf
  {\bibinfo {volume} {81}},\ \bibinfo {pages} {1359} (\bibinfo {year}
  {1998})},\ \Eprint {http://arxiv.org/abs/hep-ph/9803255}
  {arXiv:hep-ph/9803255 [hep-ph]} \BibitemShut {NoStop}%
\bibitem [{\citenamefont {Asaka}\ and\ \citenamefont
  {Shaposhnikov}(2005)}]{Asaka:2005pn}%
  \BibitemOpen
  \bibfield  {author} {\bibinfo {author} {\bibfnamefont {T.}~\bibnamefont
  {Asaka}}\ and\ \bibinfo {author} {\bibfnamefont {M.}~\bibnamefont
  {Shaposhnikov}},\ }\href {\doibase 10.1016/j.physletb.2005.06.020} {\bibfield
   {journal} {\bibinfo  {journal} {Phys. Lett.}\ }\textbf {\bibinfo {volume}
  {B620}},\ \bibinfo {pages} {17} (\bibinfo {year} {2005})},\ \Eprint
  {http://arxiv.org/abs/hep-ph/0505013} {arXiv:hep-ph/0505013 [hep-ph]}
  \BibitemShut {NoStop}%
\bibitem [{\citenamefont {Pilaftsis}\ and\ \citenamefont
  {Underwood}(2004)}]{Pilaftsis:2003gt}%
  \BibitemOpen
  \bibfield  {author} {\bibinfo {author} {\bibfnamefont {A.}~\bibnamefont
  {Pilaftsis}}\ and\ \bibinfo {author} {\bibfnamefont {T.~E.~J.}\ \bibnamefont
  {Underwood}},\ }\href {\doibase 10.1016/j.nuclphysb.2004.05.029} {\bibfield
  {journal} {\bibinfo  {journal} {Nucl. Phys. B}\ }\textbf {\bibinfo {volume}
  {692}},\ \bibinfo {pages} {303} (\bibinfo {year} {2004})},\ \Eprint
  {http://arxiv.org/abs/hep-ph/0309342} {arXiv:hep-ph/0309342} \BibitemShut
  {NoStop}%
\bibitem [{\citenamefont {Giudice}\ \emph {et~al.}(2004)\citenamefont
  {Giudice}, \citenamefont {Notari}, \citenamefont {Raidal}, \citenamefont
  {Riotto},\ and\ \citenamefont {Strumia}}]{Giudice:2003jh}%
  \BibitemOpen
  \bibfield  {author} {\bibinfo {author} {\bibfnamefont {G.~F.}\ \bibnamefont
  {Giudice}}, \bibinfo {author} {\bibfnamefont {A.}~\bibnamefont {Notari}},
  \bibinfo {author} {\bibfnamefont {M.}~\bibnamefont {Raidal}}, \bibinfo
  {author} {\bibfnamefont {A.}~\bibnamefont {Riotto}}, \ and\ \bibinfo {author}
  {\bibfnamefont {A.}~\bibnamefont {Strumia}},\ }\href {\doibase
  10.1016/j.nuclphysb.2004.02.019} {\bibfield  {journal} {\bibinfo  {journal}
  {Nucl. Phys. B}\ }\textbf {\bibinfo {volume} {685}},\ \bibinfo {pages} {89}
  (\bibinfo {year} {2004})},\ \Eprint {http://arxiv.org/abs/hep-ph/0310123}
  {arXiv:hep-ph/0310123} \BibitemShut {NoStop}%
\bibitem [{\citenamefont {Borah}\ \emph {et~al.}(2021)\citenamefont {Borah},
  \citenamefont {Dasgupta},\ and\ \citenamefont {Mahanta}}]{Borah:2021mri}%
  \BibitemOpen
  \bibfield  {author} {\bibinfo {author} {\bibfnamefont {D.}~\bibnamefont
  {Borah}}, \bibinfo {author} {\bibfnamefont {A.}~\bibnamefont {Dasgupta}}, \
  and\ \bibinfo {author} {\bibfnamefont {D.}~\bibnamefont {Mahanta}},\ }\href
  {\doibase 10.1103/PhysRevD.104.075006} {\bibfield  {journal} {\bibinfo
  {journal} {Phys. Rev. D}\ }\textbf {\bibinfo {volume} {104}},\ \bibinfo
  {pages} {075006} (\bibinfo {year} {2021})},\ \Eprint
  {http://arxiv.org/abs/2106.14410} {arXiv:2106.14410 [hep-ph]} \BibitemShut
  {NoStop}%
\bibitem [{\citenamefont {Abdullah}\ \emph {et~al.}(2018)\citenamefont
  {Abdullah}, \citenamefont {Dent}, \citenamefont {Dutta}, \citenamefont
  {Kane}, \citenamefont {Liao},\ and\ \citenamefont
  {Strigari}}]{Abdullah:2018ykz}%
  \BibitemOpen
  \bibfield  {author} {\bibinfo {author} {\bibfnamefont {M.}~\bibnamefont
  {Abdullah}}, \bibinfo {author} {\bibfnamefont {J.~B.}\ \bibnamefont {Dent}},
  \bibinfo {author} {\bibfnamefont {B.}~\bibnamefont {Dutta}}, \bibinfo
  {author} {\bibfnamefont {G.~L.}\ \bibnamefont {Kane}}, \bibinfo {author}
  {\bibfnamefont {S.}~\bibnamefont {Liao}}, \ and\ \bibinfo {author}
  {\bibfnamefont {L.~E.}\ \bibnamefont {Strigari}},\ }\href {\doibase
  10.1103/PhysRevD.98.015005} {\bibfield  {journal} {\bibinfo  {journal} {Phys.
  Rev. D}\ }\textbf {\bibinfo {volume} {98}},\ \bibinfo {pages} {015005}
  (\bibinfo {year} {2018})},\ \Eprint {http://arxiv.org/abs/1803.01224}
  {arXiv:1803.01224 [hep-ph]} \BibitemShut {NoStop}%
\bibitem [{\citenamefont {Gninenko}\ \emph {et~al.}(2015)\citenamefont
  {Gninenko}, \citenamefont {Krasnikov},\ and\ \citenamefont
  {Matveev}}]{Gninenko:2014pea}%
  \BibitemOpen
  \bibfield  {author} {\bibinfo {author} {\bibfnamefont {S.~N.}\ \bibnamefont
  {Gninenko}}, \bibinfo {author} {\bibfnamefont {N.~V.}\ \bibnamefont
  {Krasnikov}}, \ and\ \bibinfo {author} {\bibfnamefont {V.~A.}\ \bibnamefont
  {Matveev}},\ }\href {\doibase 10.1103/PhysRevD.91.095015} {\bibfield
  {journal} {\bibinfo  {journal} {Phys. Rev. D}\ }\textbf {\bibinfo {volume}
  {91}},\ \bibinfo {pages} {095015} (\bibinfo {year} {2015})},\ \Eprint
  {http://arxiv.org/abs/1412.1400} {arXiv:1412.1400 [hep-ph]} \BibitemShut
  {NoStop}%
\bibitem [{\citenamefont {Gninenko}\ and\ \citenamefont
  {Krasnikov}(2018)}]{Gninenko:2018tlp}%
  \BibitemOpen
  \bibfield  {author} {\bibinfo {author} {\bibfnamefont {S.~N.}\ \bibnamefont
  {Gninenko}}\ and\ \bibinfo {author} {\bibfnamefont {N.~V.}\ \bibnamefont
  {Krasnikov}},\ }\href {\doibase 10.1016/j.physletb.2018.06.043} {\bibfield
  {journal} {\bibinfo  {journal} {Phys. Lett. B}\ }\textbf {\bibinfo {volume}
  {783}},\ \bibinfo {pages} {24} (\bibinfo {year} {2018})},\ \Eprint
  {http://arxiv.org/abs/1801.10448} {arXiv:1801.10448 [hep-ph]} \BibitemShut
  {NoStop}%
\bibitem [{\citenamefont {Abazajian}\ \emph {et~al.}(2019)\citenamefont
  {Abazajian} \emph {et~al.}}]{Abazajian:2019eic}%
  \BibitemOpen
  \bibfield  {author} {\bibinfo {author} {\bibfnamefont {K.}~\bibnamefont
  {Abazajian}} \emph {et~al.},\ }\href@noop {} {\bibfield  {journal} {\bibinfo
  {journal} {arXiv preprint}\ } (\bibinfo {year} {2019})},\ \Eprint
  {http://arxiv.org/abs/1907.04473} {arXiv:1907.04473 [astro-ph.IM]}
  \BibitemShut {NoStop}%
\bibitem [{\citenamefont {Winkler}(2019)}]{Winkler:2018qyg}%
  \BibitemOpen
  \bibfield  {author} {\bibinfo {author} {\bibfnamefont {M.~W.}\ \bibnamefont
  {Winkler}},\ }\href {\doibase 10.1103/PhysRevD.99.015018} {\bibfield
  {journal} {\bibinfo  {journal} {Phys. Rev. D}\ }\textbf {\bibinfo {volume}
  {99}},\ \bibinfo {pages} {015018} (\bibinfo {year} {2019})},\ \Eprint
  {http://arxiv.org/abs/1809.01876} {arXiv:1809.01876 [hep-ph]} \BibitemShut
  {NoStop}%
\bibitem [{\citenamefont {Dittmaier}\ \emph {et~al.}(2011)\citenamefont
  {Dittmaier} \emph {et~al.}}]{LHCHiggsCrossSectionWorkingGroup:2011wcg}%
  \BibitemOpen
  \bibfield  {author} {\bibinfo {author} {\bibfnamefont {S.}~\bibnamefont
  {Dittmaier}} \emph {et~al.} (\bibinfo {collaboration} {LHC Higgs Cross
  Section Working Group}),\ }\href {\doibase 10.5170/CERN-2011-002} {\
  (\bibinfo {year} {2011}),\ 10.5170/CERN-2011-002},\ \Eprint
  {http://arxiv.org/abs/1101.0593} {arXiv:1101.0593 [hep-ph]} \BibitemShut
  {NoStop}%
\end{thebibliography}%

\end{document}